\documentclass[a4paper,11pt]{article}

\usepackage{jheppub}
\usepackage{amsmath,amssymb,amsfonts,graphicx}
\usepackage{braket}
\usepackage{bm}
\usepackage{amsthm}
\usepackage[caption=false]{subfig}
\usepackage{float}
\usepackage{dsfont}
\usepackage{amsmath}
\usepackage{array}
\usepackage[dvipsnames]{xcolor}

\renewcommand\bra[1]{{\langle{#1}|}}
\renewcommand\ket[1]{{|{#1}\rangle}}

\newcommand{\arcsinh}{\hspace{0.08cm}\textrm{arcsinh}}

\newtheorem{theorem}{Theorem}[section]
\newtheorem{corollary}{Corollary}[theorem]

\newtheorem{lemma}{Lemma}[section]
\newenvironment{hproof}{%
  \proof}{\endproof}
  
\DeclareMathOperator{\sn}{sn}

\def\bea{\begin{eqnarray}}
\def\eea{\end{eqnarray}}
\def\nn{\nonumber}
\def\ba{\begin{array}}
\def\ea{\end{array}}
\def\nn{\nonumber}
\def\Tr{\text{Tr}}

\title{\boldmath Holographic measurement and bulk teleportation}

\author[1]{Stefano Antonini,}
\author[2]{Gregory Bentsen,}
\author[3,4]{ChunJun Cao,}
\author[2,5]{Jonathan Harper,}
\author[2]{Shao-Kai Jian,}
\author[1,2]{Brian Swingle,}

\affiliation[1]{Maryland Center for Fundamental Physics, University of Maryland, College Park, MD 20742, USA}
 \affiliation[2]{Department of Physics, Brandeis University, Waltham, MA 02453, USA}
 \affiliation[3]{Joint Center for Quantum Information and Computer Science, University of Maryland, College Park, MD 20742, USA}
 \affiliation[4]{Institute for Quantum Information and Matter, California Institute of Technology, Pasadena, CA
91125, USA}
\affiliation[5]{Center for Gravitational Physics and Quantum Information, Yukawa Institute for Theoretical Physics, Kyoto University, Kitashirakawa Oiwakecho, Sakyo-ku, Kyoto 606-8502, Japan}
\preprint{BRX-TH-6710, YITP-22-96}
\emailAdd{santonin@umd.edu}
\emailAdd{gbentsen@brandeis.edu}
\emailAdd{ccj991@gmail.com}
\emailAdd{jonathan.harper@yukawa.kyoto-u.ac.jp}
\emailAdd{skjian@brandeis.edu}
\emailAdd{bswingle@brandeis.edu}

\abstract{Holography has taught us that spacetime is emergent and its properties depend on the entanglement structure of the dual theory. In this paper, we describe how changes in the entanglement due to a local projective measurement (LPM) on a subregion $A$ of the boundary theory modify the bulk dual spacetime. We find that LPMs destroy portions of the bulk geometry, yielding post-measurement bulk spacetimes dual to the complementary unmeasured region $A^c$ that are cut off by end-of-the-world branes. Using a bulk calculation in $AdS_3$ and tensor network models of holography, we show that the portions of the bulk geometry that are preserved after the measurement depend on the size of $A$ and the state we project onto. The post-measurement bulk dual to $A^c$ includes regions that were originally part of the entanglement wedge of $A$ prior to measurement. This suggests that LPMs performed on a boundary subregion $A$ teleport part of the bulk information originally encoded in $A$ into the complementary region $A^c$. In semiclassical holography an arbitrary amount of bulk information can be teleported in this way, while in tensor network models the teleported information is upper-bounded by the amount of entanglement shared between $A$ and $A^c$ due to finite-$N$ effects. When $A$ is the union of two disjoint subregions, the measurement triggers an entangled/disentangled phase transition between the remaining two unmeasured subregions, corresponding to a connected/disconnected phase transition in the bulk description. Our results shed new light on the effects of measurement on the entanglement structure of holographic theories and give insight on how bulk information can be manipulated from the boundary theory. They could also be extended to more general quantum systems and tested experimentally, and represent a first step towards a holographic description of measurement-induced phase transitions.}

\begin{document} 
\maketitle
\flushbottom

\newpage

\section{Introduction}

In the last few decades the Anti de-Sitter/Conformal Field Theory (AdS/CFT) correspondence \cite{Maldacena:1997re,Witten:1998qj,Gubser:1998bc,Aharony:1999ti} has emerged as a prominent framework for the formulation of a quantum theory of gravity. Motivated by the holographic principle \cite{tHooft:1993dmi,Susskind:1994vu}, AdS/CFT relates a quantum gravity theory in an asymptotically AdS spacetime to a UV-complete, non-gravitational quantum mechanical theory living on the AdS boundary. Among other favorable features, AdS/CFT establishes an explicit connection between quantum information-theoretic observables in the boundary CFT and spacetime features in the bulk dual theory (see for instance \cite{Ryu2006a,Ryu2006b,Hubeny:2007xt,Engelhardt:2014gca,Stanford:2014jda,Brown:2015bva,Brown:2019rox}). A particularly important lesson we have learned from AdS/CFT is that spacetime emerges from quantum entanglement present in the boundary theory \cite{Swingle:2009bg,VanRaamsdonk:2010pw,Maldacena:2013xja}. This feature is best understood in the context of holographic quantum error correcting codes (QECC) \cite{Almheiri:2014lwa,Harlow:2018fse} and especially in explicit tensor network toy models of holography \cite{Swingle:2009bg,Pastawski:2015qua,hayden2016holographic}.

One important question related to spacetime emergence is how manipulations of the boundary theory and its entanglement structure affect the bulk dual description. In the present paper, we start investigating this problem by studying the effects that local projective measurements (LPM) performed on subregions of the boundary theory have on the bulk spacetime. A LPM can be thought of as the continuum limit of a projective measurement performed on a lattice regularization of the boundary CFT, where lattice sites in the measured region $A$ are projected onto a product state (see Section \ref{slitsection})\footnote{More precisely, in this work we post-select region $A$ onto some specific state.}. The LPM naturally projects the measured region $A$ onto a pure state.\footnote{If we start in a pure state for the whole system, as is the case for the setups we analyze, the post-measurement state of the complementary unmeasured region $A^c$ is also pure.} How does such a LPM affect the correponding bulk geometry? Here we investigate this question using a collection of analytical and numerical tools. In Section \ref{sec:holographiccalculation} we consider a specific $AdS_3/CFT_2$ setup and  make use of a prescription developed in \cite{namasawa2016epr}---based on the AdS/BCFT proposal \cite{takayanagi2011holographic,fujita2011aspects}---to construct the bulk dual spacetime after a boundary LPM is performed and analyze the entanglement properties of the post-measurement system. An analogous form of measurement is also naturally realized in holographic tensor network constructions thanks to their discrete nature. In this case, the LPM corresponds to a projective measurement on a product basis of the external legs of the network. We explore in detail the effects of such a measurement on the entanglement structure of generic QECC (Section \ref{sec:boundmeasuremteleportation}), and then apply our results to the HaPPY code \cite{Pastawski:2015qua} (Section \ref{sec:happycode}) and random tensor networks (RTN) (Section \ref{sec:rtn}), finding results qualitatively compatible with the ones arising from our holographic construction.

What lessons can we expect to learn from this analysis? First, we are motivated by holographic questions related to entanglement wedge reconstruction \cite{Dong:2016eik,Harlow:2016vwg}, which guarantees that a boundary observer having control over a subregion $A$ of the boundary theory can potentially access bulk information contained anywhere in its entanglement wedge $W(A)$. What happens to the bulk geometry if that boundary observer chooses to projectively collapse a large fraction of their qubits? Or, by making a sufficiently complicated measurement, could that bulk observer measure (or collapse) a selected local region of the bulk geometry? In order to reconstruct bulk degrees of freedom lying deep within the entanglement wedge, we know that the boundary observer must perform a highly entangled nonlocal measurement on the boundary region $A$. In the language of QECCs, they must measure a bulk logical operator, which is mapped into a complicated non-local operator on $A$. When such a measurement is performed, the information contained in the entanglement wedge is extracted and (at least) the corresponding portion of bulk spacetime is destroyed.\footnote{Since this highly correlated measurement radically modifies the state also on $A^c$, it is possible that there is no geometrical description at all for the post-measurement state.} Our LPM, however, is a completely uncorrelated measurement, and we do not expect it to be able to extract information contained deep into the bulk.\footnote{Although we considered only local projective measurements of the type described above, we expect results similar to the ones we find in this paper to hold also for other types of measurement, including weakly correlated measurements.} Therefore, two possible scenarios are conceivable after a LPM is performed on region $A$: either the bulk information contained in $W(A)$ is lost\footnote{From a QECC point of view, we can see the LPM as an ``error'': in the first scenario described here, the error cannot be corrected, and the logical information (i.e. the bulk information) is lost.}, or it is now accessible from the complementary region $A^c$. The fact that the LPM destroys a large amount of entanglement in the boundary theory seems to support the first hypothesis. On the other hand, the QECC interpretation of holography indicates that degrees of freedom living deep into the bulk are highly protected from errors occurring on the physical qubits, including completely destructive errors such as projective measurements. Our results clarify under which conditions each one of the two scenarios above is realized.

A second broad motivation is to understand the dynamics of quantum information following projective measurement. Whereas in most cases your everyday projective measurement does nothing more than collapse a fragile quantum state, a well-placed series of measurements can instead be harnessed to teleport \cite{bennett1993teleporting}, process \cite{raussendorf2001one}, or rapidly delocalize \cite{lu2022measurement} quantum information. We see these complementary roles of projective measurement appear explicitly in our holographic setups below. For instance, the situation in which the bulk information contained in $W(A)$ before the LPM is accessible from $A^c$ after the measurement is particularly interesting. In fact, in this case the effect of the boundary measurement is to teleport bulk information from the boundary region $A$ to its complement $A^c$. In Section \ref{sec:holographiccalculation} we observe this teleportation from a bulk point of view by studying a holographic $AdS_3/CFT_2$ setup. In Section \ref{sec:boundmeasuremteleportation} we are able to explicitly study how the teleportation occurs and how much bulk information can be teleported in quantum error correcting codes. The application of these arguments to the HaPPY code in Section \ref{sec:happycode} allows us to make contact with the holographic results. The numerical results for the HaPPY code reported in Section \ref{happynumerics} support our analysis, while additional insight on the bulk teleportation is provided by random tensor network models, studied in Section \ref{sec:rtn}.

As we will discuss in Section \ref{sec:boundmeasuremteleportation}, the amount of bulk information (and the size of the bulk region) which can be teleported using a projective measurement on the boundary is upper-bounded by the entanglement entropy $S(A)$ of region $A$ for a pure state. In the large-$N$ limit of holographic theories, Newton's gravitational constant $G$ scales as $G\sim N^{-2}$, while the Ryu-Takayanagi (RT) formula for holographic entanglement entropy \cite{Ryu2006a,Ryu2006b}\footnote{See also \cite{Hubeny:2007xt} for a covariant generalization and \cite{Engelhardt:2014gca} for quantum corrections.} scales as $S(A)\sim G^{-1}\sim N^2$. Therefore, the amount of entanglement resource scales as $N^2$, while the amount of bulk information scales as $N^0$.\footnote{We are restricting to semiclassical holography, in which the bulk theory is a low-energy effective field theory. The presence of a bulk UV cutoff implies that the number of degrees of freedom in the bulk is subleading in $N$ with respect to the one in the UV-complete boundary theory. From a QECC point of view, we are restricting to a low-energy code subspace.} As a result of this large-$N$ effect, in semiclassical holography we expect to be able to measure very large regions $A$ of the boundary theory and teleport the information in the corresponding entanglement wedge $W(A)$ (which contains most of the bulk spacetime) into the small complementary regions $A^c$. Our bulk calculation in Section \ref{sec:holographiccalculation} shows that this is indeed the case. For the same reasons, we also expect the resulting post-measurement bulk-to-boundary map to remain isometric. Contrarily, in the simplest tensor network models with uniform bond dimension there is no separation of scales between the entanglement entropy of the boundary region and the amount of bulk information to be teleported. Therefore, in this case we do not expect to be able to measure arbitrarily large regions of the boundary without destroying most of the bulk information. However, by increasing the ratio of boundary to bulk degrees of freedom we are able to obtain a better approximation to the large-$N$ semiclassical limit. These expectations are confirmed by our results for the HaPPY code (Section \ref{sec:happycode}) and for RTN (Section \ref{sec:rtn}). Even when the bulk information originally contained in $A$ is teleported into $A^c$ and not destroyed, if region $A$ is too large the post-measurement bulk-to-boundary map is non-isometric and bulk reconstruction from region $A^c$ is state-dependent. This fact, which is reminiscent of the discussion about Python's lunches and the reconstruction of the black hole interior \cite{Brown:2019rox,Papadodimas:2013jku,Papadodimas:2013wnh,Engelhardt:2021mue,Engelhardt:2021qjs,Akers:2022qdl}, is particularly manifest in our RTN analysis (see Section \ref{sec:rtn}). However, whenever the amount of bulk information to be teleported does not exceed the amount of entanglement resource available for teleportation, we are able to qualitatively reproduce the results obtained in our holographic setting.

Finally, we are motivated by possibilities of simulating holographic physics in near-term experimental platforms. Our analysis predicts a specific entanglement structure for the post-measurement state of the boundary theory. The impact of measurement on the bulk geometry is particularly relevant and potentially testable for future experiments that aim to simulate certain aspects of holography using quantum computers. Since our predictions are derived using holography, this could represent new experimental evidence in support of a holographic description of spacetime. We also remark that our results in tensor network constructions can have interesting implications not only for the holographic applications we considered, but also to understand the effect of measurement on the entanglement structure of more generic quantum many-body systems. Hopefully, they will also represent a starting point to holographically describe measurement-induced phase transitions \cite{aharonov,Skinner:2018tjl,Li:2018mcv,Chan:2018upn,Choi:2019nhg,Bentsen:2021ukm,Li2,jian2021measurement}. We plan to explore this possibility in future work.

\subsection{Summary of results}

Before presenting our technical results in Sections \ref{sec:holographiccalculation}-\ref{sec:rtn}, we use the remainder of this section to summarize our main findings.

\subsubsection*{Bulk analysis in $AdS_3/CFT_2$ correspondence}

\begin{figure}[h]
    \centering
    \includegraphics[width=0.8\textwidth]{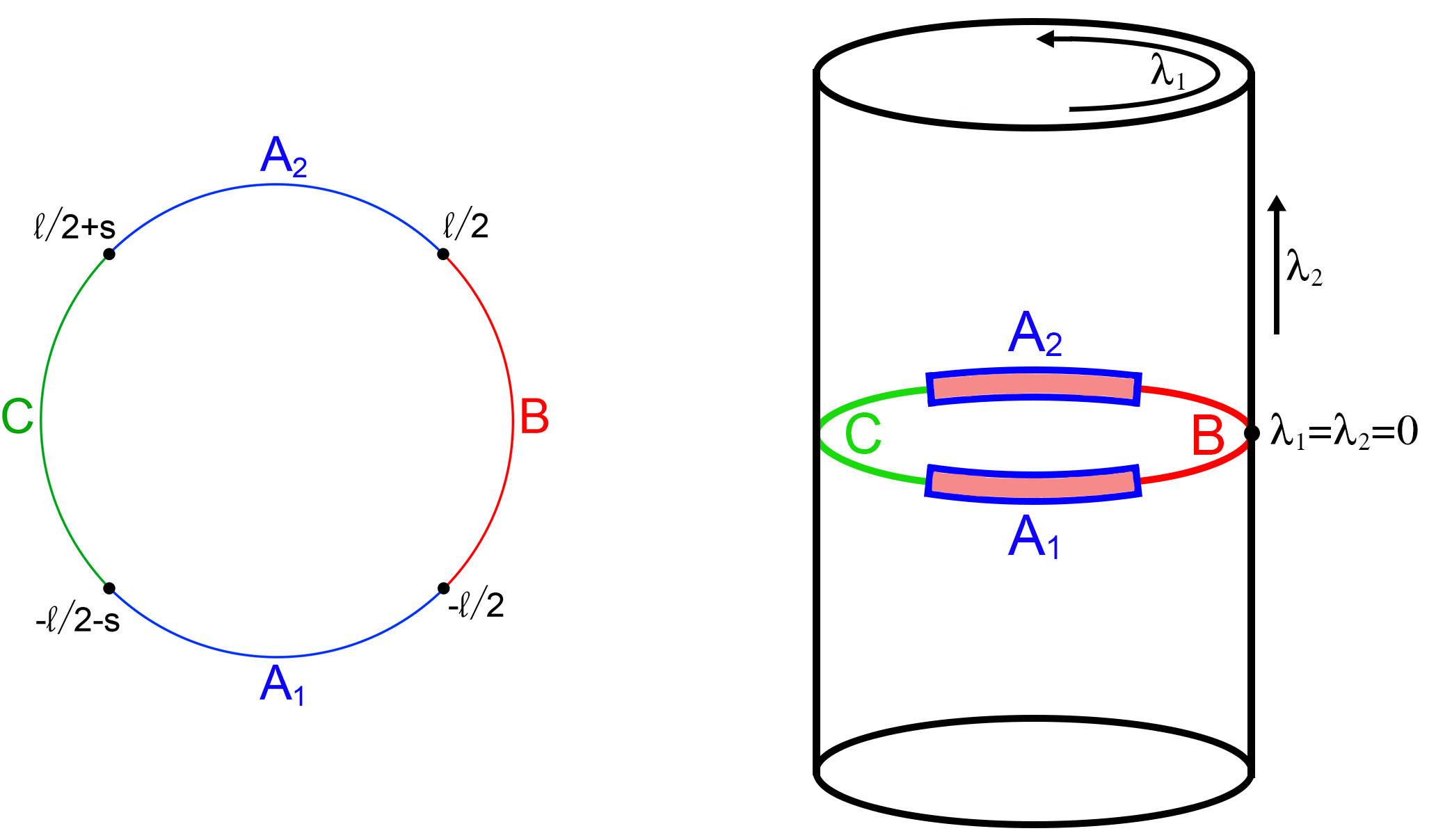}
    \caption{Left: Spatial (constant $\lambda_2$) slice of the CFT tri-partite system. The regions $A$, $B$, $C$ are represented along with the values of $\lambda_1$ at their boundaries. Right: Euclidean path integral preparing the post-measurement state of the CFT on a circle at $\lambda_2=0$. Conformal boundary conditions are imposed around the slits, which corresponds to having a post-measurement state for the subsystem $A$ given by $\ket{\psi}_A=\ket{B}_{A_1}\otimes\ket{B}_{A_2}$, where $\ket{B}$ is a Cardy state.}
    \label{splittingPI}
\end{figure}

In Section \ref{sec:holographiccalculation} we study the effect of boundary measurements in a holographic AdS/BCFT setup by considering a LPM performed on a subregion $A$ of a holographic 2D CFT in its vacuum state. Following \cite{namasawa2016epr,rajabpour2015post,rajabpour2015entanglement,miyaji2014boundary}, the corresponding post-measurement state is obtained by inserting a slit along the measured region at the time reflection symmetric slice in the Euclidean path integral preparing the CFT vacuum, and then slicing open the path integral along the same slice as illustrated in Figure \ref{splittingPI}. If region $A$ is a union of disjoint pieces, multiple slits will be present. Conformal boundary conditions must be imposed around the slits, resulting in a path integral describing a Euclidean boundary conformal field theory (BCFT). In other words, the LPM projects the measured region $A$ onto a Cardy state (or boundary state) $\ket{B}$ \cite{Cardy:1989ir,Affleck:1991tk,Cardy:2004hm,Calabrese:2009qy}. We focus our attention on a disjoint region $A=A_1\cup A_2$, where $A_1$ and $A_2$ are two identical segments located diametrically opposed from one another on the spatial circle. The complementary region $A^c$ is then split into two subregions $B$ and $C$, which we also take to be of identical size (see Figure \ref{splittingPI}).

Following \cite{namasawa2016epr}, we use conformal symmetry in two dimensions to map our infinite cylinder with two slits into a finite cylinder. The two slits are mapped to the boundaries of this cylinder. The resulting system is a well-known BCFT \cite{Cardy:2004hm}, of which we can build the holographic bulk dual using the AdS/BCFT prescription \cite{takayanagi2011holographic,fujita2011aspects}. Depending on the size of the measured region $A$ and the state we project on, the Euclidean bulk spacetime dual to the finite cylinder is given by either a portion of the BTZ black hole, or by a portion of thermal $AdS_3$ cut off by end-of-the-world (ETW) branes with tension $T$.
We study the phase transition between these two semiclassical bulk geometries, finding that the phase structure is determined by the size of region $A$ and the brane tension $T$.
Given a size and shape of region $A$, the tension $T$ controls the trajectory of the ETW branes through the bulk. The value of this parameter depends on the specific choice of boundary conditions imposed around the slits--- equivalently, on the specific choice of Cardy state we project region $A$ onto. By projecting region $A$ onto different Cardy states we can tune the brane tension $T$ and tune across the phase transition even while the size of region $A$ remains fixed. In Section \ref{sec:holographiccalculation}, we simply tune the brane tension $T$ directly in our bulk calculation, but we would like to remark that a more complete understanding of the relationship between the specific boundary measurement protocol chosen and the particular post-measurement Cardy state that we obtain remains an open question.

We then focus our attention on the bulk spatial slice located in the time-reflection symmetric plane in our original cylinder with slits. This is the slice hosting the bulk state dual to the post-measurement CFT state. In particular, we study the connectivity of such bulk slice between regions $B$ and $C$ and the entanglement entropy of region $B$. We find that the BTZ/thermal $AdS_3$ phase transition corresponds to a connected/disconnected phase transition for the bulk slice: the connected BTZ phase is favored for small $A$ and large positive tensions, while the disconnected thermal $AdS_3$ phase is favored for large $A$ and small or negative tensions. In the connected phase $B$ and $C$ share a large amount of entanglement\footnote{This result is to be expected, because entanglement is a necessary (although not sufficient \cite{Engelhardt:2022qts}) condition for spacetime connectivity.}, while in the disconnected phase they share no entanglement. This phase transition can be interpreted as a non-dynamical measurement-induced phase transition, although it is not the same as the dynamical measurement-induced phase transition recently observed in quantum many-body systems \cite{aharonov,Skinner:2018tjl,Li:2018mcv,Chan:2018upn,Choi:2019nhg,Bentsen:2021ukm,Li2,jian2021measurement}.

\begin{figure}
    \centering
    \includegraphics[width=0.4\textwidth]{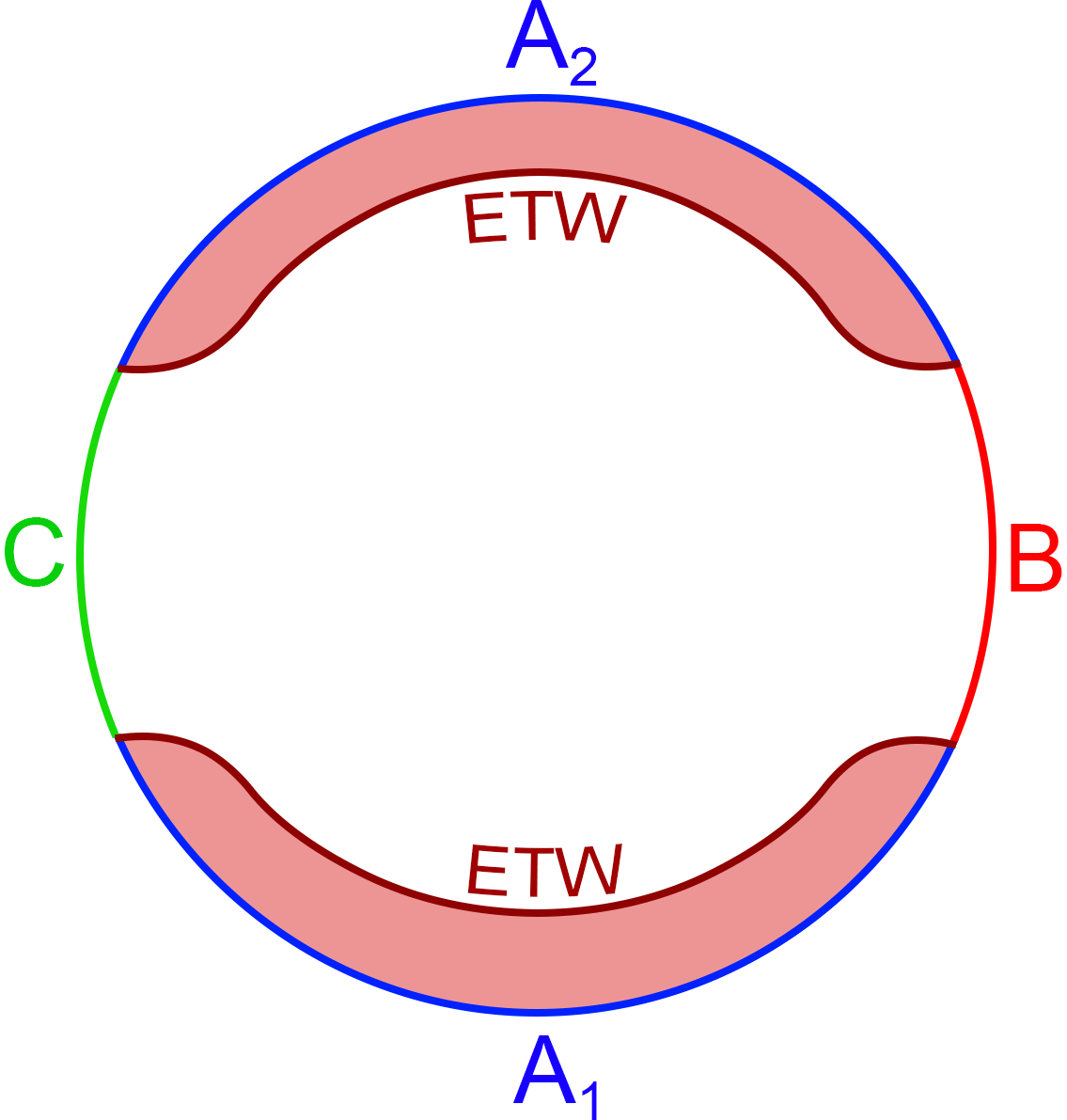}
    \caption{Schematic representation of the $\lambda_2=0$ slice in our original coordinates for the infinite cylinder with slits when a large region $A$ is projectively measured on a Cardy state with large positive boundary entropy. By choosing a sufficiently large positive value for the ETW brane tension (i.e. $T\lesssim T_c$), the slice can remain connected even if $|A|\gg |A^c|$. This is possible because the holographic theory living on the boundary is a large-$N$ gauge theory. The connectedness of the slice in this setup is evidence that the LPM teleports bulk information from $A$ to $A^c$.}
    \label{largebulk}
\end{figure}

In the tensionless case $T=0$, the connected/disconnected phase transition occurs when exactly half of the CFT is measured, i.e. $|A|=|B\cup C|$. Increasing the tension to positive values, the connected phase is more and more favored. Remarkably, as we tune the tension to values close to the critical value $T\lesssim T_c\equiv 1/R$ (where $R$ is the AdS radius), the connected phase is always dominant, even when the measured region $A$ comprises almost all of the CFT (see Figure \ref{largebulk}). This is a realization of how the LPM is able to teleport bulk information from region $A$ to its complement $A^c=B\cup C$. In fact, when $A$ is given by more than half of the CFT, its pre-measurement entanglement wedge $W(A)$ is clearly connected and contains the central region of the bulk on the time-symmetric slice. But projecting $A$ on an appropriate Cardy state corresponding to a large tension brane, the bulk information in most of $W(A)$ (including the center of the bulk) is now part of $W(A^c)$ and is accessible from the complementary region $A^c$. For a fixed size of $A$, the larger the tension the more bulk information is teleported into $A^c$. 

\subsubsection*{Boundary measurements and quantum teleportation}

In order to understand in more detail how bulk teleportation arises, in Section \ref{sec:boundmeasuremteleportation} we study local projective measurements performed on the physical qubits of quantum error correcting codes. This allows us to gain intuition about the physical processes involved in the teleportation and predict how tensor network models of holography behave under local projective measurements. The results obtained in Section \ref{sec:boundmeasuremteleportation} are confirmed by the explicit analyses of the HaPPY code (Section \ref{sec:happycode}) and random tensor networks (Section \ref{sec:rtn}).

As a first step, we study the effect of LPM on an individual [[5,1,3]] code, which encodes 1 logical qubit into 5
physical qubits. Before any measurement is performed, any two physical qubits do not have access to the logical information. We show that by measuring three out of the five physical qubits the logical information is either erased or teleported into the remaining two qubits, depending on the measurement basis we choose. We generalize this result to generic QECC and show that only so much bulk information can be teleported into region $A^c$: if the measured region $A$ is too large, most of the bulk information, including the center of the slice, is destroyed. This is because, different from semiclassical holography, the amount of entanglement resource is now limited due to the absence of a large-$N$ limit in the boundary theory.

Because these results hold for generic QECC, they provide a fundamental insight into the modification of the entanglement structure caused by the LPM and responsible for bulk teleportation. Their validity in toy models of holography---verified in Sections \ref{sec:happycode} and \ref{sec:rtn}---confirm their ability to explain how the bulk teleportation occurs.

\subsubsection*{HaPPY code}

In Section \ref{sec:happycode} we consider a tensor network representation of a bulk slice using the HaPPY code \cite{Pastawski:2015qua}. This allows us to understand in detail how the bulk teleportation arises in a simplified toy model of holography and confirms the expectations of Section \ref{sec:boundmeasuremteleportation}. In this case, the LPM is simply implemented by performing a projective measurement on a product basis of the external legs of the tensor network corresponding to a given subregion $A$. 

Our stabilizer simulations in Section \ref{happynumerics} confirm that bulk teleportation indeed occurs if the amount of bulk information to be teleported does not exceed the amount of entanglement resource available for the teleportation. In particular, by choosing an appropriate product basis for the measurement it is possible to obtain post-measurement states with a large amount of correlation between boundary qubits in $A^c$ and bulk qubits that were contained in $W(A)$ before the measurement. On the other hand, if region $A$ is too large our numerics confirm the expectation that most of the bulk information is destroyed (see Figure \ref{fig:happystackteleport}). 

In order to determine the ``location of the ETW brane'', intended as the boundary of the post-measurement entanglement wedge $W(A^c)$ of the unmeasured region $A^c$, we can study the post-measurement mutual information between bulk logical qubits and boundary physical qubits in region $A^c$---denoted by $I(x:A^c)$, where $x$ is a bulk qubit. We argue that whenever $I(x:A^c)=0$ the bulk qubit $x$ is not contained in $W(A^c)$ and is therefore cut out of the post-measurement bulk by a ``quasi-ETW brane''. With this interpretation, the results of our numerical experiments would lead us to conclude that, given a fixed size of the measured region $A$, we can obtain brane tensions ranging from $T=0$ to near-critical by choosing different measurement bases. In tri-partite systems similar to those we study in Section \ref{sec:holographiccalculation}, we find results compatible with the ones outlined above for the $AdS_3/CFT_2$ setup. 

However, there are key differences between this HaPPY code analysis and the holography results of Section \ref{sec:holographiccalculation}, because the ratio between the number of bulk logical qubits and the number of boundary physical qubits is not small enough to reproduce the holographic large-$N$ limit. As a result, the entanglement structure of the post-measurement state in the HaPPY code is more subtle: many of the bulk qubits for which $I(x:A^c)=0$ holds are still highly entangled with other bulk qubits. Therefore, it is problematic to consider such qubits as being cut out of the bulk by an actual ETW brane, but rather they are behind a more ``diffuse'' quasi-brane separation which can only be considered as pre-geometrical.\footnote{This can be interpreted as a finite-$N$ effect: in the presence of a large-$N$ limit on the boundary, all bulk qubits are effectively only entangled with boundary qubits. Therefore, if after the measurement $I(x:A^c)=0$ we can conclude that the bulk qubit $x$ is in a product state with the rest of the network, including other bulk qubits.} Additionally, some bulk qubits share a non-maximal amount of mutual information with the boundary, and it is unclear whether or not they should be considered as part of the post-measurement entanglement wedge of $A^c$. Therefore, although the HaPPY code results described above share many qualitative similarities with our holographic analysis in Section \ref{sec:holographiccalculation} there are important differences which may be interpreted as a strong finite-$N$ effect.

To address these issues, we also study a new concatenated version of the HaPPY code which mimics the large-$N$ limit in holography. This allows us to study how these ``finite-$N$'' effects may be suppressed by taking an analogous large-$N$ limit. The bulk legs of all members of the stack are combined into a single set of bulk logical qubits. The ratio between logical and physical qubits is therefore reduced by a factor of $1/M$. Heuristically, the large-$M$ limit in this construction corresponds to the holographic large-$N$ limit. The measurement is now performed on a region $A$ on each copy in the stack. Our numerical analysis for $M\leq 6$ shows that as we increase $M$, the quasi-brane ``condenses'' into an object that is geometrically better defined. 
More precisely, for most bulk qubits, $I(x:A^c)=0$ implies that $x$ shares no mutual information with other bulk qubits either and $I(x:A^c)$ is either vanishing or maximal, leaving no ambiguity in the definition of the brane location (and therefore its tension).

For $M>1$, measurements generically destroy very small regions of the bulk that lie close to the boundary. In other words, for a generic choice of measurement basis for each copy in the stack we obtain branes with large to near-critical tension. By scanning the space of possible measurement bases, we are able to tune the tension only slightly away from criticality. Intuitively, the reason is that teleportation can now happen not only within different regions of one HaPPY code, but also across different copies in the stack. Therefore, teleportation becomes far more likely and destroying the bulk with a random local projection far more unlikely as we increase $M$. We also explain the reason why seemingly analogous boundary measurements for $M=1$ and $M>1$ yield radically different results for the mutual information $I(x:A^c)$. Whether or not lower values of the tension (including vanishing and negative values) can be obtained in the $M>1$ HaPPY code remains an interesting open question to be explored in future work.

\subsubsection*{Random tensor networks}

In Section \ref{sec:rtn} we study similar questions in random tensor network constructions. First, we show that when the dimension of the bulk code subspace is much smaller than the dimension of the physical Hilbert space, the post-measurement random tensor network can be mapped to an Ising model with free boundary conditions at the measured region $A$. We find that this corresponds to having a ``critical tension brane'': all the bulk information is teleported from $A$ into $A^c$ by the LPM. Second, we consider arbitrary bond dimensions for the boundary ($D$) and bulk ($D_b$) legs and ask whether a region $b$ deep into a bulk slice, whose degrees of freedom are purified by an external reference system $R$, can be reconstructed from a boundary subregion $A^c$ after the complementary region $A$ on the boundary is measured. If this is the case and if the entanglement wedge $W(A)$ of $A$ before the measurement contained region $b$, we can conclude that the effect of the measurement is to teleport the bulk information from region $A$ to its complement $A^c$. 

For any fixed value of $D$ and $D_b$, we find that teleportation occurs whenever $A^c$ is large enough\footnote{This can be understood in analogy with the the discussion in Section \ref{sec:boundmeasuremteleportation}: if $A$ is too large and the amount of bulk information to teleport exceeds the amount of entanglement resource available for teleportation, most of the bulk information is destroyed.} and derive a sufficient condition for the post-measurement entanglement wedge $W(A^c)$ to contain $b$. Interestingly, there exists a regime in which $A^c$ is large enough for its entanglement wedge to contain the bulk region $b$, but not large enough for the bulk-to-boundary encoding map to be isometric. In this regime, the bulk reconstruction of region $b$ is state-dependent and the bulk-to-boundary map is non-isometric, similarly to what happens for the interior of old evaporating black holes \cite{Brown:2019rox,Papadodimas:2013jku,Papadodimas:2013wnh,Engelhardt:2021mue,Engelhardt:2021qjs,Akers:2022qdl}.

It is worth noting that the condition for $W(A^c)$ to contain $b$ does not depend only on the size of $b$ and $A^c$, but also on the bond dimensions of the bulk and boundary legs. Specifically, if the bond dimension of the boundary legs is much larger than the one of the bulk legs, i.e. $D\gg D_b$, it is possible to reconstruct $b$ from $A^c$ even when the latter is very small, and the encoding map is always isometric. Intuitively, the reason is that in the $D\gg D_b$ limit the Hilbert space of any small region $A^c$ on the boundary is large enough to encode most or all of the bulk information. This limiting case should correspond to semiclassical holography, in which the large-$N$ limit in the boundary theory guarantees that any small boundary subregion can encode almost all the bulk, as we have explained above. This analogy suggests that in the holographic case the bulk-to-boundary encoding should also always be isometric.

\section{Holographic dual of boundary measurement}
\label{sec:holographiccalculation}

\subsection{Description of the setup}

We begin our technical investigation by applying the tools of AdS/BCFT to study the bulk dual description of a tri-partite system after a local projective measurement is performed on one of the subsystems.
We focus on two-dimensional CFTs---and therefore make use of the $AdS_3/CFT_2$ correspondence---where the existence of an infinite-dimensional algebra of infinitesimal conformal transformations (the Virasoro algebra) implies the existence of a much wider variety of conformal maps with respect to the higher dimensional case, in which the conformal algebra is finite-dimensional. Note that the higher-dimensional generalization of the results presented in this section is not immediate: Liouville's theorem guarantees that for $d>2$ all conformal transformations are M\"obius transformations, while in the derivation of our results we make use of conformal maps other than M\"obius transformations. However, we expect the qualitative features of our analysis, and in particular the fact that boundary measurements teleport bulk information, to hold also in higher dimensions.

Consider a 2D CFT on a spatial circle with circumference $L$, defined by a Euclidean path integral on an infinite cylinder (for vanishing temperature). We label the periodic spatial direction by $\lambda_1\in [-L/2,L/2]$ (with $\lambda_1\sim \lambda_1+L$) and the Euclidean time by $\lambda_2\in (-\infty,\infty)$; we further define complex coordinates $\lambda=\lambda_1+i\lambda_2$ and $\bar{\lambda}=\lambda_1-i\lambda_2$. The vacuum state $\ket{0}$ of the CFT is obtained by slicing open the Euclidean path integral at $\lambda_2=0$. Let us split a spatial slice of the CFT into three subsystems $A$, $B$, $C$  (see Figure \ref{splittingPI}). $A$ is a disconnected region defined as $A=A_1 \cup A_2$, where $A_1$ and $A_2$ are two segments of length $s$ ($\lambda_1\in[-\ell/2-s,-\ell/2]$ for $A_1$ and $\lambda_1\in[\ell/2,\ell/2+s]$ for $A_2$). $B$ and $C$ are segments of equal length $\ell$, with $\lambda_1\in[-\ell/2,\ell/2]$ for $B$ and  $\lambda_1\in[-L/2,-\ell/2-s]\cup [\ell/2+s+L/2]$ for $C$. Imposing $|B|=|C|=\ell$ implies $L=2\ell + 2s$.

We are interested in studying the entanglement structure of the state on $BC$ after a projective measurement is performed on region $A$. In particular, our goal is to build, in analogy with \cite{namasawa2016epr} (which studied a similar setup involving a connected measured region) and using the AdS/BCFT prescription \cite{takayanagi2011holographic,fujita2011aspects}, a Euclidean bulk saddle dual to the Euclidean path integral preparing the post-measurement state at $\lambda_2=0$. We will then focus on the bulk dual of the $\lambda_2=0$ slice, study its connectivity and compute the entanglement entropy of region $B$\footnote{From now on, when we refer to region $B$ it is understood that we are referring to the $\lambda_1\in [-\ell/2,\ell/2]$ interval on the $\lambda_2=0$ slice.} using the Ryu-Takayanagi (RT) formula \cite{Ryu2006a,Ryu2006b}\footnote{For the purposes of our analysis, we can neglect quantum corrections to the RT formula \cite{Engelhardt:2014gca}.}
\begin{equation}
    S(B)=\frac{\mathcal{A}(\gamma_B)}{4G}
\end{equation} 
where $\mathcal{A}(\gamma_B)$ is the area (i.e. length in our $AdS_3/CFT_2$ setting) of an extremal codimension-2 bulk surface homologous to region $B$.

\subsection{LPM, slit prescription and AdS/BCFT}
\label{slitsection}

As we have anticipated in the introduction, the setup of our interest is one where a local projective measurement is performed on region $A$\footnote{As we will see, we are interested in the case where the outcome of the projective measurement is given by a specific Cardy state $\ket{B}$ with some definite boundary entropy. Therefore, our measurement procedure is more correctly understood as a postselection. Operationally, this means that the projective measurement in some given basis is repeated an arbitrary amount of times until the desired outcome, leaving region $A$ in the state $\ket{B}$, is obtained.}. To consistently define the measurement procedure in our continuum CFT, we can imagine implementing a lattice regularization of our theory, and then performing a projective measurement on the sites corresponding to region $A$. We focus on the case where region $A$ is projected onto a Cardy state and therefore the resulting Euclidean path integral describes a boundary conformal field theory (BCFT).

Because Cardy states have zero spatial entanglement \cite{miyaji2014boundary}, the measurement needs to be very local, i.e. the lattice sites corresponding to region $A$ are projected onto a product state \cite{namasawa2016epr}. This procedure defines the LPM of our interest. Suppose that we start in the CFT vacuum state, and that the post-measurement state on region $A=A_1\cup A_2$ is given by a product of Cardy states $\ket{\psi}_A=\ket{B}_{A_1}\otimes \ket{B}_{A_2}$, where each Cardy state can be written as a product state for the lattice sites in the appropriate region. The complete post-measurement state of the CFT on the lattice is then given by
\begin{equation}
    \ket{\Psi}=P_\psi\ket{0}
\end{equation}
where $P_\psi\equiv {}_A\ket{\psi}\bra{\psi}_A\otimes \mathds{1}_{BC}$, and $\mathds{1}_{BC}$ is the identity operator on regions $B$ and $C$. Note that after the measurement, the unmeasured part of the system $A^c=B\cup C$ is in a pure state. The Euclidean path integral preparing the state $\ket{\Psi}$ is then given by an infinite cylinder with two slits at $\lambda_2=0$ at the location of regions $A_1$ and $A_2$ \cite{namasawa2016epr,rajabpour2015post,rajabpour2015entanglement} (see Figure \ref{splittingPI}). The state on the union of the slits is fixed to be $\ket{\psi}_A$. 

The LPM is a UV measurement and the post-measurement state is therefore singular once we take the continuum limit. This can be intuitively understood if we consider that in a continuum quantum field theory it is not possible to have a product state of two subregions, because they share an infinite amount of UV entanglement. Therefore, we expect the post-measurement state to have infinite energy in the continuum limit, and in particular the stress-energy tensor is divergent at the endpoints of the slits \cite{namasawa2016epr}. In order to have a well-defined post-measurement state and to study Lorentzian time evolution, we should introduce some form of regularization.\footnote{\label{regfootnote} One possibility is to give the slits a finite height by evolving the state on $A^c$ for an amount $\varepsilon$ of Euclidean time using the Hamiltonian restricted to region $A^c$ with appropriate conformal boundary conditions. A different possibility is to evolve the state of the full CFT in Euclidean time, splitting each slit into two slits which are mapped into each other by time-reflection symmetry. The latter regularization was considered for a connected region $A$ in \cite{namasawa2016epr}. We leave the implementation of these possible regularizations in our setup to future work.} For our purposes, it is enough to take a different approach first introduced in \cite{namasawa2016epr}: we make use of the conformal symmetry to map our infinite cylinder with two slits into a finite cylinder (see Section \ref{sec:bulkdual}). The two slits are mapped to the boundaries of such cylinder. Practically, the conformal mapping procedure acts as a regularization and the resulting system is a well-known BCFT \cite{Cardy:2004hm}, of which we can build the holographic bulk dual using the AdS/BCFT prescription \cite{takayanagi2011holographic,fujita2011aspects}. 

A BCFT is a CFT defined on a manifold $\mathcal{M}$ with a boundary $\partial \mathcal{M}$\footnote{In general, the boundary $\partial \mathcal{M}$ can be the union of multiple disconnected pieces.}, where conformal boundary conditions are imposed \cite{Cardy:1989ir,Affleck:1991tk,Cardy:2004hm,Calabrese:2009qy}. For each disconnected piece of the boundary, a choice of conformal boundary conditions must be made, corresponding to a specific choice of Cardy state (or boundary state) $\ket{B}$ at the boundary. One feature of BCFTs which will be relevant for our discussion is that the entanglement entropy of subregions of the BCFT receives a finite contribution associated with the presence of the boundary: this is the boundary entropy $S_b=\log(g)$, where the ``g-function'' $g=\braket{0|B}$ depends on the choice of boundary state (i.e. of boundary conditions)\footnote{Note that boundary states are non-normalized (and in fact non-normalizable), and therefore the boundary entropy can take either sign in general.} \cite{Affleck:1991tk,Cardy:2004hm,Calabrese:2009qy}. There is one such contribution for each disconnected piece of the boundary $\partial \mathcal{M}$. In our setup, the manifold $\mathcal{M}$ is given by the infinite cylinder with slits where the Euclidean path integral is defined, and the boundary $\partial\mathcal{M}$ is given by two disconnected components (the edges of the two slits correspondent to regions $A_1$ and $A_2$) where we impose conformal boundary conditions. In general, we can impose different conformal boundary conditions at the two slits, i.e. $\ket{B_1}_{A_1}\neq \ket{B_2}_{A_2}$. In order to simplify our bulk dual analysis, we consider the case where identical boundary conditions are imposed on the two slits, which means that the same post-measurement state is obtained on both the region $A_1$ and $A_2$: $\ket{B_1}_{A_1}=\ket{B_2}_{A_2}\equiv \ket{B}$.\footnote{The general case where two different sets of boundary conditions are imposed at the two slits would involve the existence of two end-of-the-world branes with different tensions in the bulk dual spacetime. As we will see, there is a phase where the branes associated with the two slits join up to form a single, connected brane. When this is the case, having branes with different tensions would imply the existence of a codimension-2 defect on the brane, and the necessity to consider additional corner terms in the action (see \cite{Miyaji:2022cma} for a recent discussion). Since the setup involving the same Cardy state on both slits (and therefore branes with identical tensions) is sufficient for the purposes of the present paper, we leave the analysis of the more general case to future work.}

\begin{figure}
    \centering
    \includegraphics[width=0.3\textwidth]{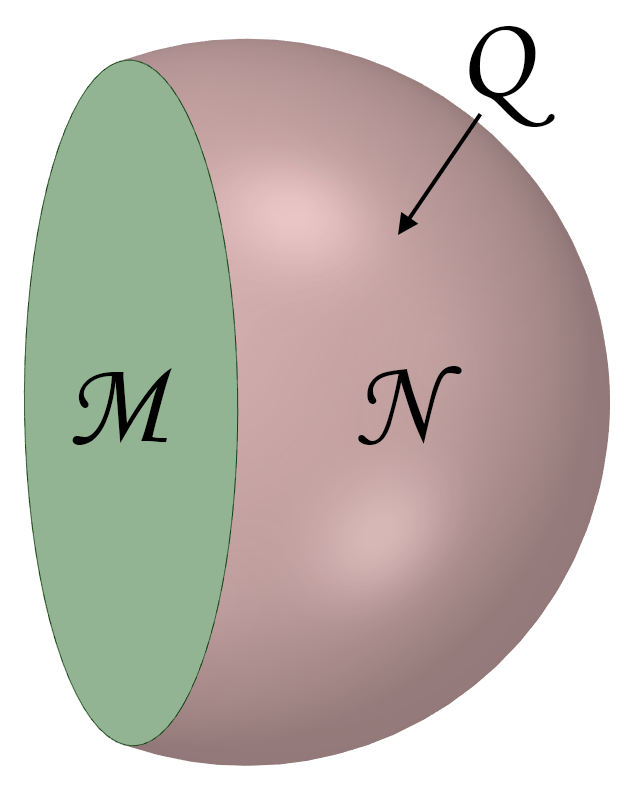}
    \caption{A schematic representation of the AdS/BCFT prescription: the BCFT is defined on a manifold $\mathcal{M}$ with a boundary, and the dual spacetime manifold $\mathcal{N}$ is cut off by an ETW brane $\mathcal{Q}$ homologous to $\mathcal{M}$.}
    \label{adsbcft}
\end{figure}

The AdS/BCFT proposal \cite{takayanagi2011holographic,fujita2011aspects} allows one to construct $(d+1)$-dimensional (Euclidean) asymptotically AdS spacetimes dual to $d$-dimensional BCFTs. In particular, the bulk spacetime manifold $\mathcal{N}$ has a boundary $\partial \mathcal{N}=\mathcal{M}\cup \mathcal{Q}$, where $\mathcal{Q}$ is a dynamical codimension-1 end-of-the-world (ETW) brane anchored at the asymptotic boundary $\mathcal{M}$ and such that $\partial{Q}=\partial{M}$.\footnote{When $\partial\mathcal{M}$ has multiple disconnected pieces, multiple ETW branes can be present.} In other words, the spacetime is cut off by an ETW brane homologous to the boundary manifold $\mathcal{M}$ where the BCFT is defined, and the brane can be seen as a bulk extension of the boundary $\partial \mathcal{M}$ (see Figure \ref{adsbcft}). Neumann boundary conditions for bulk fields (including the metric) are imposed at the location of the brane. In the simplest models, the brane is described by a single parameter, the tension $T$, which is related to the boundary entropy \cite{takayanagi2011holographic,fujita2011aspects}. Therefore, different choices of boundary states in the BCFT are holographically described by different choices of brane tensions, which in turn determine the brane trajectory in the bulk spacetime.\footnote{See \cite{Cooper:2018cmb,Antonini:2019qkt} for cosmological applications, and \cite{Kourkoulou:2017zaj,Antonini:2021xar} for lower-dimensional examples.} The brane equations of motion for such a brane of constant tension can be obtained from the Neumann boundary conditions
\begin{equation}
    K_{ab}-Kh_{ab}=-Th_{ab}
    \label{israeli}
\end{equation}
where $K_{ab}$ is the extrinsic curvature of the brane, $K=h^{ab}K_{ab}$ is its trace and $h_{ab}$ is the metric induced on the brane.

\subsection{Building the bulk dual spacetime}
\label{sec:bulkdual}

Now that we have a BCFT setup, we are ready to build a dual spacetime by means of the AdS/BCFT prescription. However, it is non-trivial to do so starting directly from the Euclidean theory on the cylinder with slits depicted in Figure \ref{splittingPI}. In fact, the singularities present at the endpoints of the slits imply that the metric of the dual spacetime will not take a simple form and will be singular, especially in the vicinity of the $\lambda_2=0$ slice we are interested in.

As we have mentioned, one way to deal with this issue is to introduce some form of physical regularization. However, we are only interested in studying the connectivity of the bulk $\lambda_2=0$ slice between regions $B$ and $C$ and in computing the holographic entanglement entropy of region $B$. This can be accomplished by conformally mapping the cylinder with slits to a cylinder of finite length ($S^1\times I$, where $I$ is a finite interval), and then employing the AdS/BCFT prescription to build its holographic dual (which, depending on how we ``fill in'' the finite cylinder, is given by either a portion of the Euclidean BTZ black hole or a portion of thermal $AdS_3$, cut off by ETW branes \cite{fujita2011aspects}). The connectivity of the $\lambda_2=0$ slice and the holographic entanglement entropy of region $B$ can then be indirectly studied in the resulting geometries\footnote{An analogous procedure was applied to a single slit setup in \cite{namasawa2016epr}.}. 

\begin{figure}
    \centering
    \includegraphics[width=.75\textwidth]{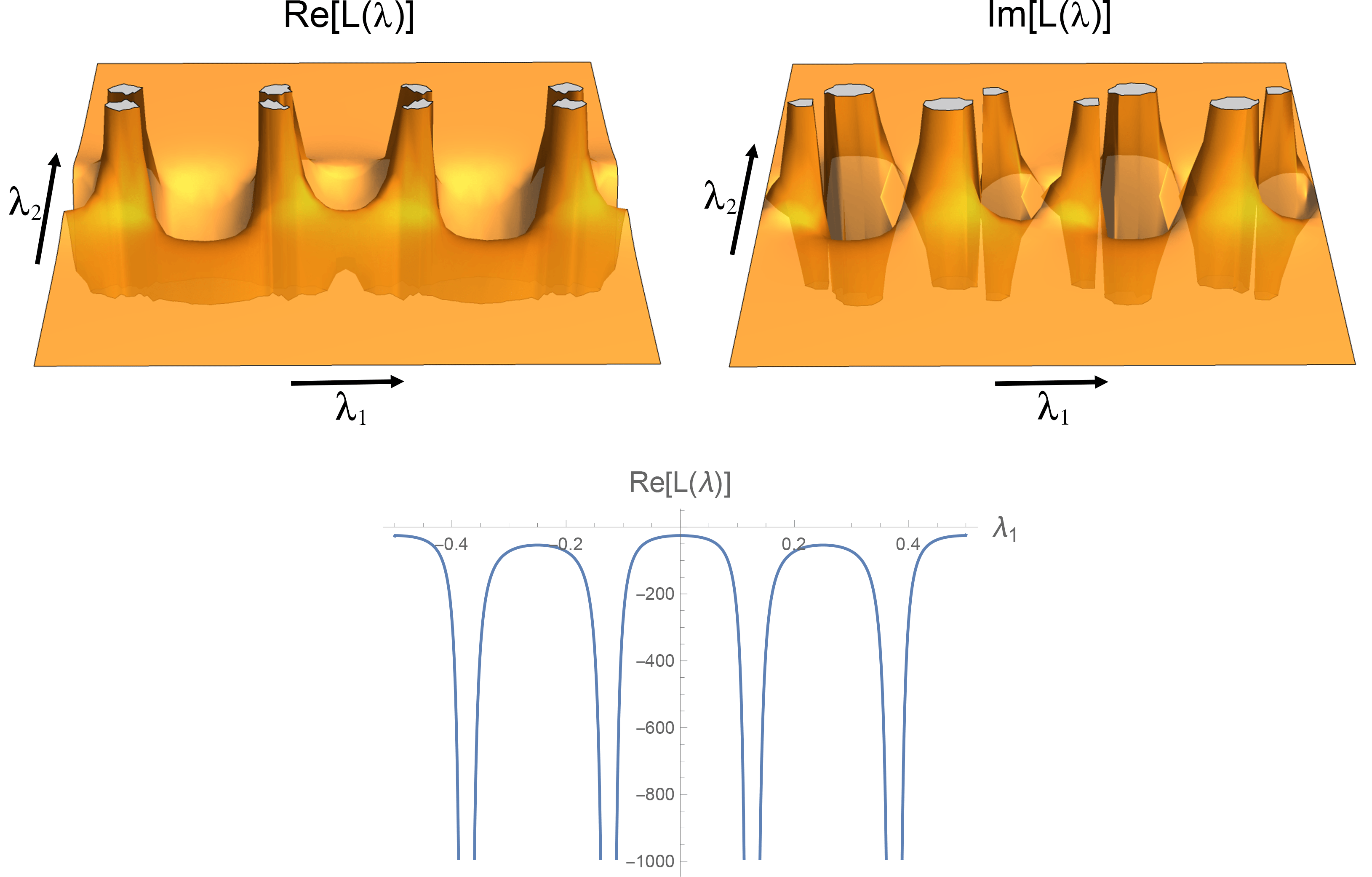}
    \caption{Top: Real (left) and imaginary (right) parts of $L(\lambda)$---defined in equation (\ref{lfunctions})---in the BTZ phase. The metric (\ref{mappedmetric}) is singular at the values of $\lambda$ corresponding to the endpoints of the slits (i.e. $\lambda_2=0$ and $\lambda_1=\{-L/2+\ell/2, -\ell/2, \ell/2, L/2-\ell/2\}$, here $\ell=0.25$, $L=1$) for any value of the radial coordinate $z_\lambda$, and approaches a constant value away from the $\lambda_2=0$ slice where the measurement is performed. Bottom: Real part of $L(\lambda)$ in the BTZ phase as a function of $\lambda_1$ on the $\lambda_2=0$ slice. The singularities at the endpoints of the slits are well visible. A similar plot for the thermal $AdS_3$ phase is reported in Figure \ref{thermallfunctions} in Appendix \ref{sec:mapback}.}
    \label{btzlfunctions}
\end{figure}

From a bulk point of view, in order for equation (\ref{israeli}) to have a solution, in general we need to solve Einstein's equations including the backreaction of the brane, and for generic shapes of the boundary manifold $\mathcal{M}$ (such as our cylinder with slits depicted in Figure \ref{splittingPI}) this is a highly non-trivial task \cite{namasawa2016epr,Nozaki:2012qd}. On the other hand, the conformal mapping procedure allows us to reduce our problem to a well-known AdS/BCFT setup studied in \cite{fujita2011aspects}, where the computation of the holographic entanglement entropy of region $B$ can be carried out without difficulties. But there is more: exploiting the fact that every solution of 3D Einstein's equations with negative cosmological constant is locally pure $AdS_3$, it is possible to establish a precise relationship between the spacetime dual to the BCFT on the cylinder with slits in $\lambda$ coordinates and a subregion of Euclidean Poincar\'e-AdS cut off by ETW branes. Such relationship allows us to numerically obtain the metric of the spacetime dual to our BCFT setup in the original $\lambda$ coordinates and verify the presence of singularities on the $\lambda_2=0$ slice (see Figure \ref{btzlfunctions} and Appendix \ref{sec:mapback}).

In order to understand how this can be achieved, consider the Euclidean Poincar\'e-$AdS_3$ metric
\begin{equation}
    ds^2=\frac{R^2}{\eta^2}\left(d\eta^2+d\xi d\bar{\xi}\right)
    \label{poincareads}
\end{equation}
where $R$ is the AdS radius, $\xi=\xi_1+i\xi_2$ and $\bar{\xi}=\xi_1-i\xi_2$ are complex coordinates describing the boundary complex plane, $\eta$ is the bulk direction with $\eta>0$, and the asymptotic AdS boundary is given by $\eta=0$. If we perform a conformal map on the asymptotic boundary
\begin{equation}
    \xi_{CFT}=f(w_{CFT}),\hspace{1cm} \bar{\xi}_{CFT}=\bar{f}(\bar{w}_{CFT})
    \label{generalconfmap}
\end{equation}
the corresponding bulk coordinate transformation is given by \cite{Roberts:2012aq,namasawa2016epr}
\begin{equation}
    \begin{aligned}
    &\xi=f(w)-\frac{2z^2(f'(w))^2\bar{f}''(\bar{w})}{8f'(w)\bar{f}'(\bar{w})+z^2f''(w)\bar{f}''(\bar{w})}\\[10pt]
    &\bar{\xi}=\bar{f}(\bar{w})-\frac{2z^2(\bar{f}'(\bar{w}))^2f''(w)}{8f'(w)\bar{f}'(\bar{w})+z^2f''(w)\bar{f}''(\bar{w})}\\[10pt]
    &\eta=\frac{4\sqrt{2}z(f'(w)\bar{f}'(\bar{w}))^{\frac{3}{2}}}{8f'(w)\bar{f}'(\bar{w})+z^2f''(w)\bar{f}''(\bar{w})}.
    \end{aligned}
    \label{bulkconftransf}
\end{equation}
The bulk metric in $(w,\bar{w},z)$ coordinates is then given by\footnote{Note that the metric (\ref{mappedmetric}) is related to the Poincar\'e metric by the diffeomorphism (\ref{bulkconftransf}): it is just a different slicing of the same manifold, which is locally pure $AdS_3$. By substituting $w=w_1+iw_2$ it is also immediate to verify that the metric (\ref{mappedmetric}) is real by construction.}
\begin{equation}
    ds^2=R^2\left[L(w)dw^2+\bar{L}(\bar{w})d\bar{w}^2+\left(\frac{2}{z^2}+\frac{z^2}{2}L(w)\bar{L}(\bar{w})\right)dwd\bar{w}+\frac{dz^2}{z^2}\right]
    \label{mappedmetric}
\end{equation}
where we defined
\begin{equation}
    L(w)\equiv -\frac{1}{2}\{f(w),w\},\hspace{1cm} \bar{L}(\bar{w})\equiv -\frac{1}{2}\{\bar{f}(\bar{w}),\bar{w}\}
    \label{lfunctions}
\end{equation}
and $\{f(w),w\}=f'''(w)/f'(w)-3/2(f''(w)/f'(w))^2$ is the Schwarzian derivative. 

The holographic dual of the finite cylinder that we will construct can be mapped to a portion of Poincar\'e-AdS cut off by ETW branes by a coordinate transformation of the form (\ref{bulkconftransf}), corresponding to a boundary conformal map of the form (\ref{generalconfmap}) (see Appendix \ref{mappingtopoincare}). Moreover, by composing such a conformal map with the ones used to map the cylinder with slits into the finite cylinder, and using the coordinate transformation (\ref{bulkconftransf}) for the composed map, the metric in the original $\lambda$ coordinates can be constructed. The asymptotic boundary of the resulting spacetime geometry will then be the cylinder with slits depicted in Figure \ref{splittingPI}, and the metric will take the form (\ref{mappedmetric}), where $f$ is given by the (composed) map relating the boundary limit of the Poincar\'e coordinate $\xi_{CFT}$ to the $\lambda$ coordinate on the cylinder with slits (see Figure \ref{btzlfunctions} and Appendix \ref{mappingtopoincare}). In principle, the location of the branes could also be mapped back to $\lambda$ coordinates, obtaining the holographic dual of the cylinder with slits. The singular nature of our setup makes this task numerically complicated and of little physical significance, so we will not carry it out in the present paper. Nonetheless, we would like to remark that for a regularized version of our setup or for other regular setups a similar procedure can be fully implemented to obtain the dual spacetime of any BCFT system (see Appendix \ref{sec:mapback}).

\subsubsection{Conformal mapping and bulk dual spacetime}

\begin{figure}
    \centering
    \includegraphics[width=0.7\textwidth]{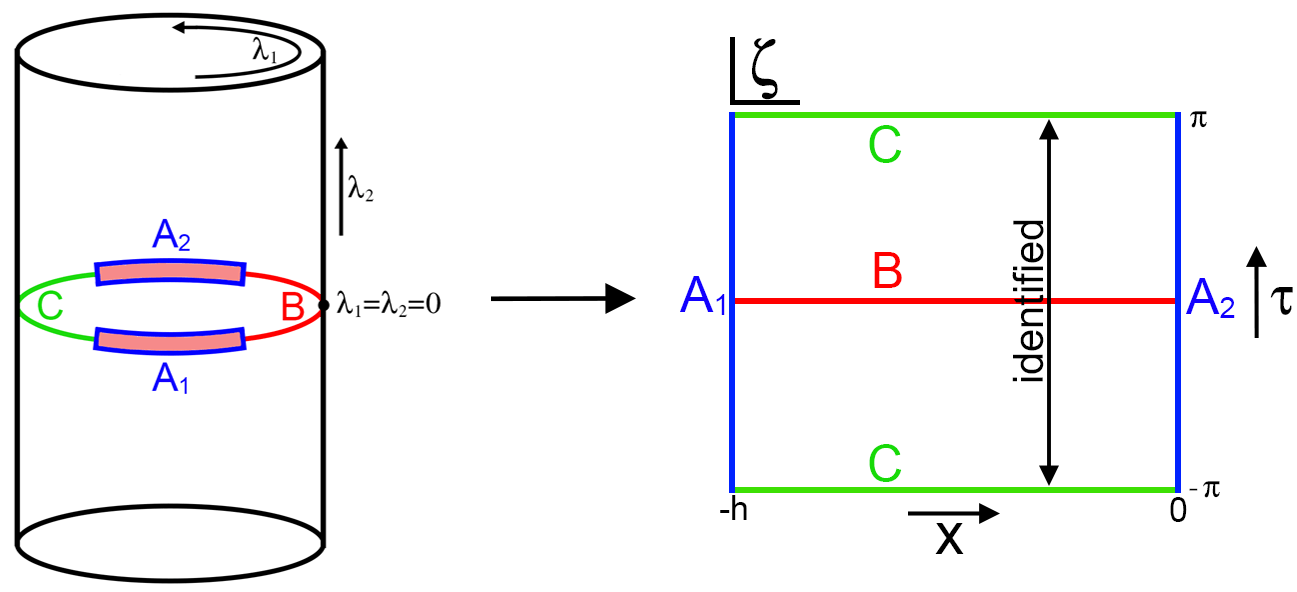}
    \caption{We conformally map the infinite cylinder with slits in $\lambda$, $\bar{\lambda}$ coordinates to the finite cylinder in $\zeta$, $\bar{\zeta}$ coordinates. The planar representation of the finite cylinder reported here is useful to understand how the two possible dual spacetimes, depicted in Figures \ref{BTZtorus} and \ref{thermaltorus}, arise.}
    \label{slitstofinite}
\end{figure}

The first step to build the holographic dual of our BCFT setup depicted in Figure \ref{splittingPI} is to map the infinite cylinder in $\lambda$ coordinates to a finite cylinder, i.e. a manifold $S^1\times I$, where $I=[-h,0]$ is a finite interval, see Figure \ref{slitstofinite}. Here we defined
\begin{equation}
    h\equiv 2\pi\frac{\mathcal{K}(k^2)}{\mathcal{K}(1-k^2)}.
    \label{hdefinition}
\end{equation}
where $\mathcal{K}(m)$ is the complete elliptic integral of the first kind and 
\begin{equation}
    k\equiv \frac{\tan\left(\frac{\pi\ell}{2L}\right)}{\tan\left[\frac{\pi}{L}\left(\frac{\ell}{2}+s\right)\right]}=\frac{\tan\left(\frac{\pi y}{2}\right)}{\tan\left[\frac{\pi (1-y)}{2}\right]}
    \label{kdefinition}
\end{equation}
with $y=\ell/L=1/2-s/L\in [0,1/2]$ (in the last equality we used $L=2\ell+2s$).
The mapping can be achieved by composing three conformal maps, whose detailed description can be found in Appendix \ref{app:maps}. Note that an analogous procedure was employed in \cite{rajabpour2015entanglement} to compute the entanglement entropy of region $B$ directly in the microscopic BCFT in specific limits. Our holographic calculation will allow us to reproduce\footnote{Up to corrections associated with bulk fields, which are not captured by our purely geometrical calculation.} the results of \cite{rajabpour2015entanglement} in the appropriate limits, and to study in more detail the entanglement phase transition present in our setup (see Section \ref{comparison}).

We can now identify the spacetime dual to the finite cylinder in $\zeta$ coordinates. Note that the Euclidean path integral on the finite cylinder is the one associated with a BCFT on an interval $[-h,0]$ at finite inverse temperature $\beta=2\pi$ (where we are identifying $x$ with the spatial direction, and $\tau$ with the Euclidean time). The corresponding dual spacetime was studied in \cite{fujita2011aspects}, and is given, depending on the value of $h$, by either a portion of the BTZ black hole or a portion of thermal $AdS_3$, cut off by ETW branes anchored at the boundaries of the boundary manifold (i.e. at the $x=-h$ and $x=0$ circles of the finite cylinder). We will review here the construction of the dual spacetimes carried out in \cite{fujita2011aspects} using a different set of coordinates, which will be useful to map the resulting spacetime to a portion of Poincar\'e $AdS_3$ (see Appendix \ref{mappingtopoincare}). Note that, although the bulk spacetimes and brane trajectories in our analysis are identical to the ones in \cite{fujita2011aspects}, the interpretation of the results is substantially different: in \cite{fujita2011aspects} the BTZ and thermal $AdS_3$ spacetimes are two possible bulk duals of a BCFT on a finite interval at finite temperature; in our case they are a tool to study the effect of a partial projective measurement on a CFT on a circle at zero temperature, whose dual spacetime is obtained after mapping back to the original $\lambda$ coordinates. 

The BTZ/thermal $AdS_3$ metric is given by \cite{namasawa2016epr,Tetradis:2011jn}:
\begin{equation}
    ds^2=\frac{R^2}{z^2}\left[2\left(1\mp \frac{b^2z^2}{2}\right)^2d\tau^2+2\left(1\pm \frac{b^2z^2}{2}\right)^2dx^2+dz^2\right]
    \label{btzthermalmetrics}
\end{equation}
where the upper signs give the BTZ black hole metric and the lower signs the thermal $AdS_3$ metric, and $b$ is a constant to be determined below. We recall that the range of the spatial coordinate $x$ on the asymptotic boundary (which is at $z=0$) is restricted to $x\in [-h,0]$. We also have $\tau\in [-\pi,\pi]$ for the Euclidean time, and $z\in (0,z_0]$ in the holographic direction, with $z_0=\sqrt{2}/b$. In the BTZ case, $z_0$ represents the black hole horizon. If we neglect the presence of the boundaries of the boundary cyilinder at $\{z=0,x=-h,0\}$, these metrics describe a solid torus (see Figures \ref{BTZtorus} and \ref{thermaltorus}).

The value of the constant $b$ is fixed, in the two cases, by smoothness of the geometry. In the BTZ case, where the $\tau$ circle is contractible in the bulk and the $x$ circle is not, the periodicity of the $\tau$ coordinate must be $P_\tau =\pi/b_{BTZ}$ in order to avoid a conical singularity. Since our $\tau$ coordinate is by construction periodic with period $P_\tau =2\pi$, this requirement fixes $b_{BTZ}=1/2$. No restriction is imposed on the periodicity of the $x$ direction in the BTZ case. In the thermal case, where the $x$ circle is contractible in the bulk and the $\tau$ circle is not, the smoothness requirement implies $P_x=\pi/b_{th}$. In order to derive the value of $b_{th}$, we have to remember that the range of the $x$ coordinate at the boundary $z=0$ is restricted to $x\in[-h,0]$. Therefore, the ETW branes must anchor at the endpoints of such interval at the boundary. In particular, the fact that the $x$ direction is contractible in the bulk implies that a single, connected ETW brane is present, anchored at $z=0$ and $x=-h,0$ for each value of $\tau$ (see Figure \ref{thermaltorus} right). It is well known \cite{fujita2011aspects,Almheiri:2018ijj,Cooper:2018cmb} that in three dimensions such connected ETW brane can only anchor at antipodal points of the contractible circle, regardless of its tension $T$\footnote{In higher dimensions, the anchorage points do not need to be antipodal, and they depend on the tension $T$ \cite{Cooper:2018cmb,Antonini:2019qkt}.}. This immediately implies that the periodicity of the $x$ coordinate must be given by twice its range at the boundary, i.e. $P_x=2h$, and therefore $b_{th}=\pi/(2h)$. The bulk range of the $x$ coordinate is therefore $x\in [-h,h]$, with $x=-h$ and $x=h$ identified. The periodicity of the $\tau$ coordinate (i.e. the inverse temperature) is still given by $P_\tau=2\pi$.

\subsubsection{ETW brane trajectories}
\label{branetrajectories}

Now that we have the bulk spacetime metric in the two possible phases, we can compute the ETW brane trajectories in the two cases. Since the metric is independent of $\tau$ and the boundaries of the asymptotic boundary manifold (i.e. the finite cylinder) sit at $x=-h,0$ for any $\tau$, the brane trajectory is also independent of $\tau$ and is given by the same curve $x(z)$ on any given fixed $\tau$ slice. The equation of motion determining the trajectory can be obtained from the Neumann boundary condition (\ref{israeli}), where the tension $T$ is a free parameter\footnote{This is true from the bulk theory point of view. As we have already pointed out, a specific choice of boundary state $\ket{B}$ in our BCFT uniquely determines the boundary entropy and therefore the value of the tension $T$. Choosing a tension in our analysis is therefore equivalent to choosing a specific Cardy state to project region $A$ on.} in the range $-T_c<T<T_c$\footnote{Note that negative tension branes violate standard energy conditions.} with the critical tension given by $T_c=1/R$. We report here the results for the brane trajectories, while their derivation is given in Appendix \ref{app:branetrajectories}.

\subsubsection*{Trajectory in the BTZ black hole}

In the BTZ black hole, there are two disconnected ETW branes. One of them (the ``left brane'') is anchored at the asymptotic boundary at $x=-h$, and the other one (the ``right brane'') is anchored at the asymptotic boundary at $x=0$. Since we imposed identical boundary conditions on the two slits in our BCFT (see Section \ref{slitsection}), the tensions of the two branes are identical, and the two trajectories are symmetric. The trajectories are given by
\begin{equation}
    x_{BTZ}^\pm (z)=\pm \arcsinh \left[\frac{RT}{\sqrt{2(1-R^2T^2)}}\frac{z}{1+\frac{z^2}{8}}\right]+c_\pm
    \label{btztrajectory}
\end{equation}
where the upper signs refer to the right brane and the lower signs refer to the left brane, and $c_+=0$, $c_-=-h$. A few comments are in order.

For vanishing tension $T=0$, the ETW branes are disks in the $\tau -z$ plane sitting at $x=-h,0$, and the range of the $x$ coordinate is $x\in [-h,0]$ for any given value of $z$. For positive tension $T>0$, the spacetime domain present in our solution is enlarged: the range of the $x$ coordinate increases from $[-h,0]$ at the asymptotic boundary ($z=0$) to $[-h-a(z),a(z)]$ as we go into the bulk, where $a(z)$ is given by the absolute value of the first term in equation (\ref{btztrajectory}). The largest range for the $x$ coordinate is obtained at the black hole horizon $z=z_0=2\sqrt{2}$ where $a_{max}\equiv a(2\sqrt{2})= \arcsinh(RT/\sqrt{1-R^2T^2})$. Analogously, for negative tension $T<0$, the spacetime domain is shrunken, and the range of the $x$ coordinate diminishes from $[-h,0]$ at the asymptotic boundary ($z=0$) to $[-h+a(z),-a(z)]$ as we go into the bulk. Note that $a_{max}$ diverges as we approach the critical value of the tension $T_c$. 

\begin{figure}
    \centering
    \includegraphics[width=0.9\textwidth]{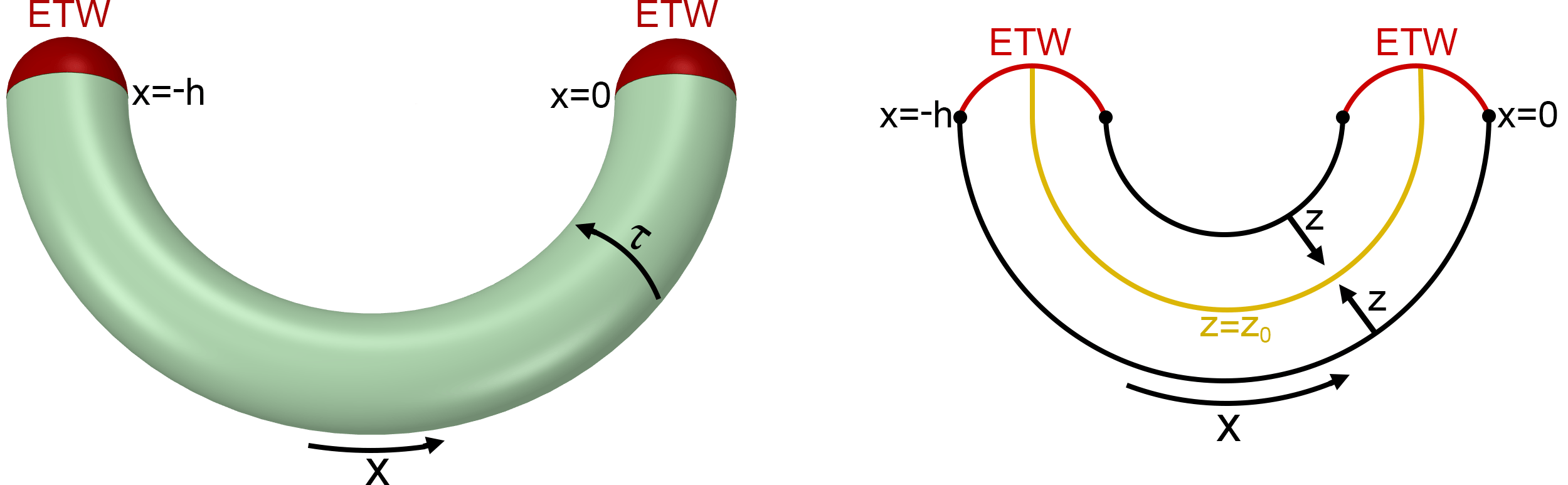}
    \caption{Left: Bulk dual of the finite cylinder in the BTZ phase. The non-contractible $x$ circle is cut off at $x=-h,0$ at the boundary. Two disconnected ETW branes extend from such boundary circles into the bulk. In the positive tension brane case depicted here, the spacetime is enlarged with respect to the tensionless case. The bulk domain is given by $\tau\in [-\pi,\pi]$, $z\in [0,2\sqrt{2}]$, $x\in [x_{BTZ}^-(z),x_{BTZ}^+(z)]$. Right: Constant $\tau$ slice of the bulk spacetime in the BTZ phase. Such a slice is defined by $\tau=c,c\pm\pi$ for some constant $c$.}
    \label{BTZtorus}
\end{figure}

As we have already explained, in the BTZ case there is no restriction on the periodicity of the $x$ coordinate. In particular, the periodicity can be chosen arbitrarily without affecting the brane trajectory, nor the on-shell value of the action (and consequently the phase structure), which depends only on the range of $x$ between the two branes (see Appendix \ref{app:actions}).\footnote{In principle, it is possible to choose a periodicity for the $x$ coordinates small enough that the two branes intersect each other for some given large tension, giving rise to a different phase structure. We exclude this case from our analysis, since it would lead to a non-smooth intersection, requiring further analysis which is beyond the scope of the present paper.} We can therefore set the periodicity to infinity, and see the BTZ phase as a filled-in cylinder; for positive tension, the bases of the cylinder are ``pulled out'' (see Figure \ref{BTZtorus}), while for negative tension they are ``pushed in''. In order to avoid intersections of the two branes for large negative tensions, we will restrict the tension parameter to the range $T_*<T<T_c$, where $T_*=-T_c\tanh (h/2)$.

\subsubsection*{Trajectory in thermal $AdS_3$}

In thermal $AdS_3$ a single, connected brane is present, emerging from the asymptotic boundary at $\{z=0,x=\pm h\}$\footnote{We remind that the periodicity of the $x$ direction in the thermal $AdS_3$ phase  is given by $P_x=2h$, and therefore $x=\pm h$ are identified. In general, $x\sim x+2h$.}, extending into the bulk until a turning point at $z=z_{max}$, and reaching the asymptotic boundary again at $\{z=0,x=0\}$. Such brane is anchored at antipodal points, as we have explained above. The brane trajectory is given by 
\begin{equation}
    x^\pm_{th}(z)=\pm\frac{h}{\pi}\arctan \left[\frac{\pi RT}{\sqrt{2}h}\frac{z}{\sqrt{\left(1-T^2R^2\right)\left(1+\frac{\pi^2z^2}{8h^2}\right)^2-\frac{\pi^2z^2}{2h^2}}}\right]+d_\pm
    \label{thermaltrajectory}
\end{equation}
where the upper signs correspond to the part of the trajectory from $\{z=0,x=0\}$ to the turning point, and the lower signs to the part of the trajectory from the turning point to $\{z=0,x= h\}$. The constants are given by $d_+=0$, $d_-=h$.

\begin{figure}
    \centering
    \includegraphics[width=0.9\textwidth]{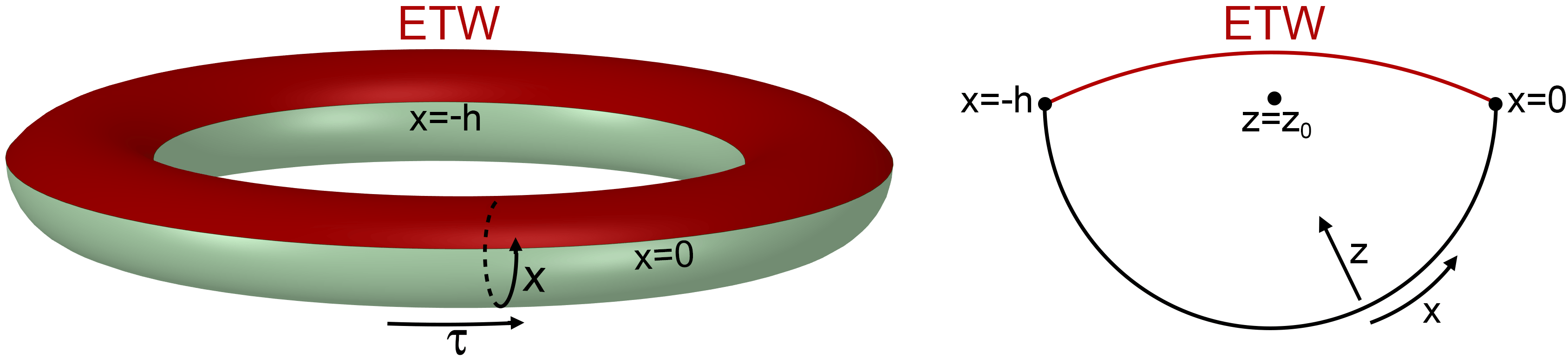}
    \caption{Left: Bulk dual of the finite cylinder in the thermal $AdS_3$ phase. The contractible $x$ circle, with circumference $2h$, is cut off at $x=-h,0$ at the boundary. A single connected ETW brane emerges from the asymptotic boundary at $\{z=0,x=-h\}$, extends into the bulk until the turning point at $z=z_{max}$, and reaches the asymptotic boundary again at $\{z=0,x=0\}$. In the positive tension brane case depicted here, the spacetime is enlarged with respect to the tensionless case, and more than half of the torus is retained in our geometry. The bulk domain is given by $\{\tau\in [-\pi,\pi]; z\in [0,2\sqrt{2}h/\pi]; x\in[-h,h]\}\setminus \{\tau\in [-\pi,\pi]; z\in [0,z_{max}]; x\in [x_{th}^+(z),x_{th}^-(z)]\}$, where we are subtracting from the full torus the portion excised by the ETW brane. Right: Fixed $\tau$ slice of the thermal $AdS_3$ geometry cut off by a (positive tension) connected ETW brane anchored at the boundary at antipodal points of the contractible circle.}
    \label{thermaltorus}
\end{figure}

The value of $z$ at the turning point is given by $z_{max}=2\sqrt{2}h\sqrt{1-R|T|}/(\pi\sqrt{1+R|T|})$, which reduces to $z_0=2\sqrt{2}h/\pi$ in the tensionless case, and vanishes for critical tension. Note that the value of $x$ at the turning point is $x=-h/2$ for negative tension and $x=h/2$ for positive tension: for $T=0$ the torus is sliced exactly in half; for $T>0$ more than half torus is retained in our geometry (see Figure \ref{thermaltorus}); for $T<0$ less than half torus is retained in our geometry. As $T\to T_c$, we recover the whole torus, and as $T\to - T_c$, the volume of the bulk geometry vanishes.\footnote{Note that, unlike the BTZ phase, in the thermal $AdS_3$ phase there is no obstruction to considering negative near-critical tensions. However, because we are able to analyze the competing BTZ phase only for $T_*<T<T_c$ as explained above, we restrict to this range also in the thermal $AdS_3$ phase.} 

\subsubsection{Action comparison and phase transition}
\label{actioncomparison}

Now that we have completely characterized the cut off bulk geometries in the BTZ and thermal $AdS_3$ cases, we can compute their on-shell Euclidean action to determine which saddle geometry is dominant in the gravitational Euclidean path integral. The Euclidean action is given by
\begin{equation}
    I=-\frac{1}{16\pi G}\int_{\mathcal{N}}d^3x\sqrt{g}\left(\mathcal{R}-2\Lambda\right)-\frac{1}{8\pi G}\int_{ETW}d^2x\sqrt{h}\left(K-T\right)+I_{GHY}+I_{CT}
    \label{action}
\end{equation}
where $G$ is Newton's constant, $\mathcal{N}$ is the spacetime manifold (whose boundary is given by the union of the asymptotic boundary $\mathcal{M}$ and the ETW brane: $\partial\mathcal{N}=\mathcal{M}\cup \textrm{ETW}$), $g$ is the determinant of the spacetime metric, $\mathcal{R}$ is the Ricci scalar, $\Lambda=-1/R^2$ is the cosmological constant, $h$ is the determinant of the metric induced on the brane, $K$ is the trace of the extrinsic curvature, $I_{GHY}$ is the Gibbons-Hawking-York term for the asymptotic boundary $\mathcal{M}$ \cite{york,gibbons}, and $I_{CT}$ are counterterm contributions (see Appendix \ref{app:actions}).

Evaluating the BTZ and thermal $AdS_3$ Euclidean actions on-shell using the metrics (\ref{btzthermalmetrics}) and the brane trajectories (\ref{btztrajectory}) and (\ref{thermaltrajectory}) after introducing a UV cutoff at $z=\varepsilon$, and then taking the $\varepsilon\to 0$ limit, we obtain (see Appendix \ref{app:actions} for a detailed derivation)
\begin{equation}
    I_{BTZ}=-\frac{Rh}{8G}-\frac{R}{2G}\arcsinh\left(\frac{RT}{\sqrt{1-R^2T^2}}\right)=-\frac{ch}{12}-\frac{c}{3}\arcsinh\left(\frac{RT}{\sqrt{1-R^2T^2}}\right)
    \label{btzaction}
\end{equation}
for the BTZ black hole and 
\begin{equation}
    I_{th}=-\frac{\pi^2 R}{8Gh}=-\frac{\pi^2 c}{12h}
    \label{thermalaction}
\end{equation}
for thermal $AdS_3$, where in the last equalities we introduce the dual CFT's central charge $c=3R/(2G)$.

In order to determine the phase diagram, we must now compare the on-shell actions for the two saddles. The dominant saddle is the one with the least action. We immediately get
\begin{equation}
    \Delta I= I_{th}-I_{BTZ}=\frac{c}{12h}\left[h^2+4h\arcsinh\left(\frac{RT}{\sqrt{1-R^2T^2}}\right)-\pi^2\right]
\label{actiondifference}
\end{equation}
which implies that the thermal $AdS_3$ phase is dominant for $h<h_c$, and the BTZ phase is dominant for $h>h_c$ where the critical value $h_c$ is given by
\begin{equation}
    h_c\equiv -2\arcsinh\left(\frac{RT}{\sqrt{1-R^2T^2}}\right)+\sqrt{4\arcsinh^2\left(\frac{RT}{\sqrt{1-R^2T^2}}\right)+\pi^2}.
    \label{hcdefinition}
\end{equation}
These results are in agreement with the ones obtained in \cite{fujita2011aspects}. 

\begin{figure}
    \centering
    \includegraphics[width=0.45\textwidth]{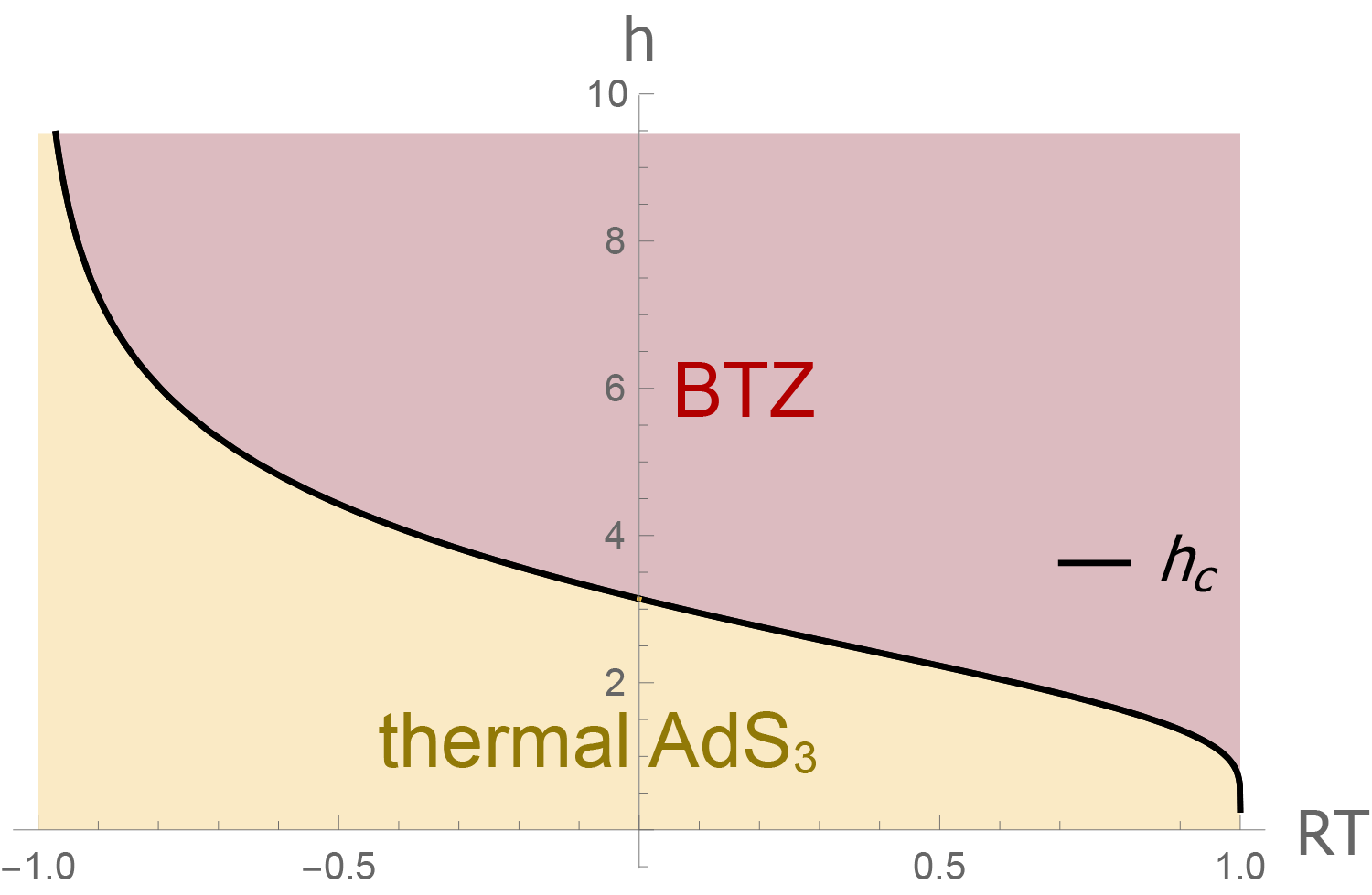}
    \hspace{0.5cm}
    \includegraphics[width=0.45\textwidth]{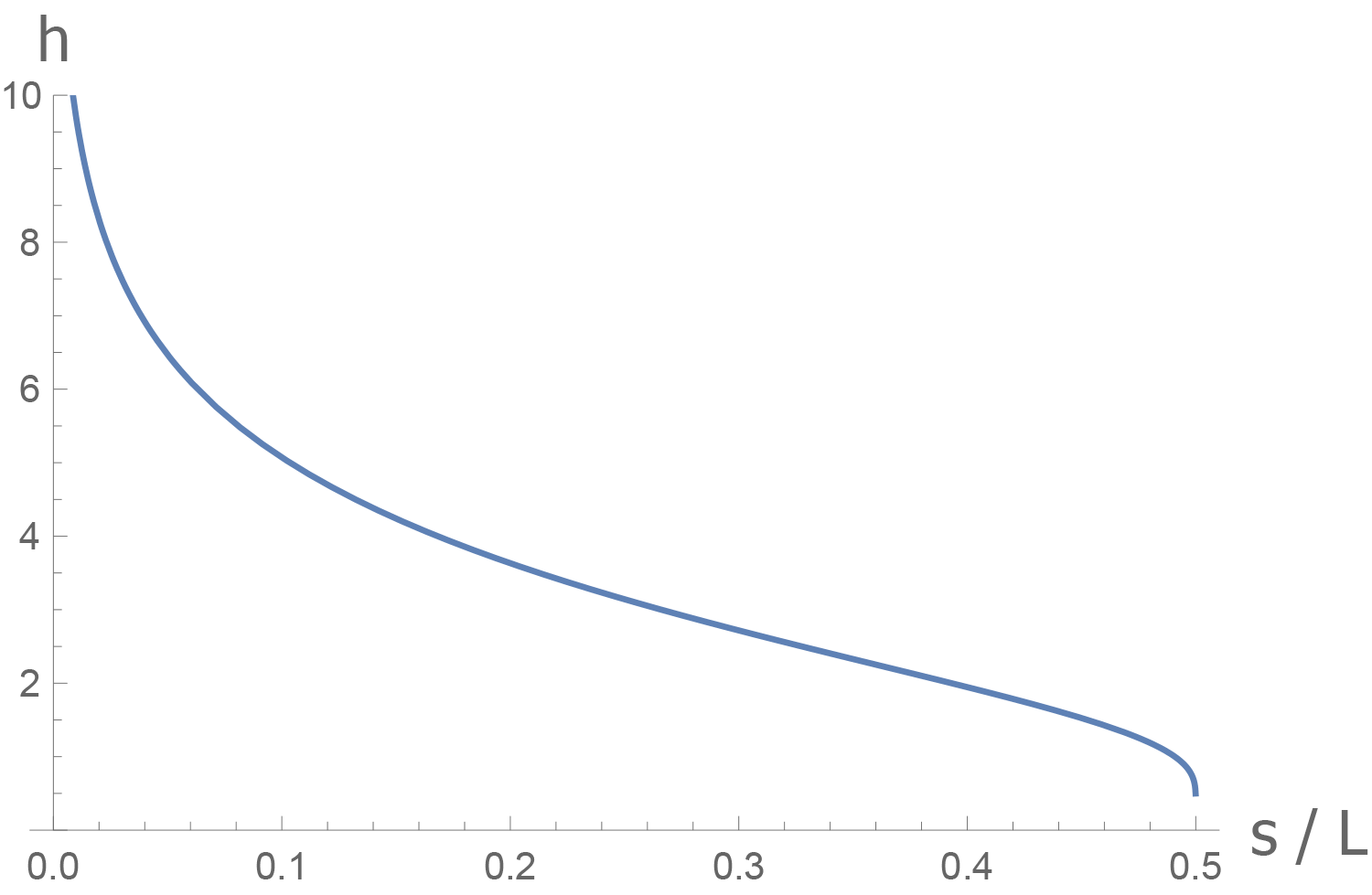}
    \caption{Left: phase diagram for the BTZ/thermal $AdS_3$ (connected/disconnected) phase transition. The critical value of $h$ is diverging for negative critical tension, it vanishes for positive critical tension and takes the value $h_c=\pi$ in the tensionless case. Right: $h$, defined in equation (\ref{hdefinition}), is a monotonically decreasing function of the size $s$ of the measured regions $A_1$ and $A_2$. When exactly half of the CFT is measured (i.e. $s/L=1/4$), $h=\pi$. In the tensionless case, this is also the (Hawking-Page) critical point.}
    \label{hcplot}
\end{figure}

Note that the critical value $h_c$ depends exclusively on the brane tension. In particular, for $T=0$ we get $h_c=\pi$\footnote{Note that this value of $h_c$ for the $T=0$ case corresponds to the periodicity of the contractible circles in the two phases ($\tau$ coordinate in the BTZ phase and $x$ coordinate in the thermal $AdS_3$ phase) to be equal, i.e. $P_x^{th}=P_\tau^{BTZ}=2\pi$. Therefore the values of $z_0$ coincide in the two cases. This is the critical point of the Hawking-Page transition \cite{fujita2011aspects}.}. From equations (\ref{hdefinition}) and (\ref{kdefinition}), and using the constraint $L=2\ell +2s$ (where we remind that $s$ is the size of the two measured regions $A_1$ and $A_2$ in the original cylinder with slits in $\lambda$ coordinates, and $\ell$ is the size of the two remaining regions $B$ and $C$), we find that $h=\pi$ corresponds to $\ell=s=L/4$: in the tensionless case, the phase transition happens when more than half of the CFT is measured. We also find that $h_c$ is a monotonically decreasing function of the tension---it is diverging for $T=-T_c$, and vanishes for $T=T_c$ (Figure \ref{hcplot} left)---whereas $h$ is a monotonically decreasing function of the size of the measured region $s$ (Figure \ref{hcplot} right). This implies that for any fixed value of the tension, the BTZ phase is favored when the measured region $A$ is small and the thermal $AdS_3$ phase is favored when $A$ is large. But it also implies that for any size of $A$ the BTZ phase is dominant for a sufficiently large value of the tension, and the thermal $AdS_3$ phase is dominant for a sufficiently small value of the tension (but we remind the additional constraint $T>T_*$ for negative tensions). 

We would like to remark that this phase transition depends on how the measurement is performed on the boundary theory, and in particular on the size and shape of region $A$ and the post-measurement state we are projecting on. Therefore, it can be viewed as a measurement-induced phase transition, although of a different kind with respect to the dynamical phase transitions induced by multiple measurements which have been observed in quantum many-body systems \cite{aharonov,Skinner:2018tjl,Li:2018mcv,Chan:2018upn,Choi:2019nhg,Bentsen:2021ukm,Li2}.

\subsection{Connectivity of the $\lambda_2=0$ slice and bulk teleportation}
\label{sec:connectivity}

One way to understand whether the measurement teleports bulk information from region $A$ to region $A^c$ in our setup is to study the connectivity of the $\lambda_2=0$ slice in the bulk spacetime dual to our cylinder with slits. To understand why, suppose that $|A|>L/2$, i.e. we are measuring more than half of the CFT. Then the pre-measurement entanglement wedge $W(A)$ is connected and contains the center of the bulk on the $\lambda_2=0$ slice. If the post-measurement slice is connected between $B$ and $C$, and therefore it contains the center of the bulk, the latter can now be reconstructed from $A^c=B\cup C$, meaning that bulk information has been teleported by the measurement. When the slice is connected, we also expect $B$ and $C$ to share entanglement. Although the singular nature of our setup prevents us from explicitly building the bulk dual of our cylinder with slits (see the discussion in Section \ref{sec:bulkdual} and Appendix \ref{sec:mapback}), we can understand whether the $\lambda_2=0$ slice is connected or not by studying the bulk geometries we built in Section \ref{sec:bulkdual}. Since the conformal mapping procedure we followed acts as a de facto regularization of the singular behavior near the endpoints of the slits and allows to build a well-defined, non-singular bulk dual spacetime, we expect the results we find to hold also in regularized versions of our setup (see footnote \ref{regfootnote}). The tensor network models studied in Sections \ref{sec:happycode} and \ref{sec:rtn}, which are naturally-regularized toy models of our setup, provide evidence in support of this expectation.

\begin{figure}
    \centering
    \includegraphics[width=0.8\textwidth]{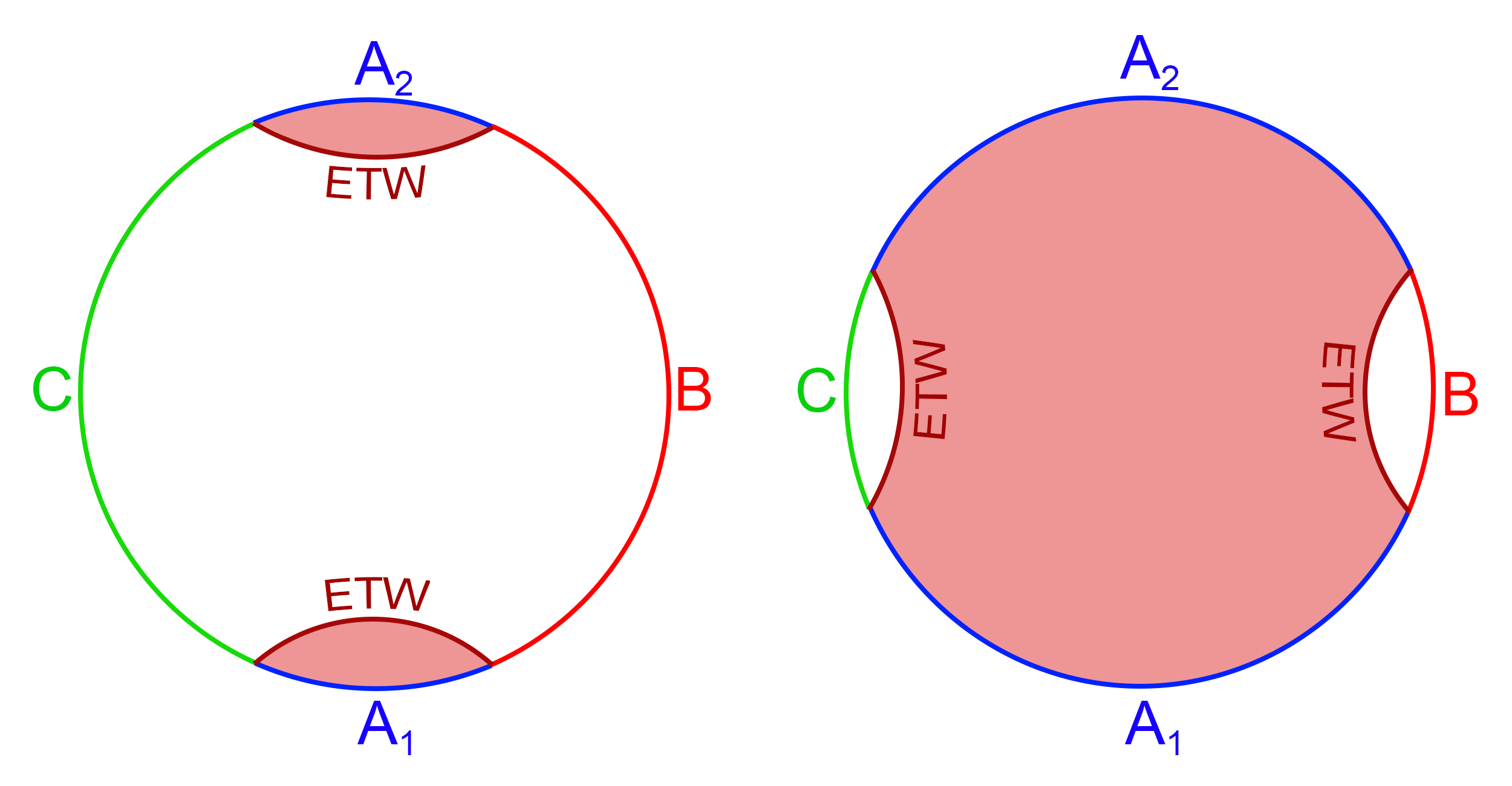}
    \caption{Schematic representation of the $\lambda_2=0$ slice in the original coordinates describing the infinite cylinder with slits. The regions shaded in red are cut off by ETW branes and are not part of the geometry. Left: In the BTZ (connected) phase the slice is connected, and regions $B$ and $C$ share a large amount of entanglement. Right: In the thermal $AdS_3$ (disconnected) phase the slice is disconnected, and $B$ and $C$ are disentangled by the measurement performed on region $A$.}
    \label{slices}
\end{figure}

In the BTZ phase each one of the two branes anchors to one boundary of the boundary finite cylinder in $\zeta=x+i\tau$ coordinates. In terms of our original cylinder with slits in $\lambda$ coordinates, this implies that there are two branes anchoring to the two slits separately, and the bulk $\lambda_2=0$ slice is connected. We therefore refer to the BTZ phase also as the ``connected phase''. On the contrary, the thermal $AdS_3$ phase has a single connected brane anchored at the two boundaries of the boundary finite cylinder. In the original cylinder with slits in $\lambda$ coordinates, this translates to a single connected brane anchored to the two slits. Therefore, the bulk $\lambda_2=0$ slice is disconnected into two separate pieces. We then refer to the thermal $AdS_3$ phase also as the ``disconnected phase''. A qualitative picture of the $\lambda_2=0$ slice in the two phases is reported in Figure \ref{slices}.

The fact that the BTZ phase can be dominant for any value of $s$ for a sufficiently large tension is a non-trivial result. In fact, it means that for sufficiently large positive $T$ we can measure a very large region $A=A_1\cup A_2$ and still obtain a connected slice between the two small regions $B$ and $C$ (see Figure \ref{largebulk}). This implies that, when we postselect region $A$ on a Cardy state with very large boundary entropy, we do not disentangle the remaining regions $B$ and $C$, no matter how large $A$ is. From an entanglement wedge reconstruction point of view, we would naively expect that the bulk information within $W(A)$ is not accessible from $A^c=B\cup C$ after measuring a large region $A$. Since the pre-measurement $W(A)$ is connected and includes the center of the bulk on the $\lambda_2=0$ slice, we would expect $W(A^c)$ and the bulk $\lambda_2=0$ slice to disconnect after the measurement. However, our results provide evidence that this is not case. If $T$ is sufficiently large, the entanglement wedge $W(A^c)$ of the complementary region $A^c$ on the boundary is connected after the measurement and contains the center of the bulk on the $\lambda_2=0$ slice. In other words, the effect of the measurement on a Cardy state with large boundary entropy is to teleport part (or all, in the critical tension case $T\to T_c$) of the bulk information contained in $A$ before the measurement into $A^c=B\cup C$. On the other hand, in the $T=0$ case regions $B$ and $C$ become disentangled when we measure more than half of the CFT, and their entanglement wedge is disconnected. In this case, the information within $W(A)$ cannot be accessed from $A^c$ after we measure $A$: bulk teleportation does not occur. Finally, for negative tensions (corresponding to negative boundary entropies in the BCFT) the measurement destroys an even larger portion of the bulk spacetime.

The fact that for large tension we are able to retain most of the bulk spacetime even when we measure a large region $A$ of the boundary and that such a spacetime is encoded in a small boundary region $A^c$ may seem problematic. However, as we have pointed out in the introduction and will describe in more detail in Section \ref{sec:boundmeasuremteleportation}, the reason why this can happen is that in semiclassical holography the dual boundary theory is a large-$N$ gauge theory, and therefore contains a parametrically larger number of degrees of freedom with respect to the dual low-energy bulk theory. Therefore, even a small boundary region $A^c$ has enough entanglement resource to store the bulk information of a large spacetime. For the same reason, in Section \ref{sec:rtn} we will also argue that the bulk-to-boundary encoding map in our holographic setup remains isometric independently of the size of $A$. The same result cannot be achieved in the original HaPPY code toy model \cite{Pastawski:2015qua}, where the amount of entanglement resource is limited and therefore we cannot teleport an arbitrarily large part of the bulk into the complementary region (see Figure \ref{fig:happystackteleport}). 

In order to observe more explicitly the connectivity of the bulk $\lambda_2=0$ slice, we can go one step further and understand what the bulk $\lambda_2=0$ slice is mapped to in our BTZ/thermal $AdS_3$ geometries. First, we note that regions $B$ and $C$ on the boundary are mapped to $z=0,\tau=0$ and $z=0, \tau=\pm \pi$ respectively in our bulk spacetimes. We can then expect that, away from the values of $\lambda$ corresponding to the two slits, the $\lambda_2=0$ slice in the bulk dual of the cylinder with slits is mapped to the bulk $\tau=0,\pm\pi$ slice in our BTZ/thermal $AdS_3$ geometries. By explicitly mapping the BTZ/thermal $AdS_3$ bulk domains to $\lambda$ coordinates as explained in Appendix \ref{sec:mapback}, we found that our expectation is confirmed by numerical results. However, at the values of $\lambda$ corresponding to the two slits, things become more complicated. In fact, the slits are branch cuts and our conformal maps ``open them up'' and map them to the $z=0$, $x=-h$ and $z=0$, $x=0$ circles in BTZ/thermal $AdS_3$ coordinates. As a result, in the BTZ geometry the image of the bulk $\lambda_2=0$ slice is given by the union of the $\tau=0,\pm \pi$ slice and two perpendicular disks given by $x=-h,0$; in the thermal $AdS_3$ geometry, it is given by the union of the $\tau=0,\pm \pi$ slice and the perpendicular $\{-\pi\leq \tau\leq 0$, $x=0,\pm h\}$ half annulus. This result has also been checked using a numerical map. Evidently, this peculiar shape of the image of the $\lambda_2=0$ slice is a consequence of the singularity of our setup. In a regularized system, the top and bottom edges of the slits do not sit on the boundary of the $\lambda_2=0$ slice. Therefore, the image of the slice in BTZ/thermal $AdS_3$ coordinates does not contain the $x=-h,0$ circles on the boundary, nor the perpendicular disks/half annulus in the bulk. Since we are interested in finding results that hold beyond our singular system, we will work under the assumption that some regularization procedure has been carried out and the $\lambda_2=0$ slice is mapped to the $\tau=0,\pm\pi$ slice in the BTZ/thermal $AdS_3$ geometries. This makes particularly evident how the $\lambda_2=0$ slice is connected in the BTZ phase and disconnected in the thermal $AdS_3$ phase (see Figures \ref{BTZRT} and \ref{thermalRT}). We will also see in Section \ref{sec:entropy} that this assumption gives a result for the entanglement entropy of $B$ consistent with previous results obtained from a purely BCFT calculation. We remark that even including the singular parts of the slice described above its connectivity between $B$ and $C$ is not spoiled, and is therefore a robust result of our analysis.

\subsection{Entanglement entropy of region $B$}
\label{sec:entropy}

We are now ready to compute the holographic entanglement entropy of region $B$. As we have seen in the previous subsection, the bulk $\lambda_2=0$ slice in the bulk dual to the infinite cylinder with slits is mapped to the $\tau=0,\pm\pi$ bulk slice in our BTZ/thermal $AdS_3$ geometries. In particular, the boundary region $B$ we are interested in is given by the $z=0,\tau=0$ segment on the asymptotic boundary of our bulk geometries. We can then apply the RT formula \cite{Ryu2006a,Ryu2006b} to such a segment to obtain the entanglement entropy. We remind that, in the presence of ETW branes, the RT surface is allowed to end on them \cite{takayanagi2011holographic,fujita2011aspects}.

\begin{figure}
    \centering
    \includegraphics[width=0.55\textwidth]{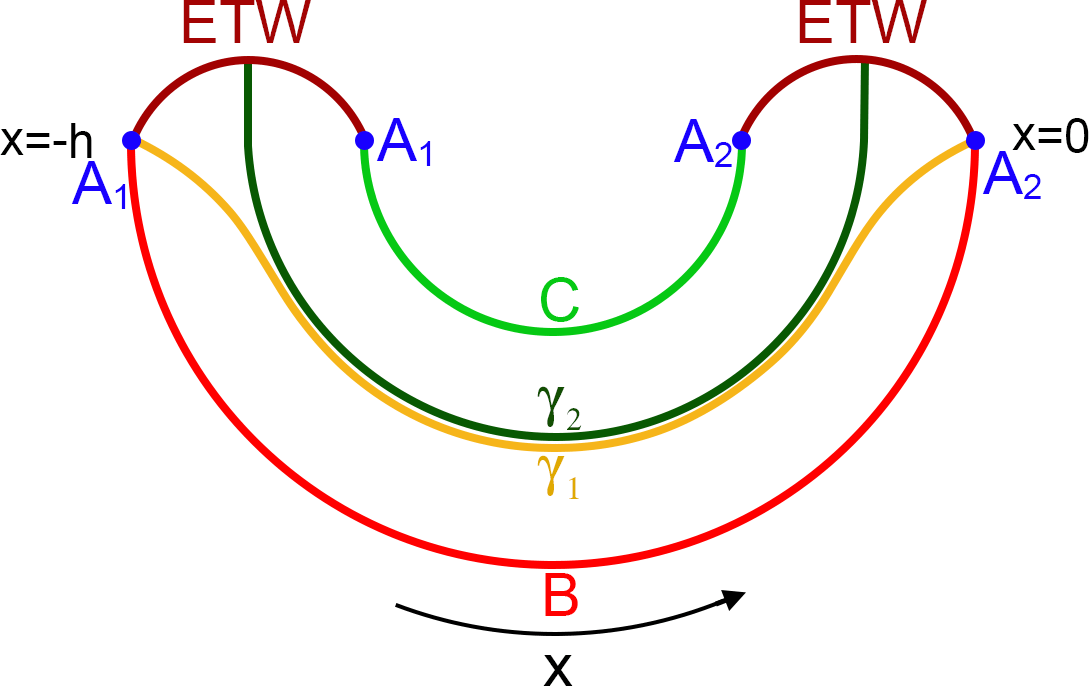}
    \caption{$\tau=0,\pm\pi$ slice of the BTZ black hole geometry cut off by ETW branes. The two candidate RT surfaces for region $B$ are depicted. $\gamma_2$ is always dominant.}
    \label{BTZRT}
\end{figure}

Let us start with the BTZ black hole phase. The $\tau=0,\pm\pi$ slice is connected, and there are two candidate RT surfaces homologous to region $B$ (see Figure \ref{BTZRT}). The first one is the ``usual'' geodesic surface $\gamma_1$ anchored at $\{z=0, x=-h\}$ and $\{z=0,x=0\}$. This is the only possible RT surface in the absence of ETW branes, and gives the entanglement entropy for a boundary subregion of length $h$ in a full BTZ black hole geometry. The second one is the surface $\gamma_2$ running along the black hole horizon $z=2\sqrt{2}$ and connecting the left and right ETW branes. It is immediate to conclude that the latter candidate is always the dominant one. In fact, the area (i.e. length) of $\gamma_1$ is divergent as it has to approach the asymptotic boundary; on the other hand, the area of $\gamma_2$ is always finite for any sub-critical value of the tension $T$. Such area is given by the proper distance between the two branes along the black hole horizon. The line element along the black hole horizon is given by
\begin{equation}
    dl=\left.\sqrt{2}R\frac{1+\frac{z^2}{8}}{z}dx\right|_{z=2\sqrt{2}}=Rdx
\end{equation}
and the area of the second candidate RT surface is
\begin{equation}
    A(\gamma_2)=R\int_{-h-a_{max}}^{a_{max}}dx=Rh+2R\arcsinh \left(\frac{RT}{\sqrt{1-R^2T^2}}\right)
\end{equation}
where we used the definition of $a_{max}$ introduced in Section \ref{branetrajectories}. The entanglement entropy of $B$ in the BTZ phase is then given by
\begin{equation}
    S_{BTZ}(B)=\frac{Rh}{4G}+\frac{R}{2G}\arcsinh \left(\frac{RT}{\sqrt{1-R^2T^2}}\right).
\end{equation}
This result shows that, whenever the BTZ phase is dominant, regions $B$ and $C$ still share entanglement after a projective measurement on region $A$ is performed. In particular, since the full state on the regions $B\cup C$ is pure by construction, the mutual information will be simply given by twice the entanglement entropy of $B$: $I(AB)=S(A)+S(B)-S(AB)=2S(B)$. This result was to be expected, given the connectivity of the post-measurement bulk slice in the BTZ phase.

\begin{figure}
    \centering
    \includegraphics[width=0.8\textwidth]{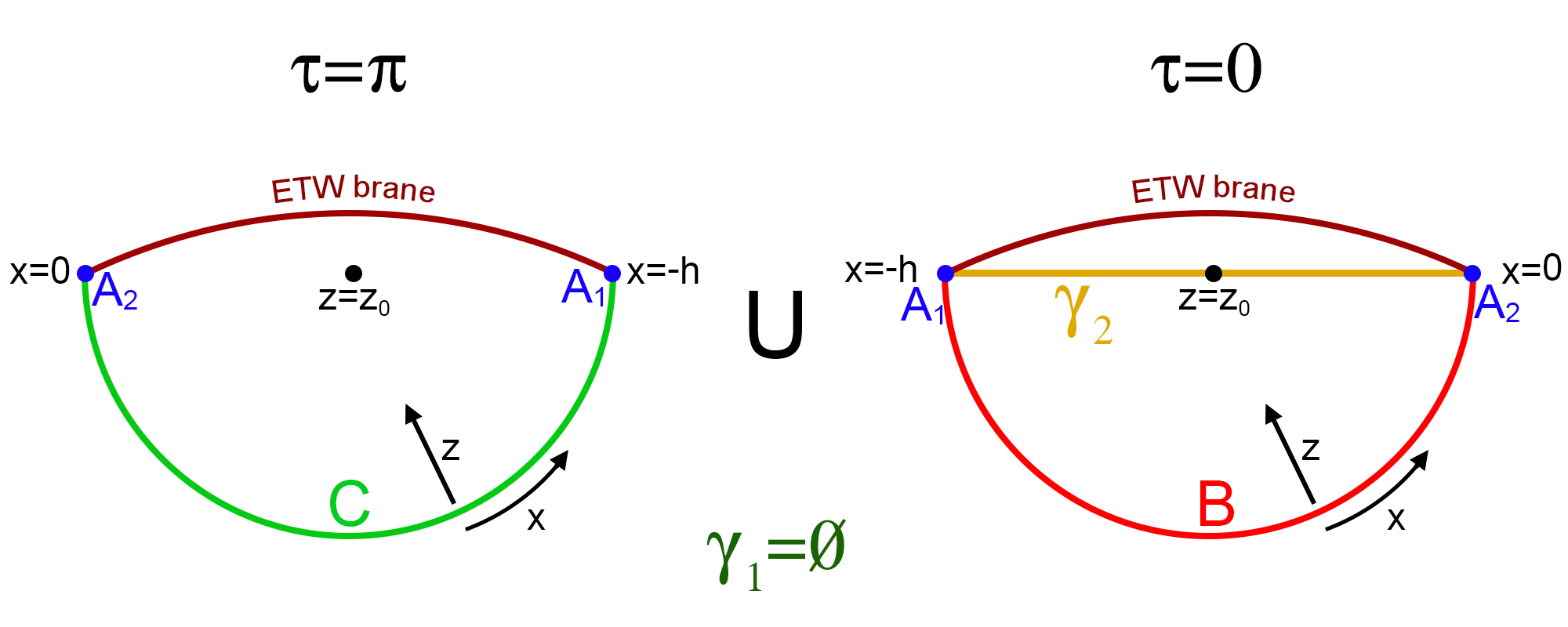}
    \caption{$\tau=0,\pm\pi$ slice of the thermal $AdS_3$ geometry cut off by ETW branes. The ``usual'' candidate RT surface $\gamma_2$ is depicted in yellow. Such surface is always subdominant, because the empty set $\gamma_1$, which has zero area, is always dominant.}
    \label{thermalRT}
\end{figure}

In the thermal $AdS_3$ phase, the $\tau=0,\pm\pi$ slice is given by two disconnected regions. Specifically, the boundary of one of such regions is given by the union of the boundary region $B$ and the $\tau=0$ section of the ETW brane; the boundary of the other region is given by the union of the boundary region $C$ and the $\tau=\pm\pi$ section of the ETW brane (see Figure \ref{thermalRT}). This implies that the empty set $\gamma_1=\emptyset$ is homologous to region $B$ and therefore it is a candidate RT surface, along with the ``usual'' geodesic surface $\gamma_2$ anchored at the boundaries of region $B$. Since it has zero area, the empty set is always the dominant candidate. We conclude that the entanglement entropy of region $B$ in the thermal $AdS_3$ phase vanishes, which implies that, whenever the thermal $AdS_3$ phase is dominant, regions $B$ and $C$ are completely disentangled by the projective measurement on region $A$. The mutual information $I(AB)$ is clearly also vanishing.

Our result for the entanglement entropy of region $B$ (and therefore $C$) is then
\begin{equation}
S(B)=S(C)=\begin{cases}
    \frac{Rh}{4G}+\frac{R}{2G}\arcsinh \left(\frac{RT}{\sqrt{1-R^2T^2}}\right) \hspace{2cm} h>h_c\\
    0\hspace{6.55cm} h<h_c
    \end{cases}
\end{equation}
where $h_c$ is the critical value of $h$ defined in equation (\ref{hcdefinition}). This result makes manifest how the BTZ/thermal $AdS_3$ (connected/disconnected) phase transition we have found in our bulk analysis corresponds to an entangled/disentangled phase transition in the dual boundary theory. The phase transition is triggered by the local projective measurement and can be viewed as a non-dynamical measurement-induced phase transition. Additional insight on how the phase transition and the associated bulk teleportation arise microscopically is provided in Section \ref{sec:boundmeasuremteleportation} using quantum error correcting codes, and explicit realizations in tensor network models of holography are described in Section \ref{sec:happycode} and Section \ref{sec:rtn}.

\subsubsection{Comparison with previous results}
\label{comparison}

We can now compare our results for the entanglement entropy with the ones obtained from a CFT replica calculation in the same setup in \cite{rajabpour2015entanglement}. In particular, we can look at two different limits analyzed in \cite{rajabpour2015entanglement}. 

\subsubsection*{$L/2\gtrsim \ell\gg s$ limit}

Let us first look at the limit in which the measured region $A$ is much smaller than the remaining region $B\cup C$. Let us take $s$ as small as possible, i.e. of the same size as the CFT UV cutoff $\varepsilon$ in the original $\lambda$, $\bar{\lambda}$ coordinates. In this regime, $h$ is very large and the BTZ black hole phase is always the dominant one\footnote{The thermal $AdS_3$ phase is dominant in this regime only for negative values of the tension which are tuned extremely close (within order $\varepsilon$) to criticality.}. Using the definitions (\ref{hdefinition}) and (\ref{kdefinition}), we can expand $h$ in powers of $s=\varepsilon$, obtaining
\begin{equation}
    h=2\log\left[\frac{L}{\pi \varepsilon}\sin\left(\frac{\pi\ell }{L}\right)\right]+\mathcal{O}(\varepsilon^0).
\end{equation}
The entanglement entropy of region $B$ is then given by
\begin{equation}
    S(B)=\frac{c}{3}\log\left[\frac{L}{\pi\varepsilon}\sin\left(\frac{\pi\ell}{L}\right)\right]+\mathcal{O}(\varepsilon^0)
    \label{largellimit}
\end{equation}
where we used $c=3R/(2G)$. The result (\ref{largellimit}) is in agreement with the one obtained in \cite{rajabpour2015entanglement}, and reproduces the well known result for the entanglement entropy of a subregion of size $\ell$ in the vacuum state of a CFT on a circle of length $L$ \cite{Calabrese:2009qy}. This is to be expected: if we measure a very small region $A$, the entanglement entropy of any other region should not be affected at leading order, and should be given by the entanglement entropy in the vacuum state of the CFT, which is the state on the $\lambda_2=0$ slice of the cylinder with slits when we take the limit of infinitesimally small slits.

\subsubsection*{$L/2\gtrsim s\gg \ell$ limit}

In the opposite limit, in which we take the unmeasured regions $A$ and $B$ to be of the same size as the UV cutoff $\varepsilon$, the thermal $AdS_3$ phase is always dominant (unless we tune the tension extremely close to the positive critical value). Therefore, the entanglement entropy is always vanishing in this limit. In the same limit, the author of \cite{rajabpour2015entanglement} found from a CFT replica calculation that the entanglement entropy takes the form
\begin{equation}
    S(B)=\left(\frac{\pi\varepsilon}{4L}\right)^{4\Delta_1}\log\left(\frac{\pi\varepsilon}{4L}\right)
    \label{powerlaw}
\end{equation}
where $\Delta_1$ is the smallest scaling dimension in the CFT. This result is vanishing as $\varepsilon\to 0$, as expected. It is not a surprise that our holographic calculation is incapable of capturing the power-law (in the cutoff) behavior expressed in equation (\ref{powerlaw}). In fact, equation (\ref{powerlaw}) is independent of the central charge $c$, which suggests that it cannot be obtained in the large-$N$ holographic limit with a purely geometrical calculation. In other words, the power-law decay is likely associated with the entanglement entropy of bulk fields living in our background geometry. A detailed analysis of the contribution of matter to entanglement entropy \cite{Engelhardt:2014gca} in our setup is beyond the scopes of the present paper and we leave it to future work.

We would like to remark that our holographic calculation not only reproduces, in the appropriate limits, the results obtained in \cite{rajabpour2015entanglement} with a CFT replica calculation, but it also provides a simple geometric interpretation of such results, while giving new insight about the phase transition, the phase structure of the system, and the effects of projecting region $A$ on different Cardy states with different boundary entropies.

\section{Boundary measurements and quantum teleportation}\label{sec:boundmeasuremteleportation}
In Section \ref{sec:holographiccalculation}, we found that it is possible to obtain different geometric configurations in holography where the end-of-the-world branes have different tensions. Most intriguingly, one can retain most of the bulk information from $A$ in $A^c$ post-measurement as long as the tensions are high enough. While we claimed that it is related to teleportation, the precise process through which such information is teleported in the holographic description remains relatively opaque. In this section, we will first approach these somewhat surprising phenomena  using a simple 5-qubit toy model for holography. We find that an analogous outcome can be identified even in this simplistic example. We then generalize this observation and understand it from the point of view of a quantum teleportation protocol, which holds for all quantum erasure correction codes with suitably factorizable code subspaces. 

\subsection{Example: five-qubit code}

 Let us first examine the example of a [[5,1,3]] code, which encodes 1 logical qubit into 5 physical qubits and can correct any single qubit error. They are Pauli stabilizer codes whose code subspace is given by the simultaneous $+1$ eigenspace of the abelian Pauli subgroup

\begin{equation}
    S=\langle XZZXI, IXZZX, XIXZZ, ZXIXZ\rangle,
\end{equation}
where a representation of the logical operators is $\bar{X}=XXXXX,\bar{Z}=ZZZZZ$. Other equivalent representations of the logical operators may be obtained by stabilizer multiplication.
The perfect code has codewords

\begin{align}
    |\bar{0}\rangle &= \sum_{s\in S} s|00000\rangle \\
    &= |00000\rangle+(|10010\rangle+cyc.)-(|11110\rangle+cyc).-(|01100\rangle+cyc.)\\
    |\bar{1}\rangle &=\bar{X}|\bar{0}\rangle\\
    &=|11111\rangle+(|01101\rangle+cyc.)-(|00001\rangle+cyc.)-(|10011\rangle+cyc.).
\end{align}

\begin{figure}
    \centering
    \includegraphics[width=0.7\linewidth]{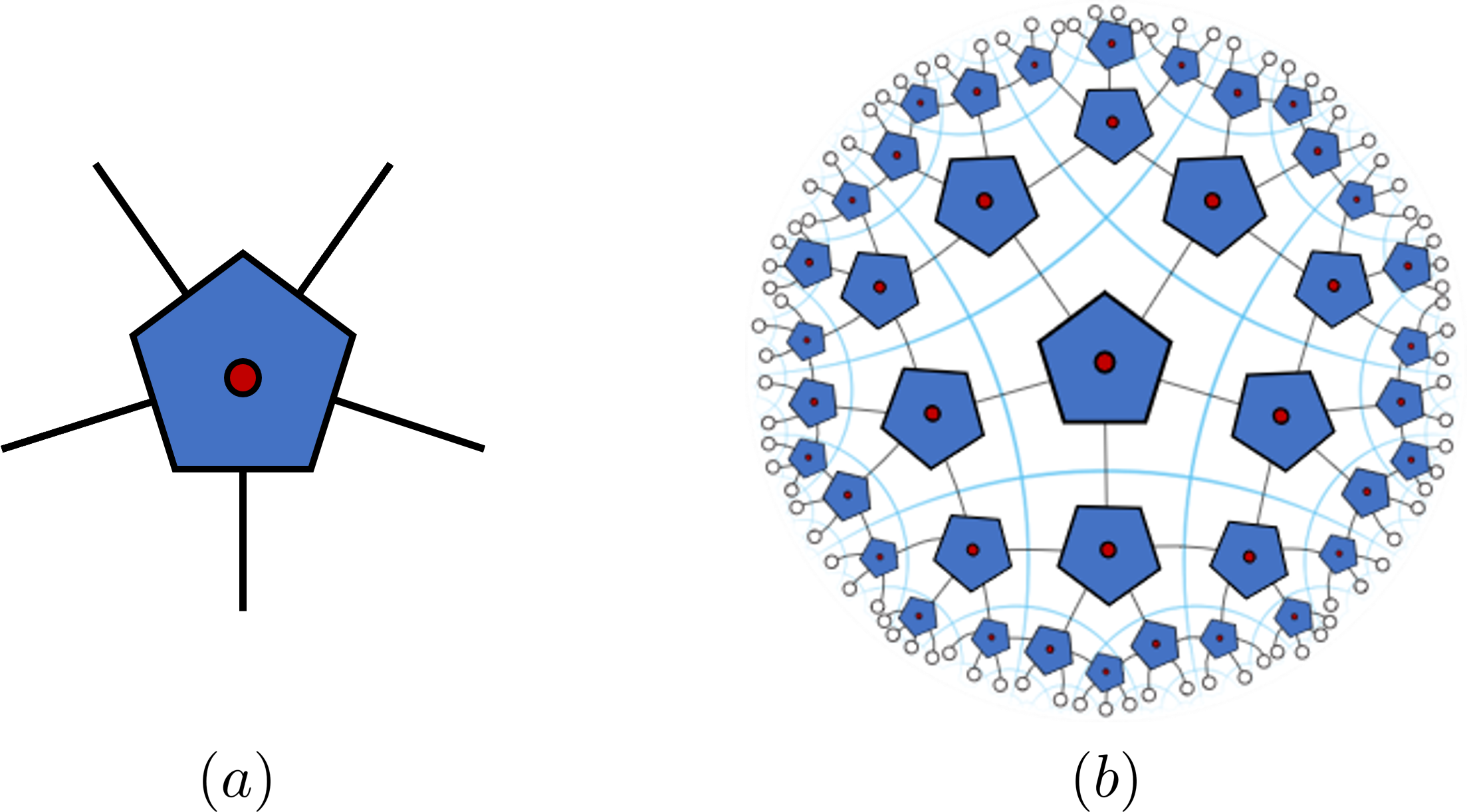}
    \caption{$[[5,1,3]]$ encoding tensor and the HaPPY pentagon code. The logical leg of the single qubit input is represented as a red dot.}
    \label{fig:HaPPY}
\end{figure}

Any 2-qubit subsystem of a state $|\bar{\psi}\rangle=a|\bar{0}\rangle+b|\bar{1}\rangle$ is maximally mixed. This implies that the encoded information is protected against any two-qubit erasures. One can construct the encoding isometry as $W_{[[5,1,3]]}=|\bar{0}\rangle\langle 0|+|\bar{1}\rangle\langle 1|$ and they each are represented graphically by a tensor (Figure~\ref{fig:HaPPY}a). We can think of this code as a simple model for holography, where the 5 qubits are analogous to 5 boundary intervals of the CFT while the bulk qubit is analogous to the central region of the bulk. The single-qubit bulk is supported on any 3 physical qubits on the boundary.

It is instructive to first look at how boundary measurements can transform such codes. Let $A$ be any set of 3 qubits which we measure in the local Pauli basis. Different outcomes then correspond to different product state projections. We can identify two distinct configurations which are analogous to the zero and critical tension scenarios in holography.

Naively, it appears that a destructive measurement on 3 qubits should completely destroy the encoded information. 
For instance, without loss of generality, assume $A$ to be the first 3 physical qubits, on which we perform the Pauli $Y$, Pauli $Z$ and Pauli $Y$ respectively.
This projects these qubits onto the eigenstates $\{|\pm y, \pm z,\pm y\rangle\}$. It can be checked that 
\begin{equation}
    \langle \pm y, \pm z,\pm y|\bar{\psi}\rangle =\langle \pm y, \pm z,\pm y|(a|\bar{0}\rangle+b|\bar{1}\rangle) \propto |\kappa\rangle
\end{equation}
where $|\kappa\rangle$ is a stabilizer state\footnote{For example, when projecting onto $|+y, +z,+y\rangle$, $|\kappa\rangle\propto i|00\rangle-|01\rangle-|10\rangle+i|11\rangle$, which is stabilized by $XX, YZ$.} independent of $a$ and $b$. Therefore,  the remaining qubits on $A^c$ retains no information of the encoded qubit whereas an observer with access to $A$ has a copy of the measurement record. This is unsurprising, because we can verify that $YZYII$ is a logical operator of the code, and by collapsing the state onto the product basis that is an eigenstate of the logical operator, we have effectively measured the encoded qubit. 

\begin{figure}
    \centering
    \includegraphics[width=0.8\linewidth]{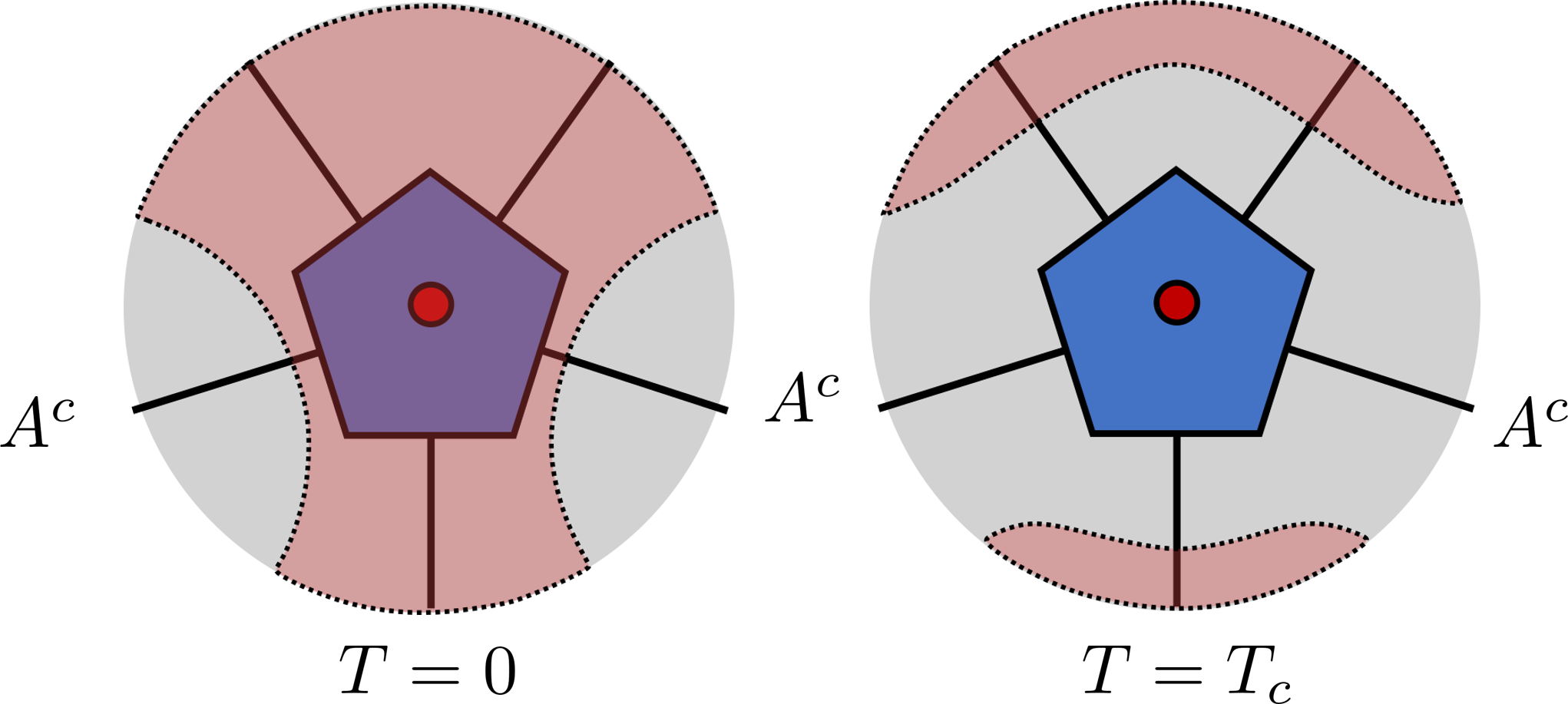}
    \caption{The perfect code as a mini-holographic code. The area shaded in red marks the portion of bulk cut out by end-of-the-world branes after a projective measurement is performed on subsystem $A$. We distinguish the analogous zero tension vs critical tension configurations on whether the bulk qubit has been teleported to $A^c$.}
    \label{fig:5qubit}
\end{figure}

This is similar to the holographic picture where $A^c$ are now bounded by $T=0$ branes which coincide with their RT surfaces. As such, the bulk wedge does not include the bulk qubit (Figure~\ref{fig:5qubit} left). 

However, as suggested by our holographic results in Section \ref{sec:holographiccalculation}, there can also be a configuration in which the bulk information is contained in a connected wedge bounded by the branes with critical tension\footnote{Since only one bulk qubit is present in this setup, only vanishing or critical tensions are possible.} (Figure~\ref{fig:5qubit} right). One such example is obtained by performing individual Pauli Z measurements. Without loss of generality, consider the all $+1$ outcome where we project $A$ onto $|000\rangle$. 

\begin{equation}
    \langle 000|\bar{\psi}\rangle \rightarrow (a|00\rangle -|11\rangle)+b(-|01\rangle-|10\rangle) = U_c|\psi\rangle|0\rangle,
\end{equation}
where $U_c$ is a two-qubit Clifford gate that only depends on the measurement outcomes\footnote{In this case $U_c=CNOT(2\rightarrow 1)CZ(1\rightarrow 2)(Z_1\otimes Z_2H_2)$, where the left of the arrow marks the control qubit, $H$ is the Hadamard gate.}.

Although $A^c$ has zero access to the encoded information prior to measurement, the encoded information is now somehow contained in $A^c$ post-measurement. This is precisely because of quantum teleportation. 

\subsection{Bulk teleportation in the five-qubit code}

More generally, we can understand the above process in the following way. Recall that the perfect code is a 2-qubit erasure correction code. There must exist an encoding unitary $U_A=U_A\otimes \mathds{1}_{A^c}$  independent of $|\bar{\psi}\rangle$ such that 
\begin{equation}
    |\bar{\psi}\rangle = U_A|\psi\rangle_{\tilde{A}_1} |\Phi_+^2\rangle_{\tilde{A}_2A^c},
    \label{eqn:erasuredec}
\end{equation}
where $A= \tilde{A}_1\cup \tilde{A}_2$ and $|\Phi_+^2\rangle$ is two copies of the Bell state $|00\rangle+|11\rangle$ that is maximally entangled across $\tilde{A}_2$ and $A^c$. In our previous example, $A=\{1,2,3\}$, $A_1=\{1\}, A_2=\{2,3\}$, $A^c=\{4,5\}$. Here $\tilde{A}_1$ and $\tilde{A}_2$ can be identified with $A_1$ and $A_2$ respectively via an isomorphism induced by the unitary encoding map $U_A$, however we note that they need not be the same for other QECCs in general.

Then the projective measurement onto any state $|\xi\rangle$ can be rephrased as 
\begin{equation}
    \langle \xi|\bar{\psi}\rangle = \langle \xi| U_A|\psi\rangle_{\tilde{A}_1} |\Phi_+^2\rangle_{\tilde{A}_2A^c} =\langle \xi_{U}|\psi\rangle_{\tilde{A}_1}|\Phi^2_+\rangle_{\tilde{A}_2A^c},
\end{equation}
where $|\xi_{U}\rangle_{{A}}=U_A^{\dagger}|\xi\rangle_A$ and is in general an entangled state over $A$. Therefore, when we project $|\bar{\psi}\rangle $ onto $|000\rangle$ in our example, it is equivalent to projecting the subsystem ${A}$ of $|\psi\rangle_{\tilde{A}_1}|\Phi^2_+\rangle_{\tilde{A}_2A^c}$ onto an entangled state $U_A^{\dagger}|000\rangle$, which teleports $|\psi\rangle$ from qubit $\tilde{A}_1$ to the subsystem $A^c$ using the EPR pairs $|\Phi_+^2\rangle$ as a resource. As usual, the teleportation is only performed up to a unitary $U_c$ that depends on the entangled basis onto which we project. In summary, by performing certain Pauli measurements on $A$, we perform a teleportation protocol that transports the bulk information $|\psi\rangle$ from $A$ to $A^c$ obscured by a unitary $U_c$, which precisely depends on the measurement outcomes in the protocol. This also explains how one can sustain critical tension branes, where the bulk wedge of $A^c$ bounded by the end-of-the-world branes can contain information that extends far beyond its pre-measurement entanglement wedge (Figure~\ref{fig:5qubit}).

In the same way, if we choose $|\xi\rangle$ to be a projection such that it is an eigenstate of the logical operator, i.e. $P_{\bar{L}}|\xi\rangle=|\xi\rangle$ where $P_{\bar{L}}\propto\bar{\mathds{1}}+\bar{L}$ for some non-identity logical operator $\bar{L}$, then

\begin{align} \label{eq:logical_measurement}
    \langle\xi|\bar{\psi}\rangle &=\langle \xi|U_A|\psi\rangle_{\tilde{A}_1}|\Phi_+^2\rangle_{\tilde{A}_2A^c} = \langle \xi|U_A(U_A^{\dagger}P_{\bar{L}}U_A)|\psi\rangle_{\tilde{A}_1}|\Phi_+^2\rangle_{\tilde{A}_2A^c}\\ \nonumber
    &\propto\langle \xi_U|(\mathds{1}+L)_{\tilde{A}_1}\otimes O_{\tilde{A}_2}|\psi\rangle_{\tilde{A}_1}|\Phi_+^2\rangle_{\tilde{A}_2A^c}.
\end{align}
Recall that for any logical operator $\bar{L}$, $U_A^{\dagger} \bar{L} U_A=L$. Therefore, we find that the boundary product state projection precisely corresponds to the projective measurement $\mathds{1}+L\propto P_L$  of the data qubit(s) on $\tilde{A}_1$. $O_{\tilde{A}_2}$ is some operator whose specific form is not relevant to our discussion. What is important is that $P_L\otimes O$ is in the form of a product. One can repeat the above argument for the other commuting measurements on the boundary, but we will find that this product form persists for all these operators and therefore we are not performing an entangled measurement by choosing $|\xi\rangle$ to be this specific form.

\subsection{Bulk teleportation in general codes}

We see that the above argument trivially generalizes to all such erasure correction codes. Indeed, any erasure correction code whose code subspace consists of tensor factors like qubits can be decoded in the form of (\ref{eqn:erasuredec}) by replacing $|\Phi_+^2\rangle_{\tilde{A}_2A^c}$ by some other entangled resource state $|\chi\rangle_{\tilde{A}_2A^c}$ and allowing $\tilde{A}_1$ to contain multiple data qubits \cite{Harlow:2016vwg}. Therefore, the exact quantum teleporation arguments hold under the boundary projection $|\xi\rangle$. Note that we only used the fact that $U_A$ encodes or decodes the logical information with respect to a particular state $|\bar{\psi}\rangle$ in the above analysis. Although $U_A$ is independent of the encoded state for exact erasure correction codes analyzed above, its state-independence is not required for the argument. Indeed, the same teleportation argument also generalizes even if $U_A$ does depend on the state. This latter scenario can usually be expected in approximate error correcting codes with linear but non-isometric encoding maps \cite{Cao:2020ksw,Akers:2022qdl}. We will examine one such scenario inspired by \cite{Akers:2022qdl} in the random tensor network construction in Section \ref{subsec:noniso}. 

Also note that in order to recreate the ``$T=0$'' configuration, we need to be able to project out the logical information supported on $A$, i.e. bulk qubits in the entanglement wedge of $A$. This is not always easy to do with $|\xi\rangle$ being a boundary product state. Indeed, we see that the logical operator measurement has to commute with a set of single-site Pauli measurements. In particular, $P_{\bar{L}}|\xi\rangle=|\xi\rangle$. For random codes, this is almost never possible, as the logical operator is just a generic entangled operator that has support on $A$; our analysis in Section \ref{sec:rtn} shows that only settings analogous to critical tensions can be achieved, confirming this expectation. However, it is possible to mimic configurations that have lower brane tensions for stabilizer codes like the one above as well as the HaPPY code in Section \ref{sec:happycode} because of their abundant symmetries. In particular, their logical Pauli operators are always transversal and can be written as tensor product of Paulis. This makes it much easier to produce a ``lower tension'' configuration. Interestingly, the very same transversal property that is desirable for fault-tolerant quantum computing \cite{EastinKnill} can also facilitate the construction of different configurations that resemble branes at different tensions.

We should remark that although we have been somewhat careless so far in comparing the amount of teleported content with brane tensions in holography, there is no real sense in which such a coarse measure we deploy for these small quantum codes can contain all the necessary ingredients to represent actual ETW branes in holography, which have a semi-classical description. Indeed, we will elaborate in Section~\ref{sec:happycode} more precisely how these analogies can diverge from our holographic expectations. Nevertheless, these intuitive connections remain a powerful approximation that we can further refine.

It is clear that the maximum number of bulk qubits one can teleport in this way is upper bounded by the entanglement resource between $A$ and $A^c$. In holography, \cite{Harlow:2016vwg} showed that the amount of entanglement resource contained in $|\chi\rangle_{\tilde{A}_2A^c}$ is given by the area of the minimal surface $S_{\rm RT}(A)=\mathcal{A}/(4G)$, with $G\propto N^{-2}$. As the amount of entanglement resource scales as $N^2$ whereas the bulk information in the code subspace is of $O(N^0)$ in a large $N$ theory, there is thus enough entanglement resource to teleport almost all bulk information in the entanglement wedge of $A$ to that of $A^c$ with boundary measurements (once we impose the boundary cutoffs appropriately). As such, this is consistent with the existence of critical tension branes which ``hug'' the boundary.

Having gleaned some insight from the above analysis, we can therefore summarize our finding for any (subsystem) quantum error-correcting code over finite dimensional Hilbert spaces. Physically, the statement is quite simple --- bulk information in the entanglement wedge of $A$ can be teleported from $A$ to $A^c$ through boundary measurements. However, one can only teleport as much information as allowed by the amount of entanglement resource available between $A$ and $A^c$. If there is more than enough entanglement resource, then trivially one can only teleport as much information as there is to teleport in the code subspace $\mathcal{C}_a$ accessible by $A$, which is bounded by the size of $\mathcal{C}_a$.
\begin{lemma}
\label{thm:teleportationlemma}
Let $\mathcal{C}=\mathcal{C}_a\otimes \mathcal{C}_{a^c}\subset \mathcal{H}$ be a code over a finite dimensional physical Hilbert space $\mathcal{H}=\mathcal{H}_A\otimes\mathcal{H}_{A^c}$ such that $\forall \bar{\rho}_a\in L(\mathcal{C}_a)$ is recoverable in $A$. Let $\{|\bar{i}\rangle_a\}$ be a complete orthonormal basis for $\mathcal{C}_a$, and consider a state 
\begin{equation}
    |\Phi\rangle = \sum_{i}|\bar{i} i\rangle_{aR}
\end{equation}
that is maximally entangled between the subsystem $a$ and a reference $R$ of equal size.
Then the projection of $A$ onto any state $|\xi\rangle$ constitutes an imperfect teleportation protocol such that 
\begin{equation}
    I(A^c:R)\leq \min \{2S(R), 2S(A^c)\}
\end{equation}
after the projective measurement.
\end{lemma}
\begin{hproof}
Trivially, $I(A^c:R)\leq S(A^c)+S(R)$. Therefore, let us write
\begin{align}
   \langle \xi |\Phi\rangle &= \langle \xi |U_A\sum_{i=1}^{|R|} |i\rangle_{\tilde{A}_1} |\chi\rangle_{\tilde{A}_2 A^c} |i\rangle_R\\ &=\sum_{j=1}^{r_{\chi}}\sum_{i=1}^{|R|}c_j\langle \xi U_A |i\rangle_{\tilde{A}_1} |\chi_j\rangle_{\tilde{A}_2}|\chi'_j\rangle_{A^c} |i\rangle_R\\ \nonumber
   &=\sum_{j=1}^{\min\{r_{\chi},|R|\}} d_{j} |\chi'_j\rangle_{A^c}|j\rangle_R,
\end{align}
where the sum over $j$ cannot be greater than the Schmidt ranks $r_{\chi}$ or $|R|$, the dimension of $R$. The encoding unitary $U_A$ is guaranteed to exist by our assumptions thanks to \cite{Harlow:2016vwg}.

We know that the capacity of teleportation is upper bounded by the entanglement resource shared between $\tilde{A}_2$ and $A^c$. Asymptotically, $S(A^c)=S(\Tr_{\tilde{A}_2}[|\chi\rangle\langle\chi|])$ upper bounds the number of distillable Bell pairs. Suppose we arranged the entanglement resource of $|\chi\rangle$ into Bell pairs, then $\log r_{\chi}\leq S(A^c)$ and it is clear that the maximal entanglement $R$ can attain post-projection is when all $d_j$ are equal and $S(R)=S(A^c)=\min \{\log r_{\chi},|R|\}$.
\end{hproof}

\begin{corollary} 
\label{corollary}
For a random code,
\begin{equation}
    I(A^c:R)\approx \min \{2S(R), 2S(A^c)\}
\end{equation}
 post-measurement, provided that $S(R),S(A^c)\gg 1$.
\end{corollary}

\begin{hproof}
For a random code, we expect $|\xi U_A\rangle = U_A^{\dagger}|\xi\rangle$ to be generically Haar random, unless we have chosen $|\xi\rangle$ specifically to undo the action of $U_A$. Let us again suppose that $|\chi\rangle$ can be approximately converted into a maximally mixed state with entropy $S(A^c)$ (e.g. $S(A^c)$ Bell pairs) through some distillation process, such that a subsystem $A^c_{1}$ with size $S(A^c)$ of $A^c$ is maximally mixed. As a result, the projected state supported on $A^cR$ is nothing but a random tensor with  $|T\rangle_{A^cR}=\langle \xi|\Phi\rangle\rightarrow |T\rangle_{A_1^c R}\otimes |\gamma\rangle_{A^c\setminus A_1^c}$ where the two subsystems of $|T\rangle_{A_1^c R}$ are close to being maximally entangled in the large bond dimension limit \cite{hayden2016holographic}. Hence, $S(R)= S(A^c_1)\approx \min\{\log|R|, \log|A^c_1|\}$. Here we assume the appropriate asymptotic properties so that the distillation argument holds, therefore $\log|A_1^c|=S(A^c)$.
\end{hproof}
We will see that this teleportation upper bound can be related to the switchover in Section \ref{sec:bulkteleportation} for HaPPY codes and in Section \ref{sec:rtn} for random tensor networks when the bulk Hilbert space is sufficiently large compared to that of the boundary.

\section{HaPPY code}\label{sec:happycode}

To make better connection with holography, we extend our analysis of the 5-qubit code with the more sophisticated HaPPY code. We will see that it also confirms our physical picture of quantum teleportation. Different from the random codes that we discuss in Section \ref{sec:rtn}, these stabilizer codes enjoy a large degree of symmetries that are manifest in the form of transversal operators. We will see that this can be helpful in constructing ``branes of different tensions''. 
\subsection{Code construction, stabilizers}
The HaPPY pentagon code (Figure~\ref{fig:HaPPY}b) was first introduced by \cite{Pastawski:2015qua} as a toy model for AdS/CFT, which captures many features of bulk (operator) reconstruction. It is a class of stabilizer codes built out of the smaller $[[5,1,3]]$ perfect codes (Figure~\ref{fig:HaPPY}a) we just analyzed. 

To construct a tensor network with contracted set of edges $E$ as shown in Figure~\ref{fig:HaPPY}b, we arrange these tensors according to a $\{5,4\}$ tiling of the Poincare disk, where each tensor lies on a tile $\mathcal{T}$ and its dangling legs are perpendicular to the sides of the tile. Then one ``glues'' together two tensor legs intersecting the same sides into a contracted edge $e$ by projecting them onto a Bell state $|\Phi_+^e\rangle=(|00\rangle+|11\rangle)/\sqrt{2}$. 
The resulting tensor network is exactly the encoding isometry
\begin{equation}
    W_{\rm HaPPY}\propto\bigotimes_{e\in E}\langle \Phi_+^e|(\bigotimes_{\mathcal T}W_{[[5,1,3]]}^{\mathcal{T}})
\end{equation}
 of the HaPPY code, which maps the space of bulk logical qubits into the space of boundary physical qubits.

One can derive the logical Pauli operators (including the logical identities which are the stabilizers) of this network via operator matching. To do so, we simply insert a representation of the desired logical operator at each tensor. If these operators match on contracted legs, the resulting operator on dangling legs is a logical operator of the HaPPY code. Equivalently, this can be phrased in terms of operator pushing, where one begins by inserting a logical operator on a single tensor, then sequentially pushing any non-trivial operator acting on contracted edges to the boundary dangling legs using the stabilizers or logical operators of nearby tensors. A more detailed account of operator pushing or matching is already thoroughly explained in \cite{Pastawski:2015qua} and in \cite{Cao:2021ibt} for all tensor networks. Therefore, we will refer the readers to these works for further information. Although in principle one can repeat the above exercise $2^{n-k}$ times to obtain the defining stabilizer group for an $[[n,k,3]]$ HaPPY code that encodes $k$ bulk qubits into $n$ physical qubits, in practice it is far easier to only produce the $n-k$ stabilizer generators of this code. This can be done efficiently with an algorithm introduced in Appendix D of \cite{Cao:2021ibt}.

\subsection{Numerical experiments}
\label{happynumerics}

\begin{figure}
    \centering
    \includegraphics[width=1\linewidth]{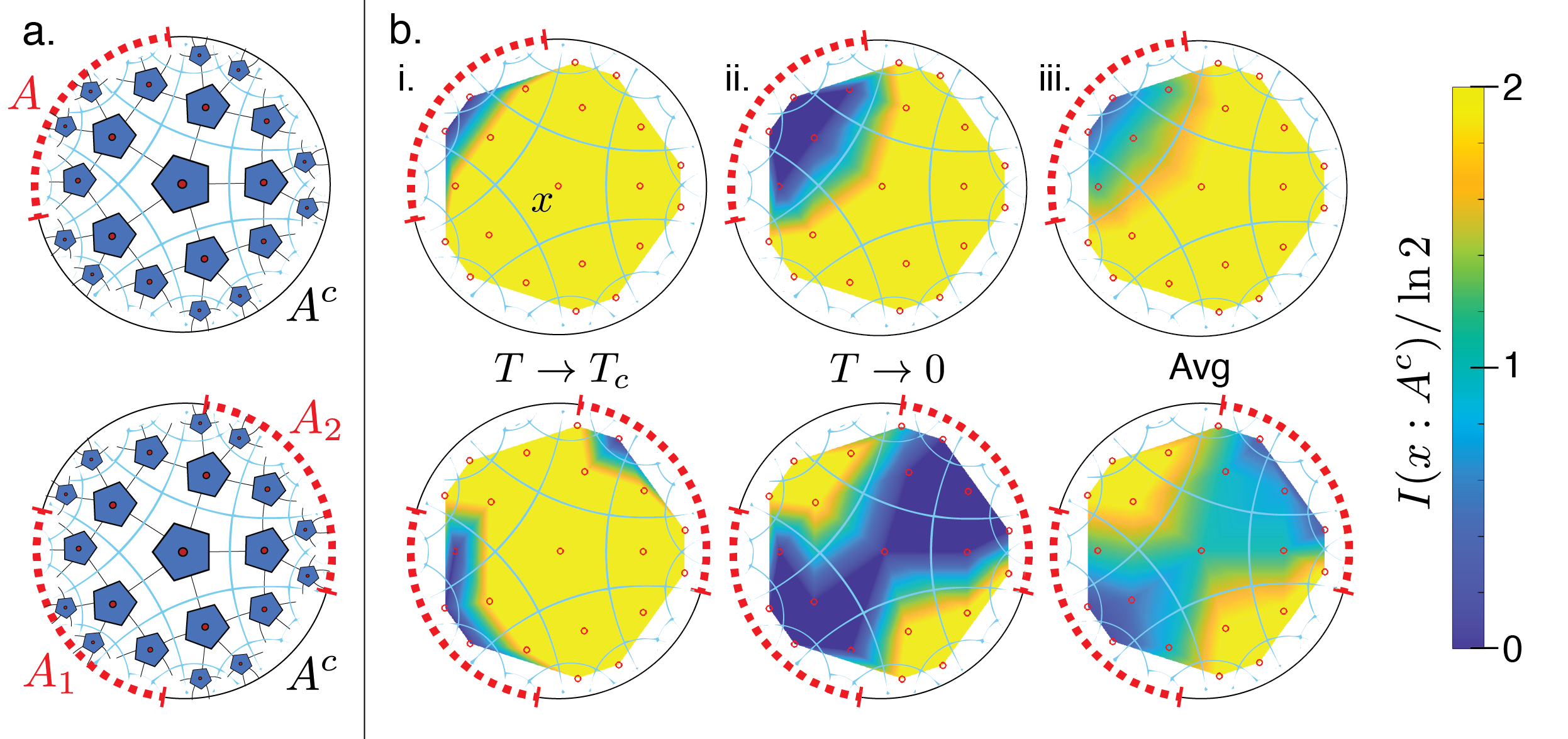}
    \caption{Bulk information retained by the $M = 1$ HaPPY code (a) after projectively measuring (dotted red) a single contiguous region $A$ (top row) or a pair of diametrically opposed regions $A = A_1 \cup A_2$ (bottom row) in local Clifford bases. Colormaps (b) show mutual information $I(x:A^c)$ between each bulk qubit $x \in b$ (red circles) and the remaining boundary region $A^c$ (solid black). By searching over local measurement bases on $A$, we can find the least destructive (i) or most destructive (ii) measurement basis, which mimic holographic ETW brane geometries with brane tension varying from nearly critical $T \rightarrow T_c$ to nearly vanishing $T \rightarrow 0$, respectively. The average mutual information (iii), sampled over 100 random choices of local boundary measurements, suggests that most choices of measurement lead to low-tension ETW branes extending deep into the bulk. The HaPPY code we implement here encodes $N = 55$ boundary qubits into $b = 21$ bulk qubits, with $|A| = 16$ for the contiguous region (top) and $|A_1| = |A_2| = 15$ for the pair of opposing regions (bottom).}
    \label{fig:happysimM1}
\end{figure}

With the teleportation intuition in mind from Section \ref{sec:boundmeasuremteleportation}, we can anticipate the behaviour of the full HaPPY code under boundary Pauli measurements. Whereas the individual 5-qubit code studied above supported only two possible brane configurations, in the full HaPPY code there are multiple bulk qubits and one anticipates that there can be a range of brane ``tensions'' under which different portions of the bulk can be recoverable in $A^c$ beyond its pre-measurement entanglement wedge.

To study this behavior directly we use numerical stabilizer methods \cite{aaronson2004improved} to simulate a finite-size HaPPY code illustrated in Figure \ref{fig:happysimM1}a which encodes 21 bulk qubits $x \in b$ into 55 boundary qubits $A \cup A^c$. We projectively measure a subregion $A$ of the system's qubits in a particular local Clifford basis and ask how many of the bulk qubits remain encoded in the remaining system $A^c$. To characterize the resulting bulk geometry, we compute the mutual information $I(x:A^c)$ between each bulk qubit $x$ and the remaining unmeasured boundary $A^c$. Bulk qubits with near-maximal mutual information can be reliably reconstructed on the region $A^c$, while qubits with near-vanishing mutual information have effectively been destroyed by the measurement and cannot be reconstructed on the boundary. In addition to this single HaPPY code ($M = 1$), we shall also consider concatenating $M > 1$ copies of these codes to more closely mimic holography. We begin by studying the single HaPPY code case $M = 1$.

\subsubsection{$M = 1$ HaPPY code}
\label{sec:happym1}

We first study the behavior of a single HaPPY code subjected to projective measurements on a subregion $A$ of boundary qubits. 
Depending on the choice of local basis and the size and shape of the region $A$, these projective measurements destroy various parts of the bulk as illustrated in Figure \ref{fig:happysimM1}b. Fixing the measured region $|A| = 16$ (b, top row), we directly search over all choices of local Pauli measurement bases to find the optimal choice of measurement basis that destroys either as few (b.i.) or as many (b.ii.) of the bulk qubits as possible. We interpret these two situations as being analogous to a holographic geometry cut off by an ETW brane with tension $T$ that can be tuned between $0 < T < T_c$ depending on the choice of measurement basis.  
In fact, by sampling over 100 randomly-chosen local measurement bases (b.iii.), we find that most choices lead to low-tension ETW branes that destroy large portions of the bulk.

We observe similar phenomena if we consider projective measurements on a pair of diametrically opposed regions $A = A_1 \cup A_2$ (bottom row of Figure \ref{fig:happysimM1}). In this case, a judicious choice of boundary measurement bases yields either a connected (i) or disconnected (ii) bulk geometry. Averaging over randomly-chosen measurement bases (iii), we find that most local measurements yield a disconnected geometry. We view this transition from connected to disconnected geometry as analogous to the phase transition studied in Section \ref{sec:holographiccalculation}, where the phase structure is determined by the tension of the brane and the size of region $A$.

\begin{figure}
    \centering
    \includegraphics[width=0.7\linewidth]{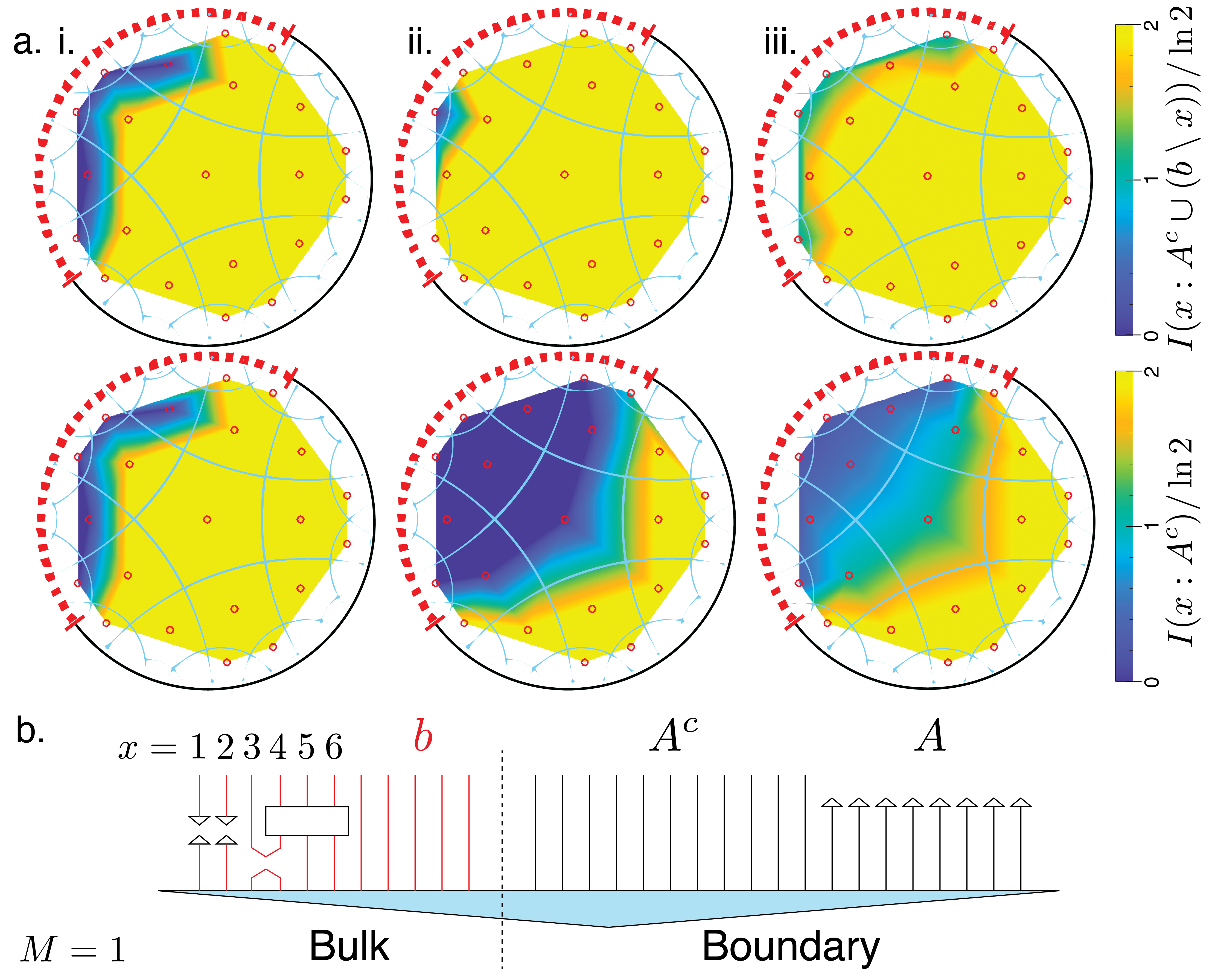}
    \caption{In the $M = 1$ HaPPY code, bulk qubits $x \in b$ can be uncorrelated with the boundary $A^c$ while remaining entangled with other bulk qubits following a LPM on the boundary $A$. To probe this phenomenon, we can ask whether a single bulk qubit $x$ can be reconstructed on any of the remaining qubits, including those in the bulk $b$. This is quantified by the bulk-inclusive mutual information $I(x:A^c \cup(b \setminus x))$ (a. top row), which we compare to the usual bulk-boundary mutual information $I(x:A^c)$ (a. bottom row). We search over LPMs to find the most destructive measurement basis for either the bulk-inclusive (i) or bulk-boundary (ii) information. We also consider the average of these quantities over $10^3$ randomly-chosen LPMs (iii). These results can be explained by a post-measurement tensor network (b) in which some of the bulk degrees of freedom (red) have become disentangled from the boundary $A^c$ but remain entangled with one another.
    }
    \label{fig:m1optimaldestruction}
\end{figure}

Although the $M = 1$ HaPPY code is a simple toy model that gives results in qualitative agreement with our earlier holographic calculations in Section \ref{sec:holographiccalculation}, it suffers from some idiosyncracies that make quantitative comparison to holography problematic. Most notably, for $M = 1$ the ratio of bulk legs to boundary legs is fixed to be of order one, which is unlike the large-$N$ holographic situation where there are $O(N^2)$ boundary legs for each bulk leg. A related problem is that bulk qubits that are disentangled from qubits in $A^c$ by the measurement can still be entangled with other bulk qubits, as shown in Figure \ref{fig:m1optimaldestruction}. Here we find instances of bulk qubits $x$ that cannot be reconstructed on the unmeasured boundary $A^c$ (as diagnosed by vanishing mutual information $I(x:A^c) = 0$) yet nevertheless remain entangled with the system. We diagnose these `ghost' bulk qubits by measuring the bulk-inclusive mutual information $I(x:A^c \cup (b \setminus x))$ (Figure \ref{fig:m1optimaldestruction}a. top row). This quantity tells us whether the bulk qubit $x$ can be reconstructed on the union of the remaining boundary and bulk qubits, and can be directly compared to the usual bulk-boundary mutual information (a. bottom row). Searching over random LPMs, we can optimize for a measurement basis that is maximally destructive of the bulk-inclusive (i) or the bulk-boundary (ii) mutual information. We can also compute the average of these quantities sampled over $10^3$ randomly chosen LPMs (iii).

\begin{figure}[h]
    \centering
    \includegraphics[width=.5\textwidth]{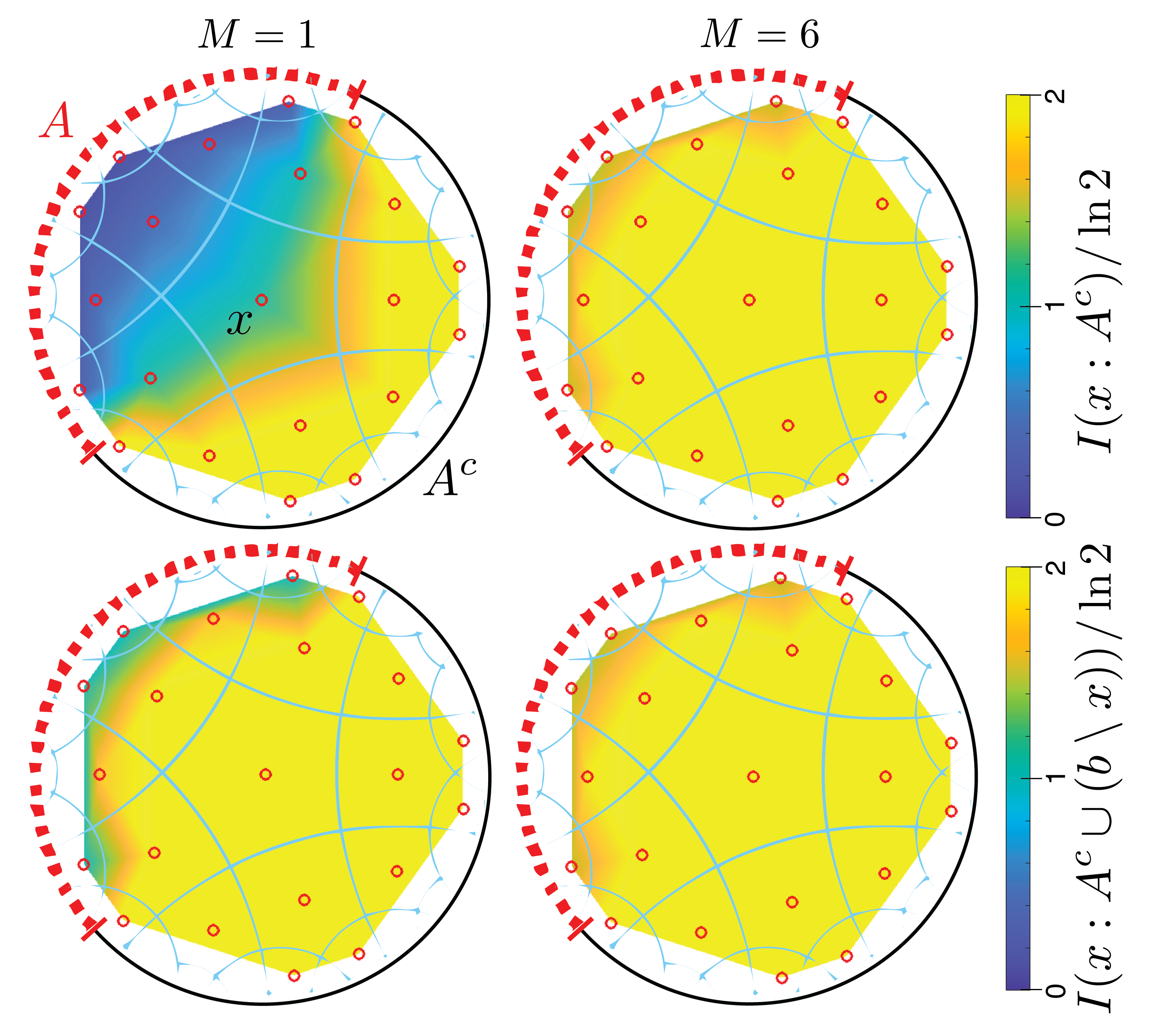}
    \caption{The precise location of the ETW brane in the HaPPY code is ``fuzzy'' in the $M = 1$ case (left column) due to residual entanglement between bulk qubits after measurement and the non-maximal mutual information between some bulk qubits and the boundary $A^c$. By contrast, the brane can be unambiguously located in the $M = 6$ case (right column). Here we plot the average bulk-boundary mutual information $I(x:A^c)$ (top row) and the average bulk-inclusive mutual information $I(x:A^c \cup (b \setminus x)))$, sampled over $10^3$ randomly-chosen LPMs performed on the boundary region $A$. Note that in the $M=6$ case the results obtained for $I(x:A^c)$ and $I(x:A^c \cup (b \setminus x)))$ are nearly identical, indicating that the ambiguity in the ETW brane definition is suppressed in this analogous large-$N$ limit.
    }
    \label{fig:fuzzbranes}
\end{figure}

We can explain this behavior with a tensor network picture shown in Figure \ref{fig:m1optimaldestruction}b. For example, whereas qubits $x = 1,2$ in this picture are completely disconnected from $A^c$, qubit $x = 3$ is entangled with other bulk qubits. In this situation, none of these bulk qubits $x = 1,2,3$ can be reconstructed on the boundary $I(x:A^c) = 0$, but qubit $x = 3$ can be reconstructed on the boundary plus bulk $I(x:A^c \cup (q \setminus x)) > 0$. A related puzzle is that some of the bulk qubits share a non-maximal mutual information with qubits in $A^c$. As a consequence, it is unclear whether these qubits are contained in the entanglement wedge of $A^c$ or not. Due to these phenomena, it is challenging to precisely locate the ``brane'' in the $M = 1$ bulk as shown in Figure \ref{fig:fuzzbranes}, and we can only interpret our results in terms of a pre-geometrical ``quasi-brane''. We expect these effects to be a consequence of finite-$N$ effects. In fact, when the number of boundary qubits is parametrically larger than the number of bulk qubits, bulk qubits should be almost uniquely entangled with boundary qubits, solving both the puzzles we have just outlined and giving an unambiguous definition of the brane location. Therefore, in the large-$N$ limit the ETW brane emerges as a better-defined geometrical object. In the next section, we shall attempt to approach such a large-$N$ regime by concatenating multiple HaPPY codes together.

\subsubsection{$M = 6$ concatenated HaPPY code}

\begin{figure}
    \centering
    \includegraphics[width=1\linewidth]{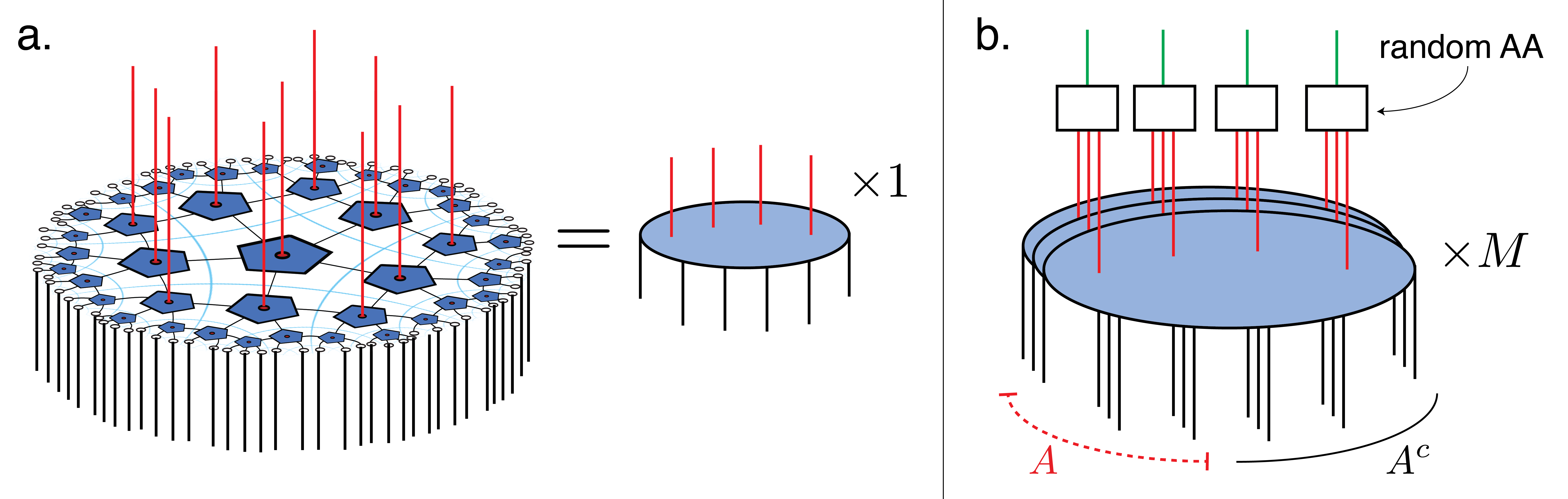}
    \caption{Concatenating HaPPY codes to mimic the effect of finite $G$. We can view the HaPPY code as an encoding map (a) from a set of input boundary qubits (black) to a smaller set of output bulk qubits (red). To increase the ratio of boundary to bulk qubits, we consider stacking $M$ HaPPY codes (b). We collect the $M$ bulk legs at each point and scramble them into a single bulk qubit (green) using a random all-to-all (AA) scrambling circuit. Projective measurements on a region $A$ (dotted red) are made separately on each layer of the stack in randomly-chosen local bases.}
    \label{fig:happystack}
\end{figure}

As mentioned above, while the single HaPPY code studied above is a nice toy example, the ratio of bulk legs to boundary legs is fixed to be of order one. This is different from the holographic situation, where the ratio of boundary to bulk degrees of freedom is controlled by the gravitational constant $G$. To more closely reflect the holographic situation, in this section we concatenate multiple HaPPY codes into a `stack' whose bulk legs are combined into a single set of bulk legs as shown in Figure \ref{fig:happystack}. This construction provides us with an additional parameter, the stack depth $M$, that controls the ratio of boundary to bulk degrees of freedom, analogous to the role of $1/G\sim N^2$ in the holographic setting. The idea of stacking holographic codes in this way is not entirely new: a slightly different 6-fold HaPPY code has been studied by \cite{Donnelly:2016qqt} in the context of edge modes and the large-$M$ concatenated holographic code was suggested by \cite{Cao:2020ksw} as a connection with large $N$ theory.

In this stack construction, there are $M$ copies of each boundary region $A$, one for each layer in the stack as shown in Figure \ref{fig:happystack}b. In the following, we apply projective measurements to all $M$ copies of a particular boundary region $A$ and study how the bulk is affected by these measurements. We compare our numerical findings to the holographic calculation presented in Section \ref{sec:holographiccalculation}.

\begin{figure}
    \centering
    \includegraphics[width=1\linewidth]{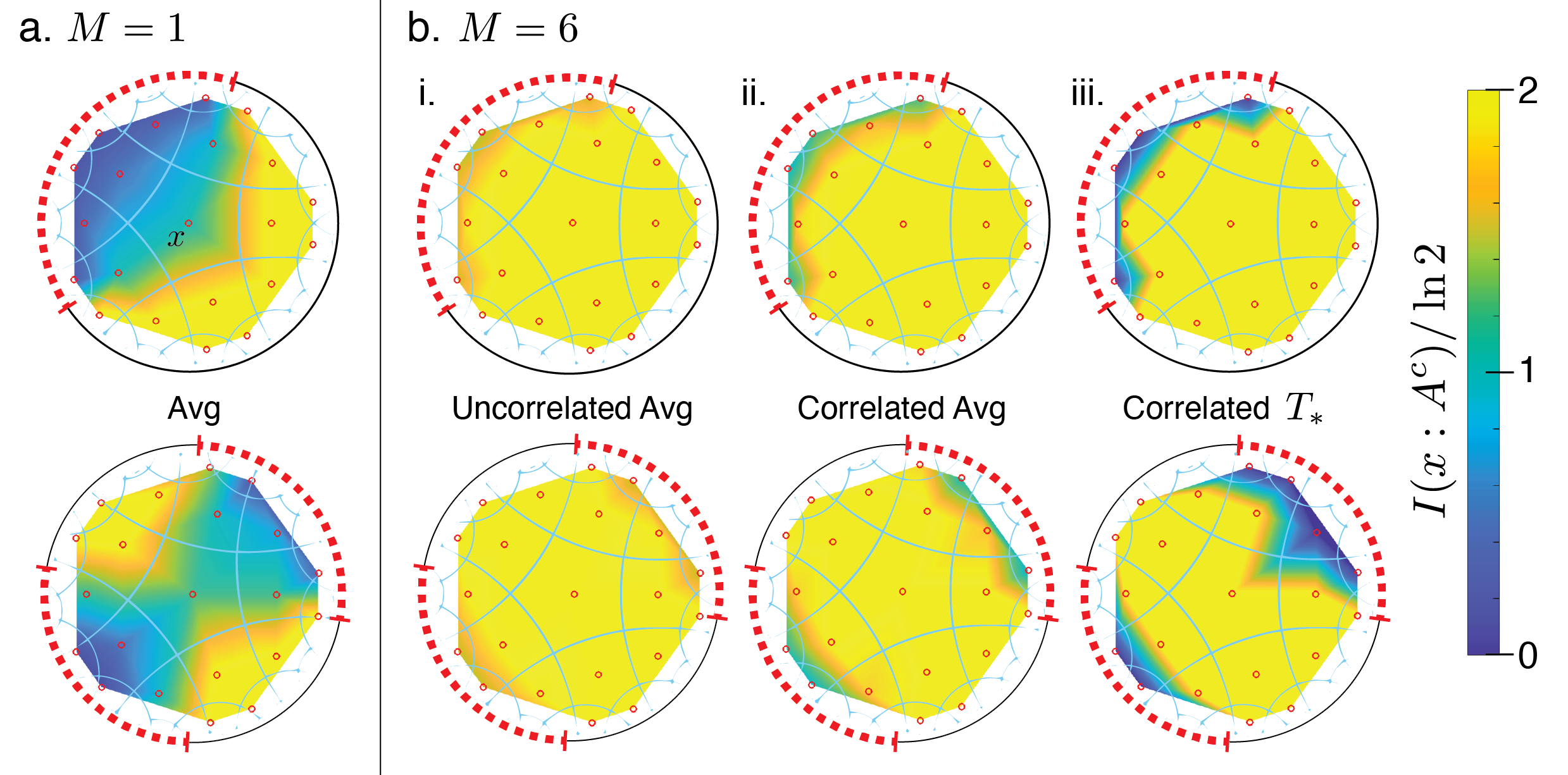}
    \caption{Bulk information retained by the depth $M = 1$ HaPPY code (a) versus the depth $M = 6$ concatenated HaPPY code stack (b) after projectively measuring (dotted red) a single contiguous region A (top row) or a pair of diametrically opposed regions $A = A_1 \cup A_2$ (bottom row). In the depth $M = 6$ stack we have some control over the ETW brane tension by choosing the local measurement bases to be classically correlated between the different layers in the stack. Completely uncorrelated random measurement bases (i) leave the bulk largely intact when averaged over 100 samples. We interpret this situation as analogous to a holographic geometry cut off by an ETW brane with near-critical tension $T \rightarrow T_c$. By contrast, measuring all $M$ legs in the same basis (ii,iii) tends to be more destructive of the bulk. Sampling over 100 of these stack-correlated measurement bases yields lower-tension branes that tend to destroy more of the bulk (ii). Finally, we search over all such stack-correlated measurement bases to find the most destructive measurement basis (iii), which corresponds to some finite minimal tension $0 < T_* < T_c$. From these experiments it is clear that both the stack depth $M$ as well as the choice of measurement basis play an essential role in determining the ETW brane tension $T$. Here we use $|A| = 24$ for the contiguous region (top) and $|A_1| = |A_2| = 15$ for the pair of opposing regions (bottom).}
    \label{fig:happystackm6}
\end{figure}

We first repeat the numerical experiments of Section \ref{sec:happym1} with a depth $M = 6$ HaPPY stack construction. Results comparing the $M = 1$ and $M = 6$ cases are shown in Figure \ref{fig:happystackm6}a,b for a single contiguous region $A$ (top row) and a pair of diametrically opposed regions $A = A_1 \cup A_2$ (bottom row). Whereas local projective measurements in the depth $M = 1$ case tend to yield low-tension ETW branes that destroy large portions of the bulk (Figure \ref{fig:happystackm6}a), the same measurements applied to the depth $M = 6$ stack tends to yield high-tension ETW branes that leave much of the bulk intact (Figure \ref{fig:happystackm6}b). We can explain the prevalence of high-tension branes in this case by appealing to the idea of ``ghost'' bulk qubits introduced above. Whereas in the $M = 1$ case these ``ghost'' qubits are irrevocably lost to the boundary region $A^c$, in the $M = 6$ case the final AA encoding circuit can effectively re-entangle these bulk qubits back to the boundary $A^c$. We therefore expect the $M = 6$ construction generally to preserve much more of the bulk than the $M = 1$ case, which is equivalent to having high-tension branes.

Evidently the stack depth $M$ appreciably changes the ETW brane tension. We can also change the brane tension by picking measurement bases that are (classically) correlated between different layers in the stack. Specifically, to perform a \emph{stack-correlated} measurement we randomly pick a stabilizer measurement basis for each boundary site in region $A$, and then projectively measure all $M$ legs in the stack in that same measurement basis. Whereas completely random uncorrelated measurement bases (Figure \ref{fig:happystackm6}b.i) lead on average to a near-critical ETW brane tension $T \rightarrow T_c$, these stack-correlated measurements are more destructive on average (Figure \ref{fig:happystackm6}b.ii) and yield a ETW brane with lower tension. This phenomenon has a simple explanation: for stack-correlated measurements, if a particular 
bulk qubit (red lines in Figure \ref{fig:happystack}b) is collapsed by the measurements performed in any one layer, then the correlated measurement bases guarantee that it will also be destroyed in every layer, leading to collapse of the final output qubit (green lines in Figure \ref{fig:happystack}b), despite the final random AA encoding step. Finally, we can search through all stack-correlated measurement bases to find the most destructive basis (Figure \ref{fig:happystackm6}b.iii), which we interpret as analogous to an ETW brane geometry with minimal tension $0 < T_* < T_c$.

\begin{figure}
    \centering
    \includegraphics[width=1\linewidth]{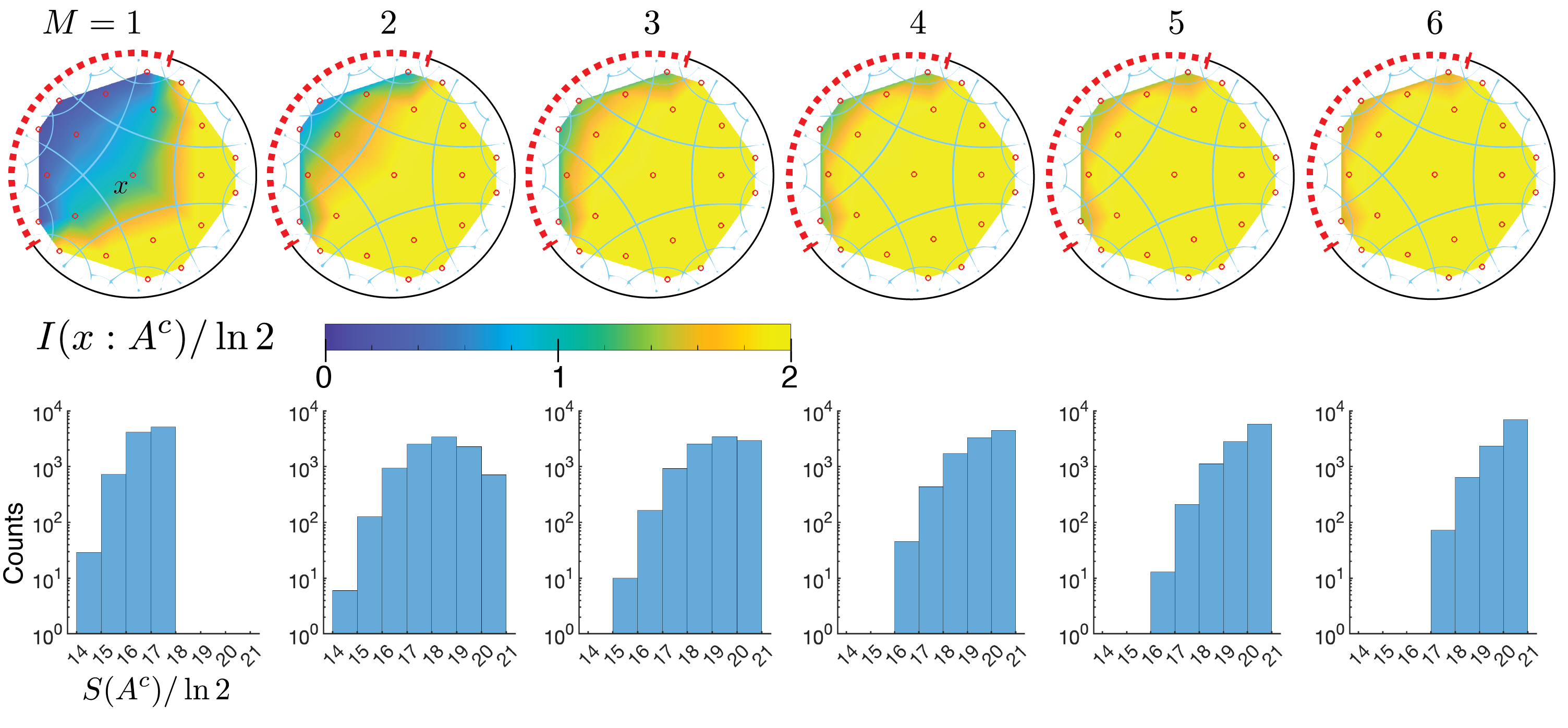}
    \caption{Brane tension $T$ versus stack depth $M$. Top row: Mutual information $I(x:A^c)$ between individual bulk qubits $x$ and the remaining boundary $A^c$, averaged over $10^4$ random product-state measurement bases on the continuous region $A$ with $|A| = 24$. Our results suggest that the brane tension $T$ tends to increase with the stack depth $M$. Bottom row: distribution of brane tensions, as measured by the entropy $S(A^c)$ of the remaining boundary qubits $A^c$, sampled over $10^4$ random product-state measurement bases.}
    \label{fig:happystackvarym}
\end{figure}

We can also ask how the brane tension depends on the stack depth $M$ as shown in Figure \ref{fig:happystackvarym}. For $M = 1$ (left), local projective measurements on the contiguous region $A$ are capable of destroying almost all qubits in the entanglement wedge of $A$, indicative of a low-tension ETW brane $T \rightarrow 0$. By contrast, bulk qubits are increasingly protected from these projective measurements at larger stack depth $M > 1$, indicative of increasingly higher brane tensions $T \rightarrow T_c$ (right). We can interpret these results following the discussion in Section \ref{sec:bulkteleportation}. At large $M$, the bulk qubits are preserved after measuring $A$ precisely because they are teleported by the measurements over to the complementary region $A^c$. Intuitively, for $M>1$ it becomes more likely to teleport bulk information when a measurement is performed, because teleportation can occur not only within a single HaPPY code layer, but also across different layers.

We note that our stabilizer simulations only yield ETW branes with positive tension $0 < T < T_c$. This is slightly different from the situation in holographic AdS/BCFT analyzed in Section \ref{sec:holographiccalculation}, where in principle the brane tension can be tuned to be negative $T^* < T < T_c$ (with $-T_c<T^*<0$) by projecting the measured region $A$ onto a boundary Cardy state with negative boundary entropy $\log (g) < 0$ \cite{Cardy:2004hm,Calabrese:2009qy}. 
Even if it is possible to obtain negative tension branes in the HaPPY code by appropriately fine-tuning the measurement basis, the fact that we did not find any such examples in our stabilizer code construction suggests that the projection onto Cardy states with negative boundary entropy is non-generic and requires some amount of fine-tuning.

\subsubsection{Bulk teleportation}
\label{sec:bulkteleportation}

\begin{figure}
    \centering
    \includegraphics[width=1\linewidth]{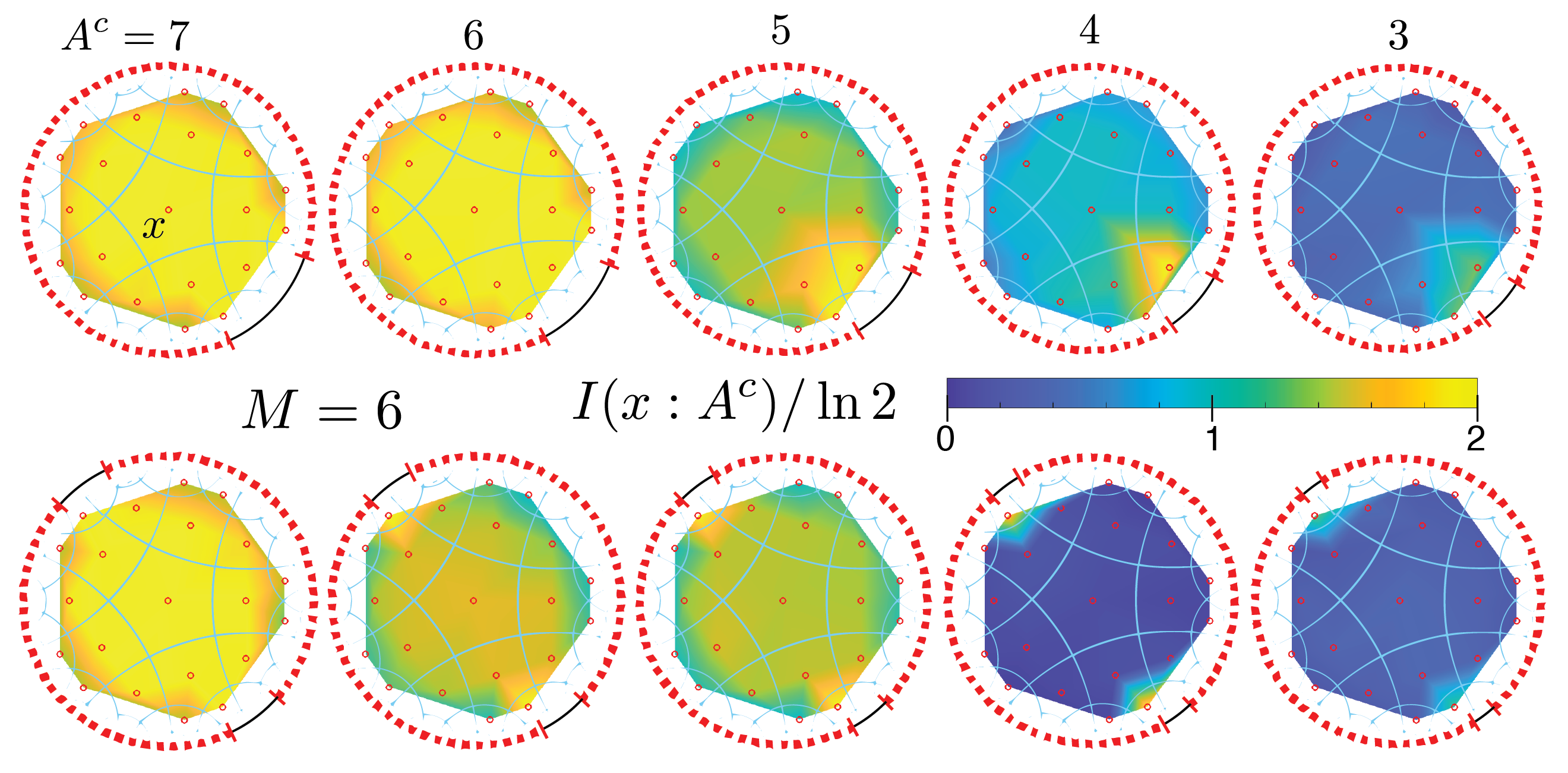}
    \caption{Teleportation fails according to Lemma \ref{thm:teleportationlemma} when there is not enough entanglement shared between $A$ and $A^c$ to teleport all of the bulk qubits. With $b = 21$ bulk qubits and stack depth $M = 6$, we naively expect this crossover to occur around $A^c \sim 3.5$. This expectation is confirmed by our numerical results. Top row: single contiguous measured region $A$. Bottom row: pair of diametrically opposed regions $A = A_1 \cup A_2$, each averaged over 100 random product-state measurement bases.
    }
    \label{fig:happystackteleport}
\end{figure}

Finally, we expect to see Lemma \ref{thm:teleportationlemma} play a role in the physics of our concatenated HaPPY codes. According to the discussion in Section \ref{sec:boundmeasuremteleportation}, the reason large portions of the bulk are able to survive following boundary measurement is because the bulk information is teleported from the measured region $A$ to the unmeasured region $A^c$. However, Lemma \ref{thm:teleportationlemma} tells us that teleportation fails for sufficiently large measured regions $A$ because there is simply not enough entanglement available between $A$ and $A^c$ to support high-fidelity teleportation of all bulk qubits. We explore this phenomenon numerically in Figure \ref{fig:happystackteleport}, where we measure large contiguous regions $A$ (top row) and a pair of large diametrically opposed regions $A = A_1 \cup A_2$ (bottom row). In both cases we find that the bulk information survives even when huge portions of the boundary qubits are measured (left), but that ultimately the bulk qubits collapse when the remaining region $A^c$ becomes small enough such that $S(A^c) < S(b)$ (i.e. the crossover in Lemma \ref{thm:teleportationlemma}). When this happens, there are simply too many bulk qubits to be teleported over to $A^c$ by the meager entanglement resource shared between $b$ and $A^c$, and the bulk qubits are destroyed by the measurement. This is an explicit realization of the predictions of Lemma \ref{thm:teleportationlemma}.

Together, these numerical results provide a qualitatively similar picture of the holographic bulk compared to the complete holographic calculation given in Section \ref{sec:holographiccalculation}. In analogy with the holographic calculation, we find that projective measurements on a boundary region $A$ generally destroy portions of the bulk spacetime. We interpret the remaining bulk degrees of freedom as a spacetime that is cut off by an ETW brane with tension $T$. We have some control over the brane tension by tuning the choice of measurement basis or the stack depth $M$. For the concatenated HaPPY code with stack depth $M = 6$, whose ``analogue large-$N$ limit'' makes it the most suitable toy model to compare to the holographic calculation of Section \ref{sec:holographiccalculation}, we generally find that local projective measurements on the boundary lead to ETW brane geometries with near-critical tension $T \rightarrow T_c$. Whether lower values for the tension---including vanishing and negative tensions---can be achieved in this setup remains an interesting open question for future work. Finally, we find that measuring sufficiently large regions $A$ leads to the complete destruction of the bulk due to Lemma \ref{thm:teleportationlemma}. We shall see that similar conclusions are also borne out in the next section, where we find near-critical ETW branes in random tensor network models.

\section{Random tensor network}\label{sec:rtn}

We can also understand many of these phenomena by modeling the holographic bulk with a random tensor network (RTN) \cite{hayden2016holographic}. In the RTN, calculating holographic entanglement entropy is mapped onto calculating the free energy of an associated classical statistical mechanics model. Under this mapping, projective boundary measurements correspond to free boundary conditions where domain walls can end. A primary benefit of this construction is that it allows us to quickly derive expressions for entanglement entropies in terms of intuitive pictures of domains of Ising spins separated by domain walls.

\subsection{Random tensor network and mapping to Ising model}

\begin{figure}
    \centering
    \includegraphics[width=0.4 \textwidth]{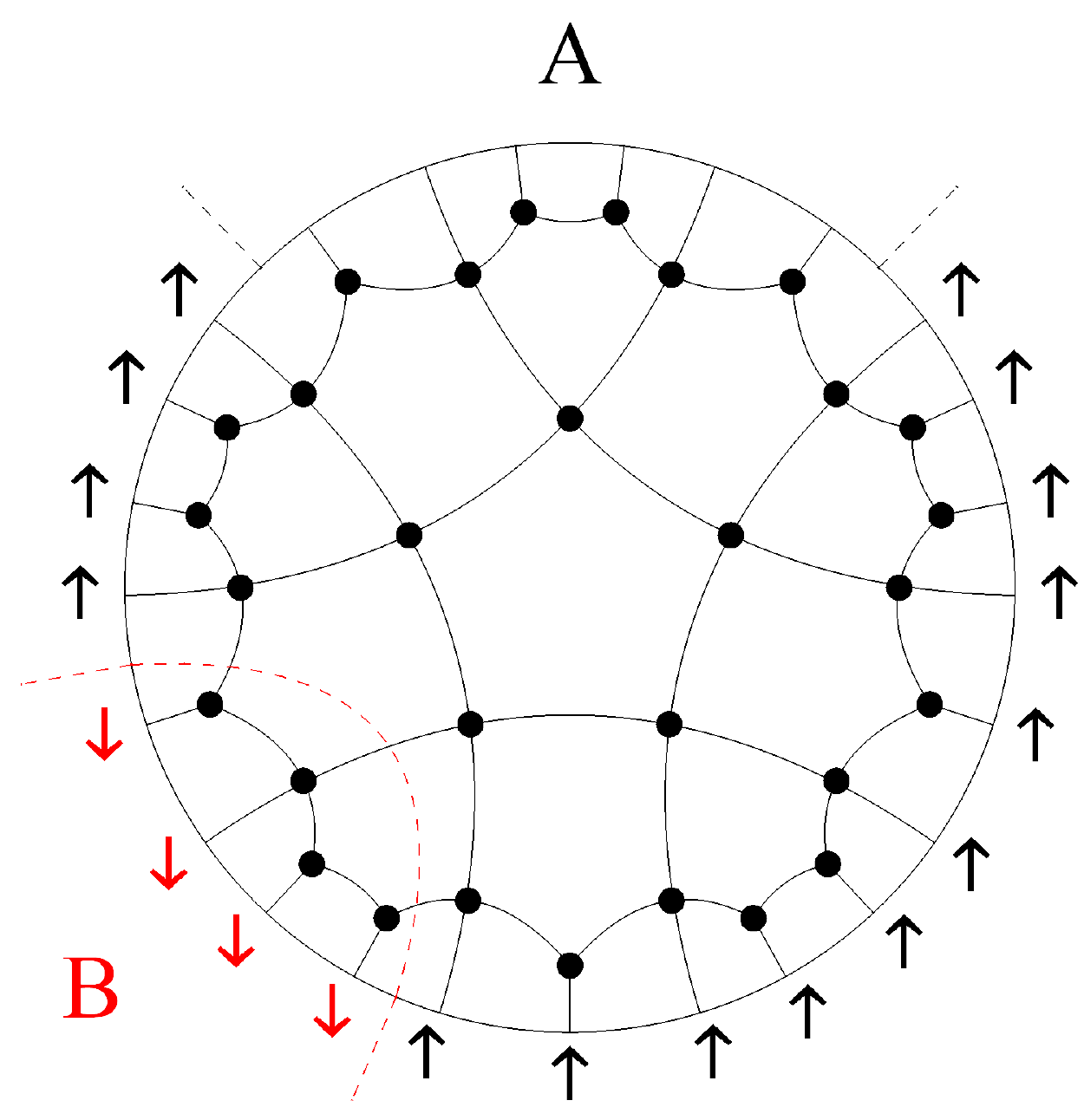}
    \caption{Schematic plot of a hyperbolic random tensor network. Each vertex (black dots) in the bulk denotes a random tensor $\ket{V_x}$. These tensors are fused together along the internal edges of the graph by projecting onto maximally-entangled EPR pairs; the remaining unfused legs define a holographic state on the boundary. Averaging over the Haar ensemble at each vertex yields an effective statistical mechanics model with a discrete classical variable located at each vertex. Under this mapping, calculating entanglement entropies in the original tensor network corresponds to computing free energies in the corresponding statistical mechanics model, subject to certain boundary conditions imposed at the holographic boundary. For example, to compute the 2nd Renyi entropy of a region $B$, we map to a classical Ising model with $-1$ boundary conditions in the region $B$ (red spins) and $+1$ in the remaining boundary (black spins). Performing local projective measurements on a region $A$ of the boundary corresponds to imposing open boundary conditions. Note that the complement of $A \cup B$ is $C$.}
    \label{fig:rtnising}
\end{figure}

In this section, we briefly review holographic random tensor networks and the mapping to an emergent statistical mechanics model~\cite{hayden2016holographic}. Similar to the HaPPY code introduced in Section \ref{sec:happycode}, we model a spatial slice of the holographic bulk using a tensor network composed of elementary tensors. Whereas previously we used the $[[5,1,3]]$ code as the elementary tensor for the HaPPY code, here instead we tessellate hyperbolic space using a network of Haar-random tensors
\begin{equation}
    |V_x \rangle = \sum_{i_k=1}^{D} V_{x; i_1,...,i_n} |i_1 \rangle \otimes ... \otimes |i_n \rangle
    \label{vertex}
\end{equation}
located at each vertex $x$ of a hyperbolic lattice of degree $n$ (shown in Figure \ref{fig:rtnising} for $n = 4$). We assume each leg $k = 1,\ldots,n$ of this tensor represents a Hilbert space of bond dimension $D$. For tensors $V_x,V_y$ sharing a common edge $xy$ in the hyperbolic graph, we fuse their corresponding legs together by projecting onto the maximally entangled state $|xy\rangle = \sum_{i=1}^{D} |i \rangle \otimes | i \rangle$ living on the edge $xy$. Upon fusing the internal legs in this way, the remaining dangling legs at the boundary define an (unnormalized) density matrix
\bea 
\rho = \Tr_P(\rho_P \bigotimes_x |V_x \rangle \langle V_x |) \label{eq:rnt_density}
\eea
where
\bea
\rho_P = %\rho_b \otimes
\bigotimes_{\langle xy \rangle} \frac{|xy\rangle \langle xy|}{D}.
\eea
We view the boundary state $\rho$ as a tensor network model for the holographic bulk, similar to scale-invariant tensor network constructions of \cite{Swingle:2009bg,hayden2016holographic}, where the bulk emerges from the entanglement between boundary degrees of freedom. 

It is also possible to add to the network a bulk density matrix, and view the tensor network as an encoding map from the bulk to the boundary. 
We will consider this case explicitly in Section~\ref{sec:reconstruction}.

Similar to previous sections, we shall be primarily interested in computing entanglement entropies of boundary subregions $B,C$ after performing projective measurements on a boundary subregion $A$. To compute the Renyi-2 entropy of a subsystem $B$, we introduce two copies of the density matrix $\rho$ and calculate the expectation value of the SWAP operator $\mathcal S_B$ acting on subregion $B$:
\bea 
    e^{-S_2(B)} = \frac{\Tr[(\rho \otimes \rho) \ \mathcal{S}_B]}{\Tr[\rho]^2}.
\eea
It is convenient to use the state-operator (Choi-Jamiolkowski) mapping to reformulate the Renyi entropy calculation in terms of a calculation of an inner product. To do so, we can associate a state $|O\rangle = \sum_{ij} O_{ij} |i \rangle \otimes | j \rangle$ to any operator $O = \sum_{ij} O_{ij} |i \rangle \langle j |$.  
For the Renyi-2 entropy, we have tensor products of two density operators, and we can perform such a mapping for both operators to get $ | \rho \rangle \rangle  = \sum_{ijkl} \rho_{ij}\rho_{kl} |i \rangle \otimes | j \rangle \otimes |k \rangle \otimes | l \rangle$.
Using this representation, the Renyi-2 entropy becomes
\bea
   e^{-S_2(B)} = \frac{ \langle \langle \mathcal S_B | \rho \rangle \rangle }{\langle \langle \mathcal I | \rho \rangle \rangle},
\eea
where $|\mathcal S_B \rangle \rangle = \bigotimes_{x \in B} |S_{x\partial} \rangle \rangle \otimes \bigotimes_{x \in C} |I_{x \partial} \rangle \rangle $, and $|\mathcal I \rangle \rangle = \bigotimes_{x \in B} |I_{x \partial} \rangle \rangle \otimes \bigotimes_{x \in C} |I_{x \partial} \rangle \rangle $ with
\bea 
    |I \rangle \rangle= \sum_{i,j} |i \rangle  \otimes | i\rangle \otimes |j\rangle  \otimes | j\rangle,
    \label{eq:ising1} \\ 
    |S \rangle \rangle = \sum_{i,j} |i \rangle  \otimes | j\rangle \otimes |j \rangle  \otimes | i \rangle,
    \label{eq:ising2}
\eea
and the subindex $x\partial$ indicates the bond connecting a bulk vertex $x$ and the boundary $\partial$.

To make progress, we will make the following approximation. 
Instead of doing the average of Renyi-2 entropy, we take the average inside the logarithm and approximate the average of the ratio with the ratio of the averaged quantities:
\bea \label{eq:renyi2}
    S_2(B) = -\log  \frac{ \overline{\langle \langle \mathcal S_B | \rho \rangle \rangle} }{ \overline{\langle \langle \mathcal I | \rho \rangle \rangle}}
\eea
where the overline indicates averaging. With a little abuse of notation, we use the same name $S_2(B)$.  
This approximation is justifed in the large bond dimension limit~\cite{hayden2016holographic}. Because the inner product is a linear operation that commutes with the average, we can first average over the Haar random state, and then take the inner product. 
The average over the random tensors will lead to an Ising-like variable~\cite{hayden2016holographic} 
\bea
    \overline{|V_x \rangle \rangle} = \frac{ |I_x\rangle\rangle + | S_x \rangle \rangle}{D^{2n} + D^n}
\eea
where $|I_x \rangle \rangle$ and $|S_x \rangle \rangle$ are the same states defined in equations (\ref{eq:ising1}) and (\ref{eq:ising2}) but living now on the vertex $x$. 
 
Thus, in the Haar-averaged ensemble, the problem reduces to the thermodynamics of a classical Ising model with an Ising spin $|\sigma_x = \pm 1 \rangle \rangle$ located on each vertex $x$ representing the two possibilities $|I_x/S_{x} \rangle \rangle$, respectively.
The internal connections in the tensor network introduce couplings
\bea
    \langle \langle xy | (|\sigma_x \rangle \rangle \otimes | \sigma_y \rangle \rangle ) = D^{\frac{\sigma_x \sigma_y +3}2},
\eea
between pairs of Ising spins $\sigma_x,\sigma_y$ connected by an edge $xy$.
To calculate the Renyi-2 entropy of a boundary subregion $B$, we  apply the SWAP operator on $B$ and the identity operator everywhere else on the boundary. 
The effect of these boundary operators is to impose fixed boundary conditions $| \sigma_x = \pm 1 \rangle \rangle$ at the boundary, i.e.
\bea \label{eq:boundary_coupling}
    \langle\langle h_{x}| \sigma_{x} \rangle \rangle = D^{\frac{h_x \sigma_x+3}2}
\eea
as shown in Figure \ref{fig:rtnising},
where $ |h_x \rangle \rangle = | I_{\partial}/ S_{\partial} \rangle \rangle $ indicates the boundary state resulting from the identity/swap operator applied on the boundary subregion $B$ and its complement (which we indicate with $C$) respectively, and $|\sigma_x \rangle \rangle$ denotes implicitly the state at a vertex $x$ connected to the boundary.

Using these results, we can express the overlap $\langle \langle \mathcal S_B | \rho \rangle \rangle$ in terms of a quantity in a classical Ising model:
\bea
   && \langle \langle \mathcal S_B | \rho \rangle \rangle = \\
   && \sum_{\sigma} \exp\left\{-\left[ - \sum_{\langle xy\rangle } \log D \left( \frac{\sigma_x \sigma_y}2 + \frac32 \right) - \sum_{x \in \partial} \log D \left( \frac{\sigma_x h_x}2 + \frac32 \right) 
   + \sum_x \log(D^{2n} + D^n) \right] \right\},  \nn 
   \label{partfunc}
\eea
where the first term is the bulk contribution, the second term is the boundary contribution, and
\bea \label{eq:renyi2_bc}
    h_{x\in B} = -1 , \quad h_{x \in C} = +1
\eea 
defines the boundary operator used to calculate Renyi-2 entropy of the subregion $B$. 
The summation in front of the exponential is over all possible Ising variables $\sigma_x$ at each vertex. 
Therefore, equation (\ref{partfunc}) represents a partition function of an Ising model living on the network. 
The coupling between two vertices is ferromagnetic with strength $-\log D$.
The Renyi-2 entropy is reflected by the boundary Ising variables $h_x$, also with a ferromagnetic coupling $- \log D$. 
Note that the Ising variable $\sigma_x$ is summed over in the bulk, whereas the Ising variable at the boundary $h_x$ is fixed.
Therefore, we sometimes refer to the latter as a boundary condition or an externally-imposed magnetic field. We obtain a similar expression for the quantity $\langle \langle \mathcal I | \rho \rangle \rangle$, except that we impose uniform boundary conditions $h = +1$ everwhere on the boundary $B \cup C$.

For a large bond dimension, the partition function is dominated by the lowest energy state. This implies that $\langle \langle \mathcal I | \rho \rangle \rangle$ is simply given by a ferromagnetic phase with  $\sigma_x = 1$ for all vertices in the bulk. 
The Renyi-2 entropy defined in equation (\ref{eq:renyi2}) is then given by
\bea
    S_{2}(B) = \min_{\sigma} \left[ -\sum_{\langle xy\rangle } \log D \left( \frac{\sigma_x \sigma_y -1 }2\right) - \sum_{x \in \partial} \log D \left( \frac{\sigma_x h_x-1}2 \right) \right].
    \label{renyii}
\eea
This classical spin model is defined on the hyperbolic network shown in Figure \ref{fig:rtnising}, and
minimization over the configurations of all possible Ising variables leads to a domain wall related to the minimal (RT) surface in the hyperbolic space~\cite{hayden2016holographic}.

\subsection{Measurement at the boundary}
\label{measbdy}

From the arguments above, we find that Renyi entropies $S_2(B)$ in the random tensor network can be calculated by studying a classical Ising model with particular boundary conditions imposed. Here we argue that projective measurements applied to a boundary region $A$ correspond to imposing free boundary conditions on region $A$~\footnote{Projective measurements in the bulk of random tensor networks have been discussed in Ref.~\cite{yang2022entanglement,li2021statistical}}. Any measurement on a region $A$, as illustrated in Figure~\ref{fig:rtnising}, can be represented by a set of Kraus operators, $K_{a,\alpha},$
such that 
\bea
    \sum_{a,\alpha} K_{a,\alpha}^\dag K_{a,\alpha} = \mathds{1}.
    \label{kraus}
\eea
Generally speaking, measurement of a quantum system collapses the state and yields a classical result called the \emph{measurement record}. It is crucial whether we choose to either retain this measurement record or throw it away -- which is tantamount to tracing over the measured region. Here for generality, we construct a model where both options are available and we can tune between them. In equation (\ref{kraus}) we introduced two different indices $a$ and $\alpha$, where the $a$ type index means the measurement outcome is discarded, and the $\alpha$ type index means the measurement outcome is recorded. Physically it means that the apparatus is able to read some types of measurement, but not all of them. 

If the Kraus operator is uncorrelated with the random tensors, the following quantity contains the boundary contributions:
\bea
    \sum_{a,b} \left(\bigotimes_{x\in A} \langle \langle h_{x} | \right) (K_{a,\alpha} \otimes K_{a,\alpha}^\ast \otimes K_{b,\alpha} \otimes K_{b,\alpha}^\ast ) \left( \bigotimes_{y\in A} |\sigma_{y} \rangle \rangle \right).
\eea
Here, $\sigma_y$ denotes the vertex that is connected to the boundary $A$ and $h_x$ denotes the possible boundary condition arisen from the calculation of Renyi-2 entropy. 
The summation over $a$ and $b$ is due to the fact that the measurement outcome is discarded for these basis states, while $\alpha$ is the recorded measurement outcome and is the same for all four copies. 
This expression is a complicated boundary term and depends not only on the measurement, but also on whether region $A$ is included in the Renyi-2 entropy calculation. 
In the following, we assume there are three non-overlapping regions $A, B, C$, and we are interested in the Renyi-2 entropy of region $B$ after measuring $A$, as shown in Figure~\ref{fig:rtnising}. 
In this case, the boundary condition in region $A$ after the measurement is $\left(\bigotimes_{x\in A} \langle \langle I_{x\partial} | \right)$, so the boundary term is simplified to be
\bea \label{eq:K-boundary}
    \sum_{a,b} \left(\bigotimes_{x\in A} \langle \langle I_{x\partial} | \right) (K_{a,\alpha} \otimes K_{a,\alpha}^\ast \otimes K_{b,\alpha} \otimes K_{b,\alpha}^\ast ) \left( \bigotimes_{y\in A} |\sigma_{y} \rangle \rangle \right).
\eea
On the other hand, the boundary couplings in regions $B$ and $C$ are the same as in~(\ref{eq:boundary_coupling}).

While the boundary coupling is completely general for any quantum channel $K$, in the following we will focus on two specific cases.  
Inspired by the boundary CFT, we consider local measurements, namely the measurement is implemented at each individual leg. 
In this case, the projected state has vanishing spatial entanglement~\cite{miyaji2014boundary,Cardy:1989ir, Cardy:2004hm}.  
We take the quantum channel for each boundary leg to be the same, i.e.,
\bea
\label{eq:local_projective}
&& K_{a,\alpha}=|a\rangle\langle a| \otimes |\alpha \rangle \langle \alpha|, \quad a=1,...,D_1, \quad \alpha=1,...,D_2, \quad  D = D_1 D_2,
\eea
where we neglect the leg index for simplicity. 
Then (\ref{eq:K-boundary}) factorizes as
\bea
    \prod_{x\in A} \sum_{a,b} \langle \langle I_{x\partial} | (K_{a,\alpha} \otimes K_{a,\alpha}^\ast \otimes K_{b,\alpha} \otimes K_{b,\alpha}^\ast )  |\sigma_{x} \rangle \rangle,
\eea
and each factor is given by one of the following quantities depending on the value of $\sigma_{x\in A}$:
\bea 
    \sum_{a,b} \langle \langle I_{x\partial} | (K_{a,\alpha} \otimes K_{a,\alpha}^\ast \otimes K_{b,\alpha} \otimes K_{b,\alpha}^\ast )  | I_{x} \rangle \rangle = D_1^2, \label{eq:measure_ising1}\\
    \sum_{a,b} \langle \langle I_{x\partial} | (K_{a,\alpha} \otimes K_{a,\alpha}^\ast \otimes K_{b,\alpha} \otimes K_{b,\alpha}^\ast )  | S_{x} \rangle \rangle = D_1.\label{eq:measure_ising2}
\eea
Thanks to the average over the Haar random state, these contributions do not depend on the measurement outcome $\alpha$, but only on the dimension $D_1$. 

Thus, the boundary term in the computation of the Renyi-2 entropy in the Ising model is changed to
\begin{small}
\bea
\begin{aligned}
    &S_2^{(A)}(B) =\\
    &\min_\sigma \left[ - \sum_{\langle xy\rangle } \log D \left( \frac{\sigma_x \sigma_y -1}2 \right) - \sum_{x \in B,C} \log D \left( \frac{\sigma_x h_x -1}2 \right) - \sum_{x \in A} \log D_1 \left( \frac{\sigma_x h_x  -1}2 \right)\right],
    \end{aligned}
    \label{newboundary}
\eea
\end{small}
where $h_{x \in B} = -1$, $h_{x \in C} = 1$ and $h_{x \in A} = 1$. 
The effect of measurement is to modify the boundary term associated with the measured region $A$. 
The boundary Ising variable is still $h_{x \in A} = 1$, whereas the coupling of this boundary term is modified and determined by the dimension of the unrecorded measurement $-\log D_1$.

If we are able to keep all the measurement outcomes, i.e. $D_1 = 1$ and $D_2 = D$, then the boundary term for $A$ vanishes, and the boundary variable $h_{x\in A}$ decouples from the bulk Ising variable $\sigma_{x \in A}$. 
In this case, we have a free boundary condition, i.e. the boundary contributions for $\sigma_{x\in A} = \pm 1$ are the same (they are both zero). 
To see the implication of this boundary condition, it is instructive to compare it to the boundary conditions (\ref{eq:renyi2_bc}) for the computation of the Renyi-2 entropy. 
Apparently, the boundary conditions in equation (\ref{eq:renyi2_bc}) actually create a domain wall between $B$ and $C$ since the boundary Ising variables are opposite. 
As the holographic interpretation of the domain wall is the RT surface, it means the RT surface ends right at the intersection point between the boundary subregions $B$ and $C$. 
On the other hand, since after measurement the boundary Ising variable in region $A$ decouples and $\sigma_{x \in A}$ has a free boundary condition, a domain wall can be created between any two adjacent points in the measured region $A$.
Holographically, it means that a RT surface can end on any point of the measured region $A$. This result is reminiscent of an ETW brane which closely ``hugs'' the boundary along region $A$. We interpret this situation as having a brane with critical tension $T \rightarrow T_c$ in the random tensor network.

A quantum information understanding of the critical tension here is that the random tensors lead to a random encoding map of bulk information into the boundary legs. 
If we use this encoding map to push the logical operator and the corresponding measurement in the bulk to the boundary---see the discussion near equation (\ref{eq:logical_measurement})---in general it would become a very complicated logical operator that cannot be factorized into local projective measurements performed on the boundary like in equation (\ref{eq:local_projective}). 
Thus, the information remains intact after a local projective measurement, and is therefore teleported to the unmeasured region $A^c=B\cup C$.\footnote{Although we do not include bulk legs explicitly in this subsection, we can imagine there is a dangling bulk leg with $\mathcal{O}(1)$ bond dimension for each tensor. The teleportation can thus be successful because the entanglement resource needed to perform it is also $\mathcal{O}(1)$. We will discuss the case of large bulk bond dimension in the next subsection.}
This effect is nicely manifested in the Ising model as a free boundary condition.
More precisely, the measurement does not have any information about the bulk Ising variable as the quantities (\ref{eq:measure_ising1}) and (\ref{eq:measure_ising2}) are the same when $D_1 =1$.

On the other hand, if we are not able to keep any measurement outcome, i.e. $D_1 = D$ and $D_2=1$, then the boundary term for region $A$ in equation (\ref{newboundary}) reduces to the case without measurement of equation (\ref{renyii}). 
This is because measuring region $A$ and discarding the outcome is equivalent to tracing region $A$ out. When computing the entanglement entropy of $B$, region $A$ must be traced out anyway and therefore it does not matter whether region $A$ is measured or not as long as the outcome of the measurement is discarded.

\subsection{Bulk reconstruction} \label{sec:reconstruction}

In the previous section, we used the effective Ising model to show that performing local measurements on a boundary subregion $A$ -- and keeping the measurement record -- is equivalent to imposing free boundary conditions on $A$. We view these open boundary conditions as representing a near-critical tension ETW brane hugging the boundary on which RT surfaces can end.

In this section, we consider the case in which dangling bulk legs are present, and the dimension of the bulk Hilbert space is comparable to the dimension of the boundary Hilbert space. 
The entanglement resource is then not enough to teleport the information when the measurement region is big enough. We investigate how a measurement performed in the boundary affects the reconstruction of bulk operators under these conditions.
In order to address this question, we view the tensor network as an encoding map from the bulk Hilbert space to the boundary Hilbert space, and perform the bulk reconstruction using this map. 
When the measured region $A$ is small, the random tensor network remains an isometry and the bulk reconstruction will always succeed via canonical reconstruction---see equation (\ref{eq:canonical-reconstruct}). 
On the other hand, when the measured region is big, the map becomes non-isometric. 
In this case, we will see a transition in the shape of the entanglement wedge---see equation (\ref{eq:ew-bound-R})---and will give a sufficient condition (\ref{eq:reconstruct-bound-R}) under which bulk operators cannot be reconstructed from the unmeasured boundary subregion $A^c$.

\subsubsection{Encoding map from random tensor network with measurement}

Here, we properly define the encoding map from the bulk Hilbert space to the boundary Hilbert space realized by the random tensor network. 
We are particularly interested in the reconstruction of operators deep into the bulk.
Consider a region deep into the bulk called $b$, such as the region shaded in gray in Figure~\ref{fig:RTN}, and assign each vertex state within that region a bulk dangling leg, i.e.
\bea
|V_x \rangle = \sum_{i_k=1}^D \sum_{i_b=1}^{D_b} V_{x; i_1,...,i_n,i_b} |i_1 \rangle \otimes ... \otimes |i_n \rangle \otimes |i_b \rangle, \quad x \in b,
\eea
which is a Haar random state at vertex $x \in b$ with $n+1$ legs, and $k=1,...,n$. 
We assume each one of the first $n$ legs has bond dimension $D$, whereas the bulk leg has bond dimension $D_b$. 
Notice that we do not require $b$ to be the whole bulk and for the complementary region $b^c$ the vertex state is defined without a bulk leg as in (\ref{vertex}).

The holographic map realized by the random tensor network without measurement is
\bea 
    V &=&   \sqrt{\mathcal N_0}\left(\bigotimes_{\langle xy \rangle} \langle xy | \right) \left( \bigotimes_x |V_x \rangle \right)  , 
\eea
where $|xy\rangle = \sum_{i=1}^{D} |i \rangle \otimes | i \rangle$ is a maximally entangled state living in the bonds connecting vertices $x$ and $y$, and $\mathcal N_0 = \prod_{x \in b} D_{b} \cdot \prod_{\langle xy\rangle } D$. 
Viewed as a big tensor constructed from small tensors, $V$ should be understood as a map from the bulk legs to the boundary legs, so it maps a state in the Hilbert space $\mathcal H_b$ of the bulk $b$ to the Hilbert space $\mathcal H_\partial$ of the entire boundary: $V: \mathcal H_b \rightarrow H_\partial$. This map is similar to the random tensor network in~\cite{hayden2016holographic} except we restrict the bulk legs to lie within a small region $b$ instead of the full bulk. 
The normalization factor $\mathcal N_0$ is chosen such that for any two states $\psi_1$ and $\psi_2$ in the bulk Hilbert space, $\overline{ \langle \psi_1 |  V^\dag V | \psi_2 \rangle} = \langle \psi_1 | \psi_2 \rangle $ holds, i.e. the map preserves inner products on average.\footnote{In the previous subsection we did not consider a normalized state as we are doing here, but we introduced the normalization factor in the computation of the Renyi-2 entropy, see equation (\ref{eq:renyi2}).}

We are interested in investigating how a measurement performed on the boundary changes the properties of the map $V$. 
More explicitly, we separate the boundary into two subregions $\partial = A \cup B$, and measure the legs in region $A$. 
Then this defines a linear map $V_{B}: \mathcal H_b \rightarrow \mathcal H_{B}$
\bea \label{eq:map_rtn}
    V_{B} = \bigotimes_{x \in A} \sqrt{D} \langle 0_x | V,
\eea
which can be represented schematically as
\begin{center}
    \includegraphics[width=0.35\textwidth]{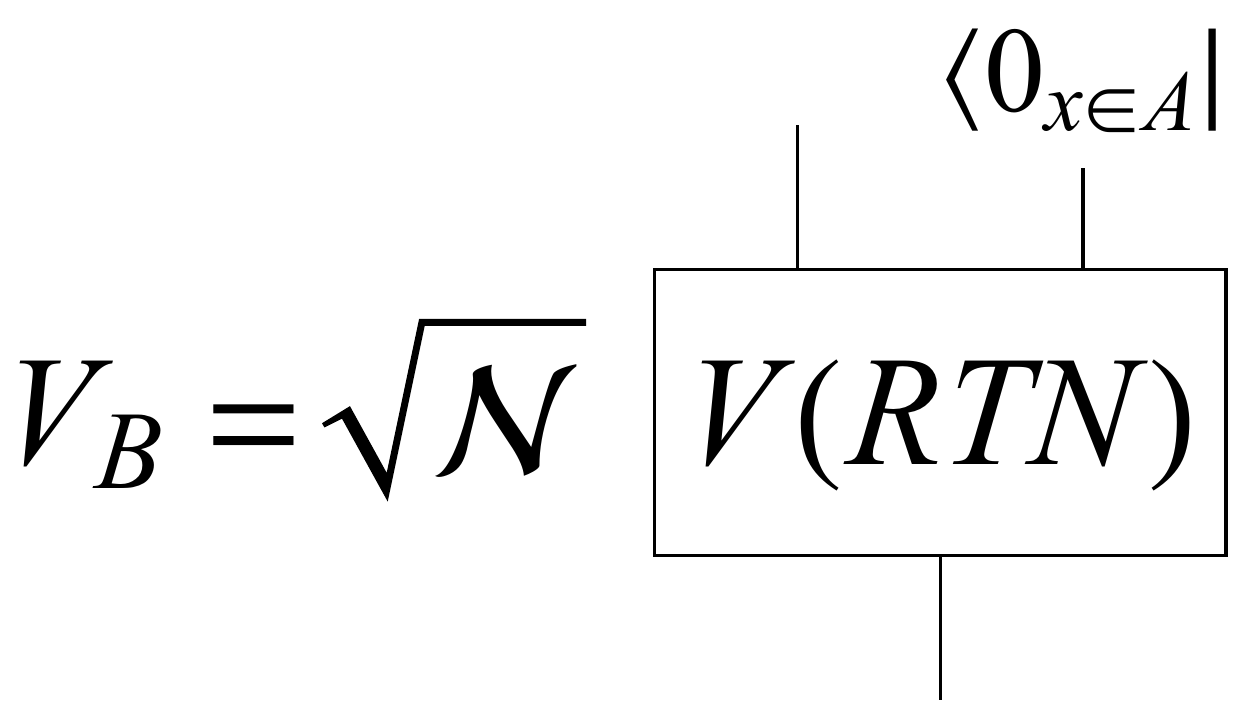}
\end{center}
where $\mathcal N =  \prod_{x\in A} D$, and RTN indicates the map $V$ constructed from a random tensor network. 
The incoming legs on the bottom are the bulk legs in $b$, and the outgoing legs on the top are the boundary legs in $B$. 
The projection outcome is not important because we are interested in the typical behavior of $V_{B}$ averaged over random tensors. 
We now introduce a reference Hilbert space $\mathcal H_R$ that has the same dimension as $\mathcal H_b$ to purify any state in the bulk. 
The full bulk-to-boundary map is then
\bea
    V_{B} \otimes \mathds{1}_R : \mathcal H_b \otimes \mathcal H_R \rightarrow \mathcal H_{B} \otimes \mathcal H_R,
    \label{measmap}
\eea
where $\mathds{1}_R$ is an identity map. 
We denote the volume of $b$ by $|R|$ and the volume of $B$ by $|B|$\footnote{Here the volume of a given region is given by the number of dangling legs in that region.} (note that $B$ is the boundary, so its volume is proportional to its length). 
In this construction, bulk dangling legs have bond dimension $D_b$ and all the other legs have dimension $D$. 
Thus, the dimension of the code space (the bulk Hilbert space) and the physical space (the boundary Hilbert space) are $D_b^{|R|}$ and $D^{|B|}$, respectively. 
In the following, we will fix the dimension of the full boundary $\partial = A \cup B$ so that increasing the size of the measured region $A$ decreases the size of the complementary region $B$.

\subsubsection{Non-isometric map from measurement}\label{subsec:noniso}

If $V_B$ is an isometry, i.e. 
\bea \label{eq:isometry}
    V_B^\dag V_B = \mathds 1_b,
\eea 
we can reconstruct any operator $W_b$ acting on $\mathcal H_b$ by an operator acting only in region $B$. 
This can be achieved by the canonical reconstruction
\bea \label{eq:canonical-reconstruct}
    W_B = V_B W_b V_B^\dag,
\eea
such that for any state in $| \psi\rangle  \in \mathcal H_b \otimes \mathcal H_R$
\bea
    W_B V_B | \psi \rangle = V_B W_b V_B^\dag V_B | \psi \rangle = V_B W_b | \psi \rangle,
\eea
where we omit the identity map $\mathds{1}_R$ acting on the reference Hilbert space for simplicity.

We first notice that if the map is non-isometric, then the canonical reconstruction~(\ref{eq:canonical-reconstruct}) is not guaranteed to work anymore. Can the measurement lead to a violation of the isometry condition (\ref{eq:isometry})? 
The answer is definitely yes. 
For any map $V_B$, we can define an associated state 
\bea
    |V_B \rangle = \sum_{\mu, m} (V_B)_{\mu, m} |\mu, m \rangle, 
\eea
where $|m \rangle $ ($|\mu \rangle$) is a complete basis of the Hilbert space $\mathcal H_b$ ($\mathcal H_B$). 
The condition~(\ref{eq:isometry}) implies that the reduced density matrix of $b$ is a maximally mixed state
\bea
    \rho_b(V_B) \equiv \Tr_B(|V_B \rangle \langle V_B|) = \frac1{D_b^{|R|}}  \mathds 1_b . 
\eea
This in turn implies the second Renyi entropy of such a maximally mixed state is maximal, i.e. $S_{2}(b)_{\rho_b(V_B)} = |R| \log D_b$. 
Thus, this is a necessary condition for $V_B$ to be an isometry.

\begin{figure}
    \centering
    \includegraphics[width=0.3\textwidth]{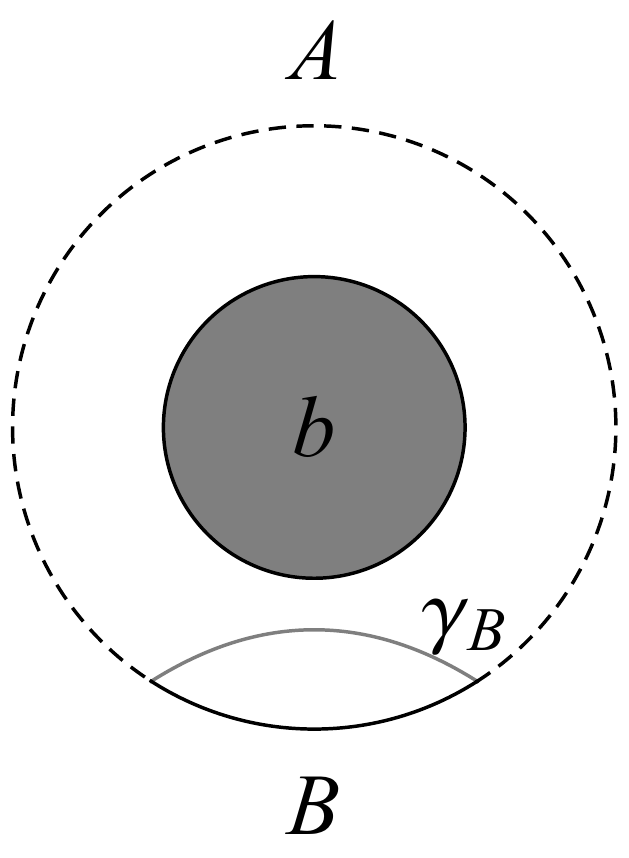}
    \caption{Schematic representation of the random tensor network model studied in Section \ref{sec:reconstruction}. $b$ indicates the bulk region with the incoming legs, $B$ is the unmeasured boundary subregion, and the dashed line along subregion $A$ on the boundary indicates measurement. $\gamma_B$ is the bulk geodesic homologous to $B$. }
    \label{fig:RTN}
\end{figure}

Similar to the previous discussion, the Renyi-2 entropy calculation can be mapped to an Ising model calculation \cite{hayden2016holographic}:
\bea
    && S_{2}(b)_{\rho_b(V_B)} = \\
    && \min_{\sigma} \Big[ -\sum_{\langle xy\rangle } \log D \left( \frac{\sigma_x \sigma_y -1 }2\right) - \sum_{x \in B} \log D \left( \frac{h_x \sigma_x -1}2 \right) - \sum_{x \in b} \log D_b \left(\frac{h_x \sigma_x-1}2 \right) \Big], \nn 
\eea
where $h_{x \in B} = 1, h_{x \in b} = -1$. 
Importantly, the magnetic field in the boundary region $B$ and in the bulk $b$ have opposite signs. 
If the measured region $A$ is small, $B$ is big and the dominant spin configuration is all spin-up ($\sigma_x = 1$). This gives rise to the maximal entropy we found in the previous discussion. 
Increasing the size of the measured region $A$, there are two competing configurations when $B$ is small enough, and the Renyi-2 entropy reads
\bea
    S_2(b)_{\rho_b(V_B)} = \min(|R|\log D_b, |\gamma_B| \log D), 
\eea
where $\gamma_B$ denotes the bulk geodesic homologous to $B$ as shown in Figure~\ref{fig:RTN}.\footnote{Here, we assume that $|B|$ is small enough such that the geodesic $\gamma_B \notin b$}
The first configuration is all spin-up, and the second has a domain wall $\gamma_B$ separating a spin-up region extending from $\gamma_B$ to the boundary $B$ from a spin-down region including $b$. 
    We can conclude that if the measured region $A$ is big enough, and so $B$ is small enough to satisfy
\bea
|\gamma_B| \log D < |R| \log D_b,
\label{noniso}
\eea
the Renyi-2 entropy is given by  $S_2(b)_{\rho_b(V_B)} = |\gamma_B| \log D$, violating the necessary condition for $V_B$ to be an isometry. We remark that in the $D\gg D_b$ limit the condition (\ref{noniso}) is never satisfied, even when $B$ is very small. Therefore the bulk-to-boundary map is always isometric. Given that this limit in which the boundary Hilbert space is parametrically larger than the bulk Hilbert space is the one relevant for holography---in which the boundary theory is a large-$N$ theory and the bulk effective field theory is not---we expect that in holographic settings such as the one described in Section \ref{sec:holographiccalculation} the bulk-to-boundary map always remains isometric, independently of the size of the measured region. 

When condition (\ref{noniso}) is satisfied, the map is non-isometric and the canonical reconstruction fails. 
In this situation, the question is whether it is still possible to reconstruct an operator $W_b$ in $B$ in a state-specific way.  
We can figure out whether this is the case by adopting an entanglement wedge point of view.\footnote{We will consider Renyi-2 entropies. However, with an abuse of terminology, we will still refer to the bulk region bounded by a boundary region and the corresponding domain wall computing the Renyi-2 entropy as the entanglement wedge.}

To this end, we consider a state $|\psi\rangle \in \mathcal H_b \otimes \mathcal H_R$.
Under the map (\ref{measmap}), we get a state in the physical Hilbert space $V_B |\psi \rangle \in \mathcal H_B \otimes \mathcal H_R$.
We look at the second Renyi entropy of $B$ calculated in $V_B |\psi \rangle$:
\bea
    && S_2(B)_{V_B|\psi\rangle} = \\
    && \min_\sigma \Big[ -\sum_{\langle xy\rangle } \log D \left( \frac{\sigma_x \sigma_y -1 }2\right) - \sum_{x \in B} \log D \left( \frac{h_x \sigma_x -1}2 \right) +  S_2(\sigma_{x \in b} = -1)_{|\psi\rangle} \Big], \nn 
\eea
where $h_{x\in B} = -1$. The last term is the second Renyi entropy of a subregion inside $b$ with $\sigma_x = -1$ calculated in the state $|\psi \rangle$.
When the measured region $A$ is large enough, there are two competing configurations:
\bea \label{eq:config}
    S_2(B)_{V_B | \psi \rangle } = \min (S_2(b)_{|\psi \rangle}, |\gamma_B| \log D).  \eea
The first configuration corresponds to all spin-down, i.e. the Ising variables in the bulk are pinned by the boundary condition $h_{x \in B} = -1$ in region $B$. 
Recalling the analogy between the domain wall and the RT surface, this means that the entanglement wedge of region $B$ is the full bulk.
On the other hand, the second configuration corresponds to a spin down region bounded by the geodesic $\gamma_B$ and the boundary $B$. 
The magnetic field in the bulk is able to create a domain wall homologous to the boundary region $B$, see Figure~\ref{fig:RTN} for an illustration. 
This means that the entanglement wedge of region $B$ is bounded by the domain wall $\gamma_B \notin b$ and does not contain region $b$. Therefore, the information in $b$ cannot be accessed by the unmeasured region $B$.

We can see that even when the map is no longer an isometry, i.e. when $|\gamma_B| \log D < |R| \log D_b$, it is still possible that the entanglement wedge of $B$ is the entire bulk for the state $|\psi \rangle$, as long as $S_2(b)_\ket{\psi} < |\gamma_B|\log D$. 
On the other hand, this calculation also suggests that when $B$ is small enough and $D_b$ is large enough
\bea \label{eq:ew-bound-R}
    S_2(b)_{|\psi \rangle} > |\gamma_B| \log D
\eea
holds and the entanglement wedge of $B$ does not contain $b$. Therefore, the operator $W_b$ cannot be reconstructed in $\mathcal H_B$. 
We then expect to be able to reconstruct such an operator in $\mathcal H_R$. 
This analysis of the entanglement wedge gives a hint that in certain circumstances the information is not intact under measurement. 

In the following, we provide a more rigorous result. Namely, we give a sufficient condition under which the operator $W_b$ can be reconstructed in $\mathcal H_R$ for a specific state. This is possible whenever the measured region $A$ is large enough.
According to complementary recovery~\cite{Akers:2022qdl}, this means that $W_b$ cannot be reconstructed in $\mathcal{H}_B$.

\subsubsection{Haar random unitary with measurement}

Before we dive into the reconstruction in random tensor network, we review a simple model that has the same properties upon measurement. 
This model is a simplified version of the random unitary model in Ref.~\cite{Akers:2022qdl}.

Again, we consider a code space $\mathcal H_b$ and a physical space $\mathcal H_B$ with dimensions $|R|$ and $|B|$, respectively. 
Note that the notation of Hilbert space dimension is different from the one used above for the tensor network.
Now, instead of (\ref{eq:map_rtn}), we consider
$\tilde V: \mathcal H_b \rightarrow \mathcal H_B, 
$ defined as $\tilde V = \sqrt{|P|} \langle 0 | U | \psi_f \rangle$:

\begin{center}
\includegraphics[width=0.35\textwidth]{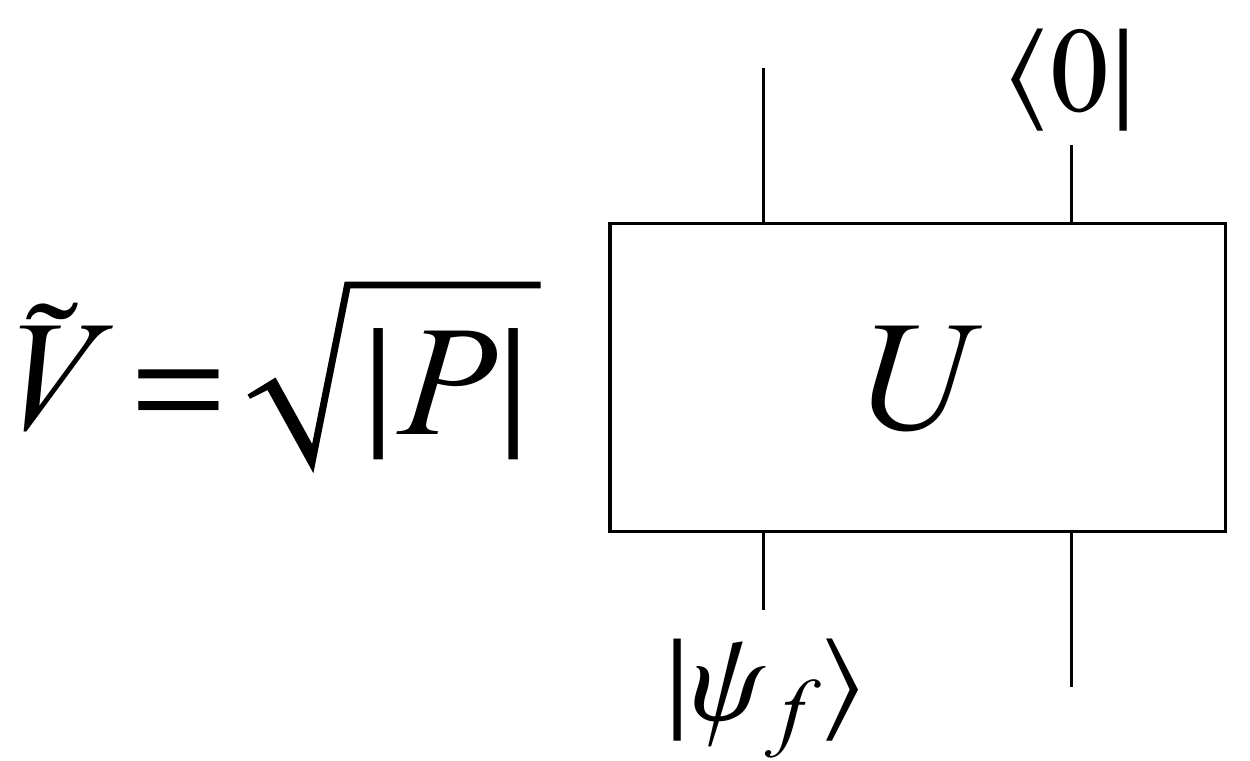}
\end{center}
where $U$ is a unitary operator. 
We measure a subset of the outgoing legs of $U$ with dimension $|P|$.
The measurement outcome is not important because we are interested in the typical properties of a random unitary drawn from the uniform Haar distribution. Therefore, we denote the measurement outcome to be $| 0\rangle$. 
Now it is clear from the definition that the dimension of the unitary is $2^L = |P||B|$, where we assume this unitary acts on $L$ qubits.
The code space dimension is $|R|$ and the mismatch between these two dimensions is compensated by the fed state $|\psi_f \rangle$. 
The details of the fixed state $|\psi_f \rangle$ are not important, as it is introduced only to match the Hilbert space dimensions of incoming and outgoing legs of the unitary operator $U$. 
When no outgoing legs are measured, $\tilde V$ is an isometry because $2^L > |R|$.
If instead we fix the total dimension $L$ and measure a large number of outgoing legs such that $|B| < |R|$, then the linear map is no longer isometric. 
We are interested in understanding how the measurement affects the reconstruction procedure. Like we did in the previous subsections, we also include another reference Hilbert space $\mathcal H_R$---which has the same dimension as $\mathcal H_b$---and consider the linear map
$\tilde V \otimes \mathds{1}_R: \mathcal H_b \otimes \mathcal H_R \rightarrow \mathcal H_B \otimes \mathcal H_R$, where $\mathds 1_R$ is the identity map on the Hilbert space $\mathcal H_R$.

Consider a unitary operator $W_b$ acting on the incoming legs of $\tilde V$ in the code space and a state $|\psi \rangle \in \mathcal H_b \otimes \mathcal H_R$. We want to understand under which conditions $W_b$ can be reconstructed in $\mathcal{H}_B$ or $\mathcal H_R$. 
We will make use of a useful state-specific reconstruction theorem introduced in \cite{Akers:2022qdl}, which we report here for completeness.
\\

\begin{theorem}
Let $\mathcal H_b$, $\mathcal H_B$, and $\mathcal H_{A}$ be finite dimensional Hilbert spaces, $L: \mathcal H_b \rightarrow \mathcal H_B \otimes \mathcal H_{A}$ a linear map, $W_b$ a unitary operator on $\mathcal H_b$, and $|\psi \rangle \in \mathcal H_b$. Then the following statements are equivalent:

(1) Existence of a unitary operator $W_B$ on $\mathcal H_B$ such that 
\bea
    || W_B L | \psi \rangle - L W_b |\psi \rangle|| \le \epsilon_1
\eea

(2) The decoupling condition
\bea
    || \Tr_B[LW_b |\psi \rangle \langle \psi | W_b^\dag L^\dag] - \Tr_B[L |\psi \rangle \langle \psi | L^\dag] || \le \epsilon_2 .
\eea
(1) implies (2) with $\epsilon_2 \le \sqrt{2(\langle \psi | L^\dag L | \psi \rangle)+ \langle \psi | W_b^\dag L^\dag L W_b | \psi \rangle} \epsilon_1$ and (2) implies (1) with $\epsilon_1 \le \sqrt{\epsilon_2}$.\footnote{For a linear map $L$, the Schatten $p$-norm is $|| L ||_p = (\Tr[(X^\dag X)^{p/2}])^{1/p}$.}
\label{akersth}
\end{theorem}

When the number of measured legs is small, we expect that the information remains intact and accessible from $B$. 
We can apply Theorem \ref{akersth} to see when the operator can be reconstructed on the unmeasured legs in region $B$. 
Consider the quantity
\bea
    && \int dU || \Tr_B[ \tilde V W_b | \psi \rangle \langle \psi | W_b^\dag \tilde V^\dag ] - \Tr_B[\tilde V | \psi \rangle \langle \psi | \tilde V^\dag ] ||_1 \\
    &\le& \sqrt{|R|} \int dU || \Tr_B[\tilde V W_b | \psi \rangle \langle \psi | W_b^\dag \tilde V^\dag ]   - \Tr_B[ \tilde V | \psi \rangle \langle \psi |  \tilde V^\dag ]||_2 \\
    &\le& \sqrt{ |R| \int dU || \Tr_B[\tilde V W_b | \psi \rangle \langle \psi | W_b^\dag \tilde V^\dag ]   - \Tr_B[\tilde V | \psi \rangle \langle \psi |  \tilde V^\dag]||_2^2 }
\eea
where two inequalities are applied to simplify and bound the 1-norm~\cite{Akers:2022qdl}. 
Since we are interested in the typical behavior of $\tilde V$, we use $\int dU $ to denote the integration over the unitary group with Haar measure (we remind that $\tilde V = \sqrt{|P|} \langle 0 | U | \psi_f \rangle$). Now we can calculate the right-hand side using the result of the Haar random average:
\bea \label{eq:B-reconstruction} 
    \int dU || \Tr_B[\tilde V W_b | \psi \rangle \langle \psi | W_b^\dag \tilde V^\dag ] - \Tr_B[\tilde V | \psi \rangle \langle \psi | \tilde V^\dag]||_2^2 
    &\le & \frac{4}{|B|}.
\eea
Using this result, we can see that when 
\bea
    \int dU || \Tr_B[\tilde V W_b | \psi \rangle \langle \psi | W_b^\dag\tilde V^\dag ]   - \Tr_B[\tilde V | \psi \rangle \langle \psi | \tilde V^\dag] ||_1 &\le& \sqrt{ \frac{4|R|}{|B|} } \le \epsilon, 
\eea
there exists an operator $W_B$ in $\mathcal H_B$ such that $||\tilde V W_b | \psi \rangle - W_B \tilde V |\psi \rangle || \le \epsilon'$.

On the other hand, when a big portion of the boundary is measured and the above inequality does not hold, one is not able to reconstruct $W_b$ in $B$.
In this situation the operator can be reconstructed in $R$ instead. 
In order to understand why this is the case, consider the quantity
\bea \label{eq:R-reconstruction}
    && \int dU || \Tr_R[\tilde V W_b | \psi \rangle \langle \psi | W_b^\dag \tilde V^\dag ] - \Tr_R[\tilde V | \psi \rangle \langle \psi | \tilde V^\dag ]  ||_1 \nn \\
    &\le& \sqrt{ |B| \int dU || \Tr_R[\tilde V W_b | \psi \rangle \langle \psi | W_b^\dag \tilde V^\dag ] - \Tr_R[\tilde V | \psi \rangle \langle \psi | \tilde V^\dag ] ||_2^2 } .
\eea
And similarly, we can apply the Haar random average to get 
\bea
     \int dU || \Tr_R[\tilde V W_b | \psi \rangle \langle \psi | W_b^\dag \tilde V^\dag ] - \Tr_R[\tilde V | \psi \rangle \langle \psi |  \tilde V^\dag ] ||_2^2 
    &\le & 4  e^{-S_2(b)_{|\psi\rangle}}.
\eea
Then the operator $W_b$ can be constructed up to a small error in the Hilbert space $\mathcal H_R$ (i.e. $ ||\tilde V W_b | \psi \rangle - W_R \tilde V |\psi \rangle || \le \epsilon'$) if
\bea
    \int dU || \Tr_R[\tilde V W_b | \psi \rangle \langle \psi | W_b^\dag\tilde V^\dag ] - \Tr_R[\tilde V | \psi \rangle \langle \psi | \tilde V^\dag ] ||_1 \le \sqrt{\frac{4|B|}{e^{S_2(b)_{|\psi \rangle}}}} \le \epsilon.
\eea
When this condition is satisfied, the operator can be reconstructed in $\mathcal H_R$ and by the no-cloning theorem it cannot be reconstructed in $\mathcal H_B$. 
Therefore, this simple model provides an example of how bulk information can be lost (assuming we do not have access to the reference $R$, which is introduced for convenience) when the number of measured boundary legs is too large.

\subsubsection{Reconstruction in non-isometric random tensor network}

In this subsection, we consider the post-measurement reconstruction problem in a random tensor network. 
In particular, we are interested in understanding the case in which the measurement can destroy the bulk information.
To this end and in analogy with the previous subsection, we derive a sufficient condition under which the operator reconstruction in $\mathcal H_R$ is guaranteed to be possible. 
Let us consider a state $|\psi\rangle \in \mathcal H_b \otimes \mathcal H_R$ and an operator $W_b$ acting on $\mathcal H_b$.
We can derive a bound similar to the one found in equation (\ref{eq:R-reconstruction}):
\bea \label{eq:inequality}
    && \overline{ || \Tr_R[ V_B W_b | \psi \rangle \langle \psi | W_b^\dag V_B^\dag ] - \Tr_R[ V_B | \psi \rangle \langle \psi |  V_B^\dag ]  ||_1} \nn \\
    &\le& \left( D^{|B|} \overline{ || \Tr_R[ V_B W_b | \psi \rangle \langle \psi | W_b^\dag V_B^\dag ] - \Tr_R[ V_B | \psi \rangle \langle \psi |  V_B^\dag ||_2^2 } \right)^{1/2}, 
\eea
where instead of Haar random average over a single unitary, the overline now indicates the average of all random states $|V_x \rangle$. 
This bound can again be mapped to some spin configuration.
For later convenience, we define the following free energy of the Ising model
\bea \label{eq:free_energy}
    F(\sigma_x, h_x) =    \sum_{\langle xy\rangle } \log D \left( \frac{\sigma_x \sigma_y -1 }2\right) + \sum_{x \in B} \log D \left( \frac{h_x \sigma_x -1}2 \right) ,
\eea
where $h_x$ denotes the magnetic field in $B$. 
The information of the input state $|\psi \rangle$ will give a contribution
\bea \label{eq:free_energy1}
    e^{-F'(b_+)} = \Tr_{b_-} \left[(\Tr_{b_+}[W_b \Psi_{b} W_b^\dag])^2 \right] + \Tr_{b_-}\left[(\Tr_{b_+}[ \Psi_b])^2 \right] - 2 ||\Tr_{b_-}[W_b |\psi \rangle \langle \psi |]||_2^2
\eea
where $b_\pm = \{ x: \sigma_x = \pm 1, x \in b\}$, $b = b_+ \cup b_-$, and $\Psi_b = \Tr_R [|\psi \rangle \langle \psi |] $.

Let us assume that in region $\Omega$ the spin is up, and in the complement the spin is down, i.e. $\Omega = \{x: \sigma_x = 1\}$.
This is enough to specify a configuration, namely there is a one-to-one map between $\Omega$ and a spin configuration $\{ \sigma_x \}$. 
The bound (\ref{eq:inequality}) then can be expressed as
\bea \label{eq:bound-R-spin}
    && \overline{ || \Tr_R[ V_B W_b | \psi \rangle \langle \psi | W_b^\dag V_B^\dag ] - \Tr_R[ V_B | \psi \rangle \langle \psi |  V_B^\dag ] ||_2^2} = \sum_{\Omega} e^{-F(\sigma_{x\in \Omega}=1, h_{x \in B} = -1) - F'(\Omega \cap b)} . \nn \\
\eea
For the argument of the free energy~(\ref{eq:free_energy}), we implicitly have $\sigma_{x \in \Omega} = 1$ and $\sigma_{x \notin \Omega} = -1$. 

Inspired by equation (\ref{eq:config}), we consider two configurations: one without a domain wall and one with a domain wall given by $\gamma_B\notin b$. 
In the second configuration $b_- = \emptyset$ and $b_+ = b $. 
It is not hard to show that $e^{-F'(b)} = 0$ using equation (\ref{eq:free_energy1}). 
Therefore, this configuration does not give any contribution.

For the first configuration, we have $\Omega = \emptyset$. 
As there is no domain wall, the free energy~(\ref{eq:free_energy}) is zero. 
On the other hand, $b_+ = \Omega \cap b = \emptyset$ and we arrive at 
\bea
    e^{-F'(\emptyset)} = 2 \left( e^{-S_2(b)_{|\psi\rangle}} - ||\Tr_b[W_b|\psi \rangle \langle \psi |]||_2^2 \right)
    &\le & 2  e^{-S_2(b)_{|\psi\rangle}}.
\eea
Then according to~(\ref{eq:inequality})
\bea
    \overline{ || \Tr_R[ V W_b | \psi \rangle \langle \psi | W_b^\dag V^\dag ] - \Tr_R[ V | \psi \rangle \langle \psi |  V^\dag ]  ||_1} &\le& \sqrt{\frac{2 D^{|B|}}{ e^{S_2(b)_{|\psi\rangle }}}}.
\eea

This provides a sufficient condition to reconstruct the operator $W_b$ in $\mathcal H_R$ for the specific state $|\psi \rangle$: 
\bea \label{eq:reconstruct-bound-R}
    S_2(b)_{|\psi\rangle} \gg |B| \log D.
\eea
We want to emphasize that this in turn implies that the operator $W_b$ cannot be reconstructed in $\mathcal H_B$ as a consequence of the no-cloning theorem. 

In Section \ref{measbdy}, we showed that when the bond dimension of the bulk legs is $D_b \sim \mathcal O(1)$, the brane ``hugs'' region $A$. 
It implies that the bulk information is intact after the measurement, and is teleported to the unmeasured region $A^c = B$ for any size of region $A$. 
This is possible because $ |R| \log D_b \ll |B| \log D$, namely there is enough entanglement resource to teleport the bulk information.

In the present subsection, we considered the bulk bond dimension $D_b$ to be comparable with the boundary bond dimension $D$ and provided a sufficient condition~(\ref{eq:reconstruct-bound-R}) under which the reconstruction of operator $W_b$ from $B$---or equivalently the bulk teleportation---fails. 
Since $S_2(b)_{|\psi \rangle } \le |R| \log D_b$, this happens when $|R| \log D_b \gg |B| \log D$. 
In this case, the brane is not longer near-critical and the bulk information fails to be protected against the measurement (assuming we have the access to the unmeasured region $B$ but not to the reference $R$). 
To connect this result to Corollary \ref{corollary}, we notice that the capacity of the teleportation is bounded by $I(A^c: R) \le \min(2S(R), 2S(A^c)) = \min(2 |R| \log D_b, 2 |B| \log D) $. 
Thus, when $|R| \log D_b \gg |B| \log D$, it is not possible to teleport the information to $A^c = B$. An analogous result was also found when measuring too many boundary legs in the HaPPY code in Section \ref{happynumerics}.

\section{Discussion}

In this paper we investigated the effects of local projective measurements of boundary subregions on the bulk dual spacetime. In many cases we found that the result of the LPM is to teleport bulk information contained before the measurement in the entanglement wedge $W(A)$ of the measured region $A$ into the entanglement wedge $W(A^c)$ of the complementary region $A^c$. In the holographic $AdS_3/CFT_2$ setup of Section \ref{sec:holographiccalculation}, region $A$ is the union of two disconnected subregions and its pre-measurement entanglement wedge is connected. In this case, teleportation can be observed in the bulk and is manifested in the connectedness of the post-measurement $W(A^c)$. Using tensor network models of holography, we were able to study in more detail how the bulk teleportation happens in simpler, discrete models (see Section \ref{sec:boundmeasuremteleportation}). In particular, we found that in the HaPPY code (Section \ref{sec:happycode}) bulk teleportation occurs and the amount of bulk teleported and preserved after the measurement depends on the product basis chosen to perform the projective measurement. In random tensor networks (Section \ref{sec:rtn}) we found that whenever the dimension of the code subspace is much smaller than the dimension of the physical Hilbert space, the bulk information is completely teleported into $A^c$. On the other hand, when the bond dimension of bulk legs is comparable to the one of boundary legs, bulk information can be teleported only if $A$ is not too large. 

In general, the amount of bulk information that can be teleported in a specific setup is upper-bounded by the amount of entanglement resource between the measured region and its complement. For the pure states that we consider here, the amount of entanglement resource is upper-bounded by the area of the minimal surface anchored to the boundary of $A$. Although bulk teleportation can be achieved both in semiclassical holography and in tensor network models, in the former the bulk information encoded in a very large boundary region $A$ can be teleported into a very small complementary region $A^c$. In fact, the large-$N$ limit of the boundary theory guarantees that the amount of entanglement resource is parametrically larger than the amount of bulk information to teleport. On the other hand, in tensor network constructions the bulk teleportation capacity is limited by the finite amount of entanglement resource. Additionally, in random tensor networks we observed that, for a given value of the bond dimensions of bulk and boundary legs, even when teleportation occurs the bulk-to-boundary map in the post-measurement state is non-isometric if $A$ is too large. 

We would like to remark that the idea of moving entanglement by means of projective measurements has been explored before in small-$N$ setups \cite{altman}. We also emphasize that our results could be interpreted as an example of measurement-induced entanglement \cite{mie}, where the amount of induced entanglement is very large (of order $1/G$).

\subsubsection*{Open questions}

Several questions directly related to our analysis still need to be answered:
\begin{itemize}
\item It would be interesting to study a regularized version of our $AdS_3/CFT_2$ setup of Section \ref{sec:holographiccalculation} using one of the regularization schemes proposed in footnote \ref{regfootnote}, and verify whether any deviations from our results occur.
\item From our present work it is not yet clear, from an operative point of view, how to obtain specific post-measurement Cardy states with a given boundary entropy (and therefore a brane with a given tension) using specific measurement protocols. Our tensor network constructions can help us gain insight on this topic by explicitly showing how different measurement protocols lead to different amounts of bulk teleportation in a controlled, discretized setup.
\item The analysis of Section \ref{happynumerics} shows that for the $M>1$ HaPPY code near-critical to critical brane tensions can be achieved. Whether lower tensions (including vanishing and negative tensions) can be obtained is still an open question.
\item It is not yet clear what specific quantity in tensor network models can be related to the brane tension in holography, and therefore to the boundary entropy of the dual Cardy states.
\end{itemize}

\subsubsection*{Future directions}

Our work also opens the door to several new research directions to be explored on a longer term. First, generalizations of the analysis presented in this paper to thermofield double setups could give new insight into the emergence of connected spacetimes in wormhole configurations and have implications for holographic cosmologies \cite{Cooper:2018cmb,Antonini:2019qkt} and some aspects of the duality between the SYK model and JT gravity \cite{Kourkoulou:2017zaj,Antonini:2021xar}. A paper containing results in this direction is in preparation \cite{measurementtfd}. Second, it would be interesting to investigate the effects of measuring highly entangled nonlocal operators on the boundary subregion $A$. This kind of entangled measurement should be able to access the bulk information contained in the entanglement wedge $W(A)$, but it is not clear whether a bulk description of the complementary boundary region $A^c$ is preserved after the entangled measurement is performed. In tensor network models it should be possible to answer this question using techniques analogous to the ones introduced in this paper; on the contrary, since now the post-measurement state of region $A$ has spatial entanglement and is therefore not a Cardy state \cite{miyaji2014boundary}, the AdS/BCFT prescription cannot be applied and it is not clear how entangled measurements can be described in a bulk holographic calculation. 

From a more general point of view, our bulk calculation predicts a specific entanglement structure for the post-measurement state of holographic theories. For example, in the case of the tri-partite system studied in Section \ref{sec:holographiccalculation} it predicts an entangled/disentangled phase transition with a non-trivial phase structure depending on the post-measurement state and the size of the measured region. It would be interesting to understand whether a similar phase structure is present in more general quantum many-body systems or if it is a specific feature of holographic theories. Understanding the impacts of measurements is also paramount in experimental settings, especially those that aim to simulate holographic theories.  The development of quantum computers in the next few decades will hopefully provide a way to test our results by preparing a state for the boundary theory, performing a projective measurement, and studying the entanglement of the post-measurement system. Such insights can also be valuable for understanding aspects of bulk physics that are difficult to explore by theory alone. 

Finally, intriguing is the possibility to extend our construction to holographically describe measurement-induced phase transitions (MIPT). In this paper, we analyzed the properties of a single time slice right after the measurement is performed and found that, for specific setups such as the one studied in Section \ref{sec:holographiccalculation}, a non-dynamical MIPT arises. A first step towards a holographic realization of dynamical MIPT such as the ones observed in quantum many-body systems \cite{aharonov,Skinner:2018tjl,Li:2018mcv,Chan:2018upn,Choi:2019nhg,Bentsen:2021ukm,Li2, jian2021measurement} is to study real time evolution of the bulk spacetime (including the ETW branes) after the measurement is performed. It will then be possible to study the effect of multiple measurements performed with a given temporal rate, which is the key ingredient of MIPT.

\acknowledgments
We would like to thank Brianna Grado-White, Daniel Harlow, Matthew Headrick, Alexey Milekhin, and Wayne Weng for helpful discussions and comments. This work is partially supported by the U.S. Department of Energy, Office of Science, Office of Advanced Scientific Computing Research, Accelerated Research for Quantum Computing program ``FAR-QC'' (S.A.) and by the AFOSR under grant number FA9550-19-1-0360 (B.S.). G.B. is supported by the GeoFlow consortium under U.S. Department of Energy 
Grant DE-SC0019380. C.C. acknowledges the support by the U.S. Department of Defense and NIST through the Hartree Postdoctoral Fellowship at QuICS. The work of J.H. is supported in part by the Simons Foundation through \emph{It from Qubit: Simons Collaboration on Quantum Fields, Gravity, and Information} and in part by MEXT-JSPS Grant-in-Aid for Transformative Research Areas (A) ``Extreme Universe” No. 21H05187. 
S.-K.J. is supported by the Simons Foundation via the It From Qubit Collaboration.

\appendix

\section{Building the bulk dual spacetime: technical details}

\subsection{Conformal mapping: cylinder with slits to finite cylinder}

\label{app:maps}

\begin{figure}
    \centering
    \includegraphics[width=\textwidth]{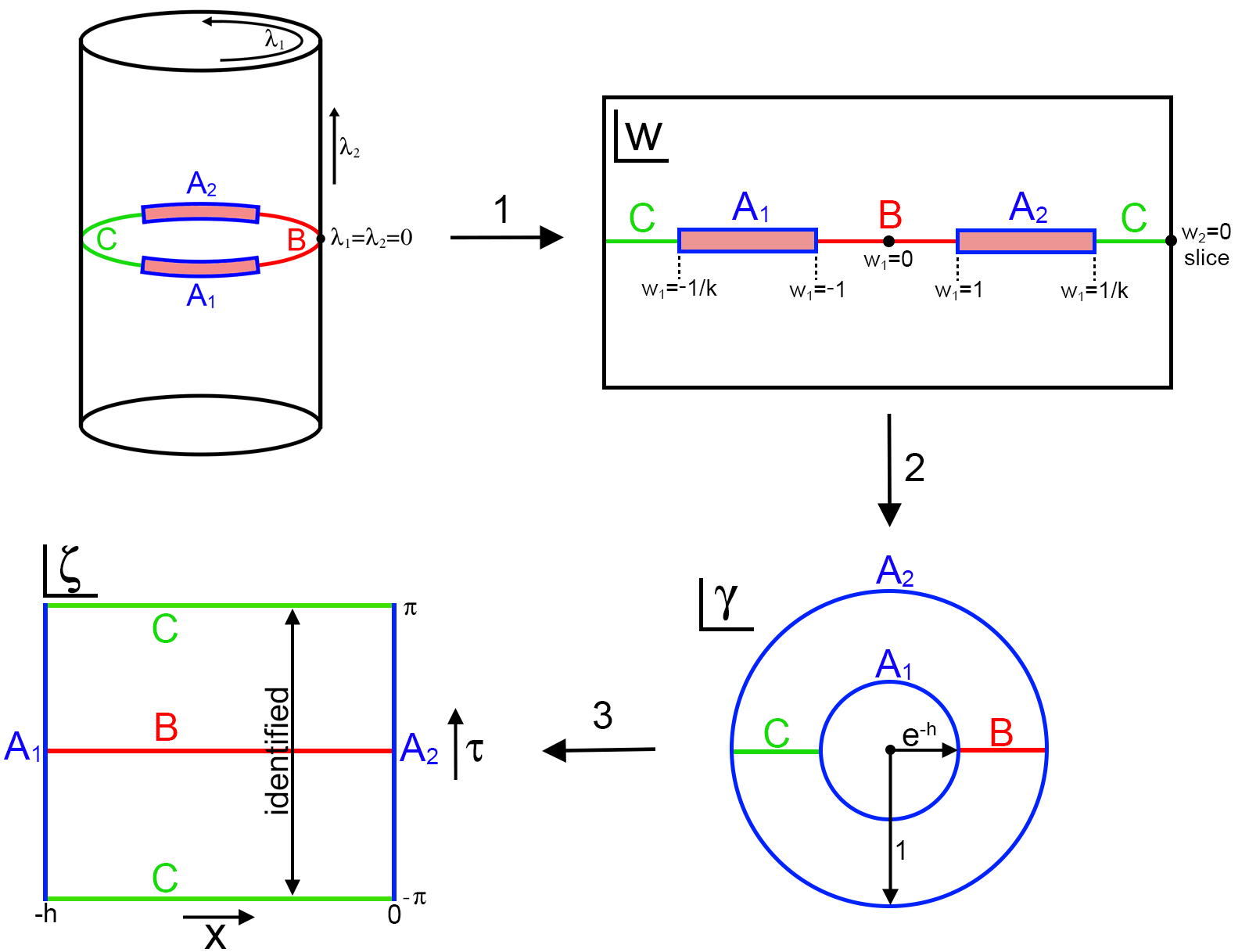}
    \caption{Conformal maps from the infinite cylinder in $\lambda$, $\bar{\lambda}$ coordinates with slits at $\{-\ell/2-s<\Re(\lambda)<-\ell/2,\Im(\lambda)=0\}$ and $\{\ell/2<\Re(\lambda)<\ell/2+s,\Im(\lambda)=0\}$ to the finite cylinder in $\zeta$, $\bar{\zeta}$ coordinates.}
    \label{maps}
\end{figure}

In this appendix, we explain how to map the infinite cylinder in $\lambda$, $\bar{\lambda}$ coordinates with slits at $\{-\ell/2-s<\Re(\lambda)<-\ell/2,\Im(\lambda)=0\}$ and $\{\ell/2<\Re(\lambda)<\ell/2+s,\Im(\lambda)=0\}$  (see Figure \ref{splittingPI}) to a finite cylinder, i.e. a manifold $S^1\times I$, where $I$ is a finite interval. In order to do that, we compose three conformal maps, summarized in Figure \ref{maps}. 

\begin{enumerate}
    \item \textit{Cylinder with slits to plane with slits}

The first conformal map is given by
\begin{equation}
    w(\lambda)=\frac{\tan\left(\frac{\pi}{L}\lambda\right)}{\tan\left(\frac{\pi\ell}{2L}\right)}.
    \label{cyltoplane}
\end{equation}
The transformation (\ref{cyltoplane}) maps our infinite cylinder with slits in $\lambda$, $\bar{\lambda}$ coordinates to the infinite complex plane in $w$, $\bar{w}$ coordinates with slits at $\{-1/k<\Re(w)<-1,\Im(w)=0\}$ (region $A_1$) and $\{1<\Re(w)<1/k,\Im(w)=0\}$ (region $A_2$) (see Figure \ref{maps}), where $k$ is defined in equation (\ref{kdefinition}).
\item The second conformal map is given by \cite{rajabpour2015entanglement,confdictionary}
\begin{equation}
    \gamma(w)=\exp\left[\frac{h}{2}\left(\frac{\textrm{sn}^{-1}(w,k^2)}{\mathcal{K}(k^2)}-1\right)\right]
    \label{planetoannulus}
\end{equation}
where $\textrm{sn}^{-1}(x,m)$ is the inverse Jacobi SN elliptic function, $\mathcal{K}(m)$ is the complete elliptic integral of the first kind, and $h$ is defined in equation (\ref{hdefinition}).

The transformation (\ref{planetoannulus}) maps the infinite complex plane with slits in $w$, $\bar{w}$ coordinates to the annulus in $\gamma$, $\bar{\gamma}$ coordinates centered at the origin, with inner radius $r_{in}=\exp(-h)$ and outer radius $r_{out}=1$. The slit at $\{-1/k<\Re(w)<-1,\Im(w)=0\}$ (region $A_1$) is mapped to the inner boundary circumference, while the slit at $\{1<\Re(w)<1/k,\Im(w)=0\}$ (region $A_2$) is mapped to the outer boundary circumference (see Figure \ref{maps}). The upper-half plane (UHP) is mapped to the upper half of the annulus, while the lower-half plane (LHP) is mapped to the lower half of the annulus; accordingly, the upper edges of the slits are mapped to the upper halves of the boundary circumferences, while the lower edges of the slits are mapped to the lower halves of the boundary circumferences. 
\item The third conformal map is given by
\begin{equation}
 \zeta(\gamma)=\log(\gamma)   
 \label{annulustocylinder}
\end{equation}
and it maps the annulus in $\gamma$ coordinates to the finite cylinder in $\zeta=x+i \tau$ coordinates. In particular the $\tau$ coordinate is compactified with periodicity $2\pi$: $\tau\sim \tau+2\pi$\footnote{$\log(\gamma)$ has a branch cut on the real axis running from $-\infty$ to $0$. We choose the branch $\arg(\gamma)\in [-\pi,\pi]$, corresponding to $\Im(\zeta)\in [-\pi,\pi]$.}, and the range of the $x$ coordinate is $-h\leq x\leq 0$; the topology of the manifold is $S^1\times I$. The inner boundary circumference of the annulus (region $A_1$) is mapped to the $x=-h$ circle, while the outer boundary circumference (region $A_2$) is mapped to $x=0$ circle. The upper half of the annulus is mapped to the $0\leq \tau \leq \pi$ region, while the lower half of the annulus is mapped to the $-\pi\leq \tau \leq 0$ region. Finally, region $B$ (of which we will be interested in computing the entanglement entropy) is mapped to the $\tau=0$ line, while region $C$ is mapped to the $\tau=\pm\pi$ line.
\end{enumerate}

\subsection{Brane trajectories}
\label{app:branetrajectories}

In this appendix we report the main steps needed to derive the brane trajectories (\ref{btztrajectory}) and (\ref{thermaltrajectory}). The Neumann boundary conditions at the brane read
\begin{equation}
    K_{ab}-Kh_{ab}=-Th_{ab}.
    \label{neumannappendix}
\end{equation}
The extrinsic curvature is defined as
\begin{equation}
    K_{ab}=\nabla_\mu n_\nu e^\mu_ae^\nu_b
\end{equation}
where $\nabla_\mu$ is the covariant derivative with respect to the 3-dimensional spacetime coordinates $x^\mu$, $n^\mu$ is the unit normal vector to the brane, and $e^\mu_a=dx^\mu/dy^a$, with $y^a$ coordinates on the brane. The induced metric on the brane is given by 
\begin{equation}
    h_{ab}=g_{\mu\nu}e^\mu_ae^\nu_b
\end{equation}
and the trace of the extrinsic curvature is $K=h^{ab}K_{ab}$.

Parametrizing the brane trajectory by the bulk coordinate $z$, the coordinates on the brane are $y^a=\tau,z$, and we get the unit normal vector
\begin{equation}
    (n_\tau,n_x,n_z)=\pm N (0,1,-x'(z))
    \label{normalvector}
\end{equation}
where the upper sign refers to the right brane in the BTZ phase and to the part of trajectory from $\{z=0,x=0\}$ to the turning point in the thermal $AdS_3$ phase, while the lower sign refers to the left brane in the BTZ phase and to the part of trajectory from the turning point to $\{z=0,x=\pm h\}$ in the thermal $AdS_3$ phase. The normalization factor $N$ reads
\begin{equation}
\begin{cases}
 N_{BTZ}=\frac{\sqrt{2}R}{z}\frac{1+\frac{z^2}{8}}{\sqrt{1+2\left(1+\frac{z^2}{8}\right)^2(x'(z))^2}}\\[20pt]
    N_{th}=\frac{\sqrt{2}R}{z}\frac{1-\frac{\pi^2 z^2}{8h^2}}{\sqrt{1+2\left(1-\frac{\pi^2 z^2}{8h^2}\right)^2[x'(z)]^2}}
    \end{cases}
\end{equation}
in the BTZ and thermal $AdS_3$ phases respectively, where we used the values $b_{BTZ}=1/2$ and $b_{th}=\pi/(2h)$ for the $b$ parameter appearing in the metrics (\ref{btzthermalmetrics}). We also have $e^x_\tau=e^z_\tau=e^\tau_z=0$, $e^\tau_\tau=e^z_z=1$, $e^x_z=x'(z)$.

The extrinsic curvature can now be evaluated\footnote{It is convenient to use $e^\mu_an_\mu=0$, which implies $K_{ab}=e^\mu_ae^\nu_b\nabla_\mu n_\nu=-e^\mu_an_\nu\nabla_\mu e^\nu_b$.}, yielding $K_{\tau z}=K_{z\tau}=0$,
\begin{equation}
\begin{cases}
    K^{BTZ}_{\tau\tau}=\pm \frac{2Nx'(z)}{z}\left(1-\frac{z^4}{64}\right)\\[10pt]
    K^{th}_{\tau\tau}=\pm \frac{2Nx'(z)}{z}\left(1-\frac{\pi^4z^4}{64h^4}\right)
    \end{cases}
\end{equation}
and
\begin{equation}
\begin{cases}
    K^{BTZ}_{zz}=\pm N\left[-x''(z)+\frac{2[x'(z)]^3}{z}\left(1-\frac{z^4}{64}\right)-\frac{x'(z)}{z}\left(1+\frac{2(z^2-8)}{z^2+8}\right)\right]\\[10pt]
    K^{th}_{zz}=\pm N\left[-x''(z)+\frac{2[x'(z)]^3}{z}\left(1-\frac{\pi^4z^4}{64h^4}\right)-\frac{x'(z)}{z}\left(1+\frac{2(\pi^2z^2+8h^2)}{\pi^2z^2-8h^2}\right)\right]
    \end{cases}
\end{equation}
where we used the same sign convention described above for the normal vector (\ref{normalvector}).

The off-diagonal components of the Neumann boundary conditions (\ref{neumannappendix}) are identically satisfied. Evaluating the $zz$ component, we get
\begin{equation}
    \begin{cases}
        \frac{dx^{\pm}_{BTZ}}{dz}=\pm \frac{RT}{\sqrt{2}\left(1+\frac{z^2}{8}\right)\sqrt{\left(1+\frac{z^2}{8}\right)^2\big{/}\left(1-\frac{z^2}{8}\right)^2-R^2T^2}}\\[20pt]
         \frac{dx^{\pm}_{th}}{dz}=\pm \frac{RT}{\sqrt{2}\left(1-\frac{\pi^2z^2}{8h^2}\right)\sqrt{\left(1-\frac{\pi^2z^2}{8h^2}\right)^2\big{/}\left(1+\frac{\pi^2z^2}{8h^2}\right)^2-R^2T^2}}.
    \end{cases}
    \label{dertrajectories}
\end{equation}
In order to solve the first order ODEs above, it is useful to consider the change of coordinates\footnote{Note that these coordinate transformations lead from the form (\ref{btzthermalmetrics}) of the BTZ black hole and thermal $AdS_3$ metrics to the ones used in \cite{fujita2011aspects}, up to a constant rescaling of the $x$ and $\tau$ coordinates.}
\begin{equation}
    z=\frac{4}{\bar{z}}\left(1-\sqrt{1-\frac{\bar{z}^2}{2}}\right)
    \label{btzchangecoord}
\end{equation}
for the BTZ phase and
\begin{equation}
    z=\frac{4h^2}{\pi^2\tilde{z}}\left(1-\sqrt{1-\frac{\pi^2\tilde{z}^2}{2h^2}}\right)
    \label{thermalchangecoord}
\end{equation}
for the thermal $AdS_3$ phase. In the new coordinates $\bar{z}$ and $\tilde{z}$ the equations (\ref{dertrajectories}) can be easily integrated. Using the inverse transformations 
\begin{equation}
    \bar{z}=\frac{z}{1+\frac{z^2}{8}}
\end{equation}
and 
\begin{equation}
    \tilde{z}=\frac{z}{1+\frac{\pi^2z^2}{8h^2}}
\end{equation}
we obtain the results (\ref{btztrajectory}) and (\ref{thermaltrajectory}). Imposing the boundary conditions $x_{BTZ}^+(0)=0$, $x_{BTZ}^-(0)=-h$, $x_{th}^+(0)=0$, $x_{th}^-(0)=h$, we get the values of the integration constants: $c_+=0$, $c_-=-h$, $d_+=0$, $d_-=h$\footnote{Note that, in the thermal $AdS_3$ phase, the $x$ coordinate is periodic with period $P_x=2h$. Therefore, the constants $d_\pm$ are defined only up to a $2nh$ additive constant, which must be the same for $d_+$ and $d_-$. Here we chose to stick to our convention where the brane anchors at $x=0,\pm h$.}. It is easy to show that the brane trajectories (\ref{btztrajectory}) and (\ref{thermaltrajectory}) (in the BTZ and thermal $AdS_3$ phases respectively) automatically satisfy the $\tau\tau$ component of the Neumann boundary conditions (\ref{neumannappendix}).

\subsection{Action evaluation}
\label{app:actions}

In this appendix we derive the expression of the on-shell actions (\ref{btzaction}) and (\ref{thermalaction}) for our solutions including ETW branes. In order to do so, we introduce a UV cutoff at $z=\varepsilon$, evaluate the gravitational action on shell, find the appropriate counterterms, and then take the $\varepsilon\to 0$ limit. We will neglect the presence of the Gibbons-Hawking-York term $I_{GHY}$ for the asymptotic boundary in the action because it gives a vanishing contribution to the on-shell action in the $\varepsilon\to 0$ limit.

\subsubsection*{On-shell BTZ action}

Let us start by evaluating the bulk action
\begin{equation}
    I_{bulk}^{BTZ}=-\frac{1}{16\pi G}\int_{\mathcal{N}}d^3x\sqrt{g}(\mathcal{R}-2\Lambda)
    \label{btzbulkaction}
\end{equation}
where $\mathcal{R}=-6/R^2$ and $\Lambda=-1/R^2$. The range of our bulk coordinates is $\tau\in [-\pi,\pi]$, $z\in [0,z_0]$ (with $z_0=2\sqrt{2}$) and $x\in [x_{BTZ}^-(z),x_{BTZ}^+(z)]$, where $x_{BTZ}^-(z)$ and $x_{BTZ}^+(z)$ are the brane trajectories (\ref{btztrajectory}) for the left and right branes. In order to evaluate the integral (\ref{btzbulkaction}) it is useful to split the $x$ domain into three intervals: $x\in [x_{BTZ}^-(z),-h]\cup [-h,0] \cup [0,x_{BTZ}^+(z)]$. We then obtain
\begin{equation}
    I_{bulk}^{BTZ}=\frac{Rh}{2G}\frac{1}{\varepsilon^2}+\frac{\sqrt{2}R^2T}{G\sqrt{1-R^2T^2}}\frac{1}{\varepsilon}-\frac{Rh}{8G}-\frac{R}{2G}\arcsinh\left(\frac{RT}{\sqrt{1-R^2T^2}}\right)-\frac{R^2T}{2G(1-R^2T^2)}+\mathcal{O}(\varepsilon)
\end{equation}
where we used $x_{BTZ}^+(z)=-x_{BTZ}^-(z)-h$ (by symmetry and the fact that the two branes have the same tension) and the change of coordinates (\ref{btzchangecoord}).

Evaluating the ETW brane action 
\begin{equation}
I_{ETW}^{BTZ}=-\frac{1}{8\pi G}\int_{ETW}d^2x\sqrt{h}\left(K-T\right)
\end{equation}
for the left and right branes (which give identical contributions) with $K=2T$ on-shell and remembering that $h_{ab}=g_{\mu\nu}e^\mu_ae^\nu_b$, we obtain
\begin{equation}
    I_{ETW}^{BTZ}=-\frac{R^2T}{\sqrt{2}G\sqrt{1-R^2T^2}}\frac{1}{\varepsilon}+\frac{R^2T}{2G(1-R^2T^2)}.
\end{equation}
The total action is then 
\begin{equation}
      I^{BTZ}_{bulk}+I^{BTZ}_{ETW}=\frac{Rh}{2G}\frac{1}{\varepsilon^2}+\frac{R^2T}{\sqrt{2}G\sqrt{1-R^2T^2}}\frac{1}{\varepsilon}-\frac{Rh}{8G}-\frac{R}{2G}\arcsinh\left(\frac{RT}{\sqrt{1-R^2T^2}}\right)+\mathcal{O}(\varepsilon).
     \label{appBTZactiontot}
\end{equation}
We must now introduce counterterms to eliminate the divergent terms before taking the $\varepsilon\to 0$ limit. The quadratically divergent term can be taken care of using a counterterm of the form
\begin{equation}
    I_{CT1}^{BTZ}=C_1^{BTZ}\int_{\mathcal{M}}d^2x\sqrt{\tilde{h}}
    \label{btzct1}
\end{equation}
where $\mathcal{M}$ is the asymptotic boundary (placed at $z=\varepsilon$) and $\tilde{h}$ the determinant of the metric induced on the asymptotic boundary (which can be obtained by fixing $z=\varepsilon$ in the BTZ metric (\ref{btzthermalmetrics})). Evaluating (\ref{btzct1}) on shell, we find that we must impose $C_1^{BTZ}=-h/(4GR)$ in order to cancel the divergent term, and no additional finite contributions arise. The linearly divergent term can be eliminated using a corner term at the intersection between the ETW brane and the asymptotic boundary:
\begin{equation}
    I_{CT2}^{BTZ}=C_2^{BTZ}\int_{\partial \mathcal{M}}d\tau\sqrt{\gamma}.
    \label{btzct2}
\end{equation}
Note that there are two such corners (one at $x=-h$ for the left brane and one at $x=0$ for the right brane), which will give two identical contributions. The metric induced on the corners can be obtained by fixing $z=\varepsilon$, $x=-h,0$ in the BTZ metric (\ref{btzthermalmetrics}) for the left and right corner respectively. The linearly divergent term is eliminated by setting $C_2^{BTZ}=-RT/(8\pi G\sqrt{1-R^2T^2})$\footnote{Note that with this choice of constant we must include the contribution from both corners. Alternatively, since the two corners give identical contribution, we could double the constant and compute the on-shell value of the counterterm for only one corner.}, and again no additional finite contributions arise. Adding the counterterms (\ref{btzct1}) and (\ref{btzct2}) to equation (\ref{appBTZactiontot}) and taking the $\varepsilon\to 0$ limit, the BTZ black hole on-shell action finally reads
\begin{equation}
    I_{BTZ}=-\frac{Rh}{8G}-\frac{R}{2G}\arcsinh\left(\frac{RT}{\sqrt{1-R^2T^2}}\right)
\end{equation}
which is the result (\ref{btzaction}) and agrees with the results of \cite{fujita2011aspects}.

\subsubsection*{On-shell thermal $AdS_3$ action}

In the thermal $AdS_3$ phase the domain is given by $\{\tau\in [-\pi,\pi]; x\in [-h,h]; z\in [0,2\sqrt{2}h/\pi]\}\setminus \{\tau\in [-\pi,\pi]; x\in [x_{th}^+ (z),x_{th}^- (z)]; z\in [0,z_{max}]\}$ where $x_{th}^- (z)$ and $x_{th}^+ (z)$ are the brane trajectories (\ref{thermaltrajectory}) and $z_{max}=2\sqrt{2}h\sqrt{1-R|T|}/(\pi\sqrt{1+R|T|})$. Here we are subtracting from the full Euclidean thermal $AdS_3$ spacetime (the first curly bracket) the part of spacetime excised by the connected ETW brane (the second curly bracket).

Using the change of coordinates (\ref{thermalchangecoord}) we can evaluate the bulk action, obtaining
\begin{equation}
    I_{bulk}^{th}=\frac{Rh}{2G}\frac{1}{\varepsilon^2}+\frac{\sqrt{2}R^2T}{G\sqrt{1-R^2T^2}}\frac{1}{\varepsilon}-\frac{\pi^2R}{4Gh}+\mathcal{O}(\varepsilon).
\end{equation}
The brane action now has only one contribution from the connected brane, which gives
\begin{equation}
    I_{ETW}^{th}=-\frac{R^2T}{\sqrt{2}G\sqrt{1-R^2T^2}}\frac{1}{\varepsilon}
\end{equation}
and therefore
\begin{equation}
     I_{bulk}^{th}+I_{ETW}^{th}=\frac{Rh}{2G}\frac{1}{\varepsilon^2}+\frac{R^2T}{\sqrt{2}G\sqrt{1-R^2T^2}}\frac{1}{\varepsilon}-\frac{\pi^2R}{4Gh}+\mathcal{O}(\varepsilon).
\end{equation}
The two needed counterterms are again a boundary term and a corner term (again, there are two corners at $x=-h$ and $x=0$):
\begin{equation}
    I_{CT1}^{th}=C_1^{th}\int_{\mathcal{M}}d^2x\sqrt{\tilde{h}}; \hspace{2cm} I_{CT2}^{th}=C_2^{th}\int_{\partial \mathcal{M}}d\tau\sqrt{\gamma}.
\end{equation}
Evaluating them on-shell, we find that, in order to eliminate the divergent terms, the two constants take the values
\begin{equation}
C_1^{th}=-\frac{1}{8\pi G R}; \hspace{2cm} C_2^{th}= -\frac{RT}{4\pi G\sqrt{1-R^2T^2}}.
\end{equation}
Unlike the BTZ case, the first counterterm $I_{CT1}^{th}$ also gives a finite contribution in the $\varepsilon\to 0$ limit:
\begin{equation}
    I_{CT1}^{th}=-\frac{Rh}{2G}\frac{1}{\varepsilon^2}+\frac{\pi^2R}{8Gh}.
\end{equation}
The second counterterm $I_{CT2}^{th}$ gives no finite contribution. Finally, adding the counterterms and taking the $\varepsilon\to 0$ limit, the thermal $AdS_3$ on-shell action reads
\begin{equation}
    I_{bulk}^{th}+I_{ETW}^{th}=-\frac{\pi^2R}{8Gh}
\end{equation}
which is the result (\ref{thermalaction}) and agrees with the results of \cite{fujita2011aspects}.

\section{Mapping back to the original cylinder with slits}
\label{mappingtopoincare}

In this appendix we will find a change of coordinates to map our BTZ and thermal $AdS_3$ spacetimes with ETW branes to portions of Poincar\'e-AdS bounded by ETW branes. Once we have such domains, it is in principle possible to map them back and find the holographic dual of our cylinder with slits in $\lambda$, $\bar{\lambda}$ coordinates in the two phases. This can be done by using of the coordinate transformation (\ref{bulkconftransf}) where $f$ is given by the composition of the conformal maps leading us from the original cylinder with slits to the final Poincar\'e coordinates. The fact that our theory on the cylinder with slits is singular prevents us from carrying out this last step consistently, but we will explain how it can be implemented in general for more regular setups.

\subsection{Mapping the domains to Poincar\'e $AdS_3$}

\subsubsection*{BTZ phase}

\begin{figure}
\centering
\includegraphics[width=.75\textwidth]{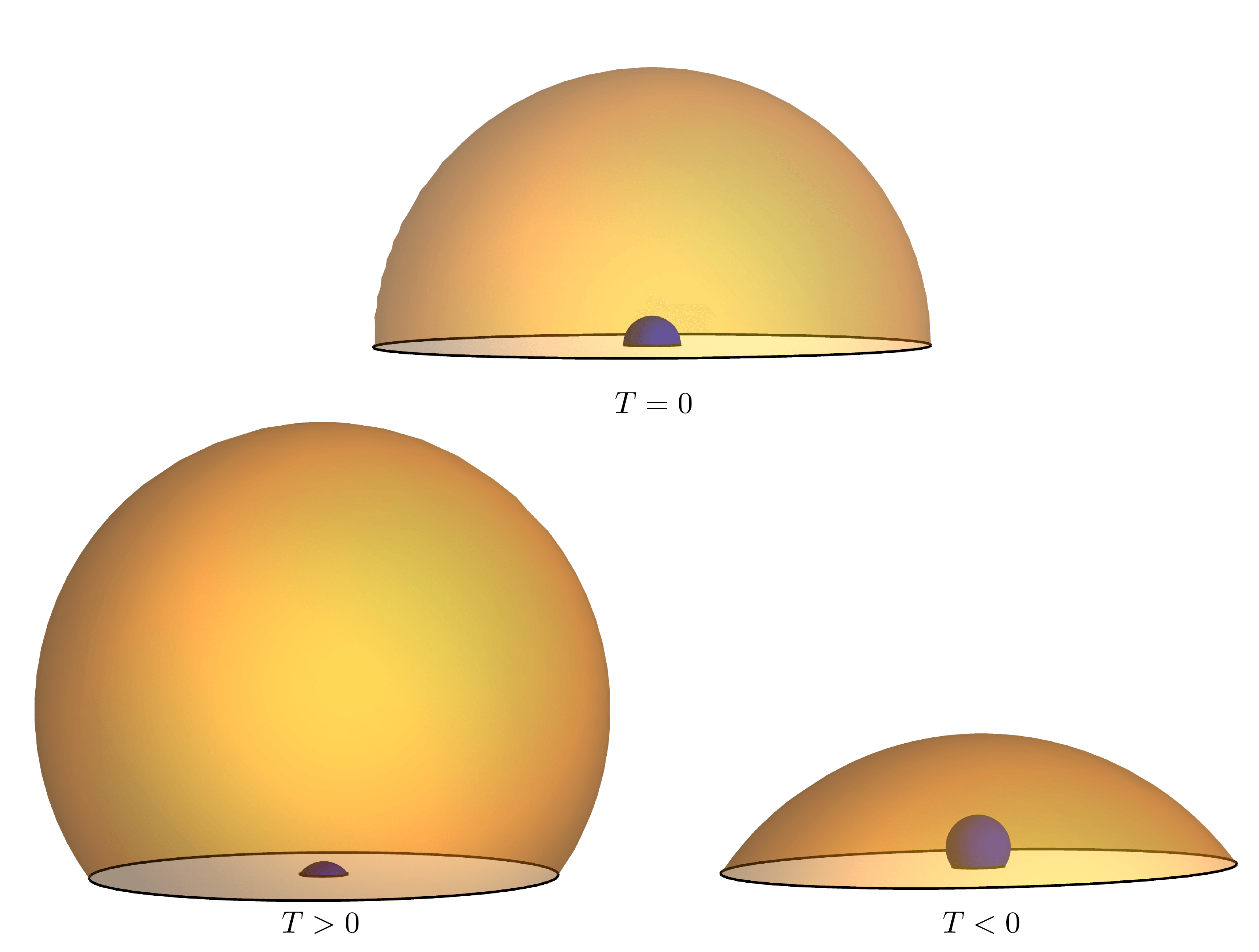}  
\caption{\label{fig:BTZ_annulusc} Bulk domain for the BTZ phase in Poincar\'e coordinates. For zero tension the domain is given by the region in between two concentric hemispheres of radii $r_{in}=\textrm{e}^{-h}$ (left BTZ brane) and $r_{out}=1$ (right BTZ brane). The addition of positive (negative) tension causes the bulk domain to enlarge (shrink): both radii increase, but the center of the inner sphere shifts to negative (positive) values of $\eta$, while the center of the outer sphere shifts to positive (negative) values of $\eta$.}
\end{figure}

In the BTZ black hole phase, the coordinate transformation we are looking for takes the form (\ref{bulkconftransf}), with the boundary conformal map given by
\begin{equation}
    \xi_{CFT}=f_{BTZ}(\zeta)=\textrm{e}^{2b_{BTZ}\zeta}
    \label{btztopoincare}
\end{equation}
where we remind that $\zeta=x+i\tau$ and $b_{BTZ}=1/2$. Note that the map (\ref{btztopoincare}) is just the inverse of the map (\ref{annulustocylinder}): the boundary manifold is mapped back from the finite cylinder in $\zeta$ coordinates to the annulus in $\gamma(=\xi_{CFT})$ coordinates (see Figure \ref{maps}). Using the map (\ref{btztopoincare}), the coordinate transformation (\ref{bulkconftransf}) reads
\begin{equation}
\begin{cases}
\xi=\frac{8-z^2}{8+z^2}\textrm{e}^\zeta\\
\bar{\xi}=\frac{8-z^2}{8+z^2}\textrm{e}^{\bar{\zeta}}\\
\eta=\frac{4\sqrt{2} z}{8+z^2}\textrm{e}^{\frac{\zeta+\bar{\zeta}}{2}}
\end{cases}
\label{BTZcoord}
\end{equation}
It is easy to check that starting from the Poincar\'e-AdS metric (\ref{poincareads}) and using the transformation (\ref{BTZcoord}), we obtain the metric (\ref{mappedmetric}) where the functions $L$, $\bar{L}$ defined in equation (\ref{lfunctions}) are given by $L(\zeta)=\bar{L}(\bar{\zeta})=b_{BTZ}^2$. Setting $\zeta=x+i\tau$, we finally recover the BTZ metric (\ref{btzthermalmetrics}). Therefore, our BTZ black hole cut off by ETW branes is mapped to a subregion of Poincar\'e-AdS whose asymptotic boundary is given by an annulus, with ETW branes anchored at the inner and outer circles. 
The bulk domain (see Figure \ref{BTZtorus}) in Poincar\'e coordinates is given by
\begin{equation}
    \xi\bar{\xi}+\left(\eta+\frac{RT\textrm{e}^{-h}}{\sqrt{1-R^2T^2}}\right)^2\geq \frac{\textrm{e}^{-2h}}{1-R^2T^2} \hspace{0.5cm} \land \hspace{0.5cm} \xi\bar{\xi}+\left(\eta-\frac{RT}{\sqrt{1-R^2T^2}}\right)^2\leq \frac{1}{1-R^2T^2}.
    \label{BTZpoincaredomain}
\end{equation}
For $T=0$, the bulk domain is given by the region between two concentric hemispheres centered at the origin and with radii $r_{in}=\textrm{e}^{-h}$ (left brane in BTZ coordinates), $r_{out}=1$ (right brane in BTZ coordinates). Increasing the tension to positive values, the radii of both spheres increase, and the center of the inner sphere moves to negative values of $\eta$, while the center of the outer sphere moves to positive values of $\eta$. Since $\eta\geq 0$, this implies that the bulk domain is enlarged, and covers all of Poincar\'e-AdS as we approach the critical value of the tension $T_c$. As we decrease the tension to negative values, the radii of both spheres again increase from their $T=0$ values, but the center of the inner sphere moves to positive $\eta$ and the center of the outer sphere to negative $\eta$: the bulk domain is shrunk (see Figure \ref{fig:BTZ_annulusc}).

\subsubsection*{Thermal $AdS_3$ phase}

\begin{figure}
    \centering
    \includegraphics[width=0.6\textwidth]{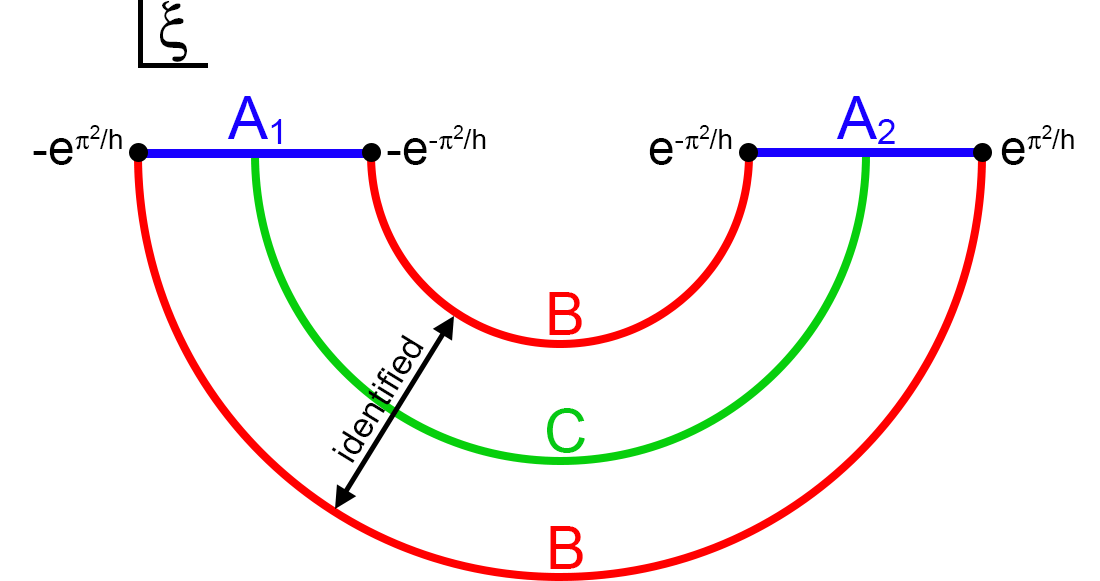}
    \caption{In the thermal $AdS_3$ phase, the boundary finite cylinder in $\zeta$ coordinates is mapped to a boundary half annulus in Poincar\'e coordinates. The inner and outer radii of the annulus are identified.}
    \label{halfannulus}
\end{figure}

\begin{figure}
\centering
\includegraphics[width=.75\textwidth]{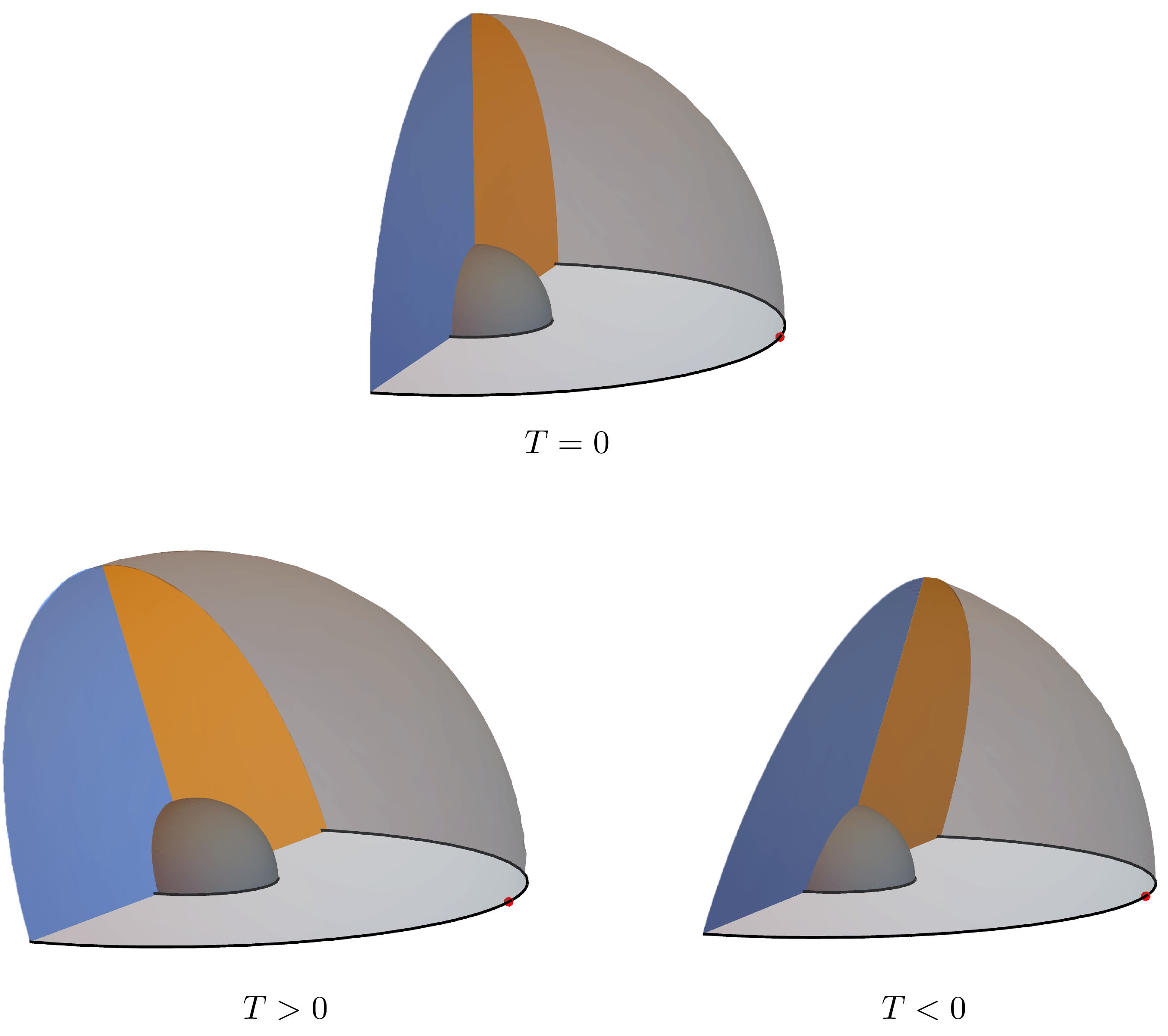}  
\caption{\label{fig:thermal_annulusc} Bulk domain for the thermal $AdS_3$ phase in Poincar\'e coordinates. For zero tension the domain is given by half of the region in between two concentric hemispheres centered in the origin and of radii $r_{out}=\exp(2\pi^2/h)$ and $r_{in}=\exp(-2\pi^2/h)$. The surfaces of the two hemispheres are identified. The brane consists of the half annulus in the $\Re(\xi)-\eta$ plane cutting in half the region in between the hemispheres. The portion of the brane shaded in orange corresponds to the portion of the brane extending from $\{z=0,x=0\}$ to the turning point in thermal $AdS_3$ coordinates (i.e. $x_{th}^+(z)$), while the portion of the brane shaded in blue represents the portion of the brane extending from the turning point to $\{z=0,x=h\}$ (i.e. $x_{th}^-(z)$). The addition of positive (negative) tension causes the bulk domain to increase (decrease) as the brane tilts towards positive (negative) values of $\Im(\xi)$. The brane sits in the plane identified by $\Im(\xi)=(T/\sqrt{1-R^2T^2})\eta$. For critical positive tension, the whole region in between the emispheres is recovered, and for negative critical tension the bulk domain approaches the empty set.}
\end{figure}

In the thermal $AdS_3$ phase, the coordinate transformation is again of the form (\ref{bulkconftransf}), where now the boundary conformal map is given by
\begin{equation}
    \xi_{CFT}=f_{BTZ}(\zeta)=\textrm{e}^{2ib_{th}\zeta}
    \label{thermaltopoincare}
\end{equation}
with $b_{th}=\pi/(2h)$. The boundary manifold is now mapped to a half annulus in $\xi_{CFT}$, $\bar{\xi}_{CFT}$ coordinates centered in the origin and with $\Im (\xi)\leq 0$ (see Figure \ref{halfannulus}). 
The inner circle with radius $r_{in}=\exp(-\pi^2/h)$ and the outer circle with radius $r_{out}=\exp(\pi^2/h)$ are identified. The boundaries of the finite cylinder in $\zeta=x+i\tau$ coordinates at $\Re (\zeta)=-h$, $\Re(\zeta)=0$ where the connected brane anchors (corresponding to the $A_1$ and $A_2$ slits respectively in the original cylinder with slits) are mapped to $-\exp(\pi^2/h)<\xi<-\exp(-\pi^2/h)$ and $\exp(-\pi^2/h)<\xi<\exp(\pi^2/h)$, respectively. Using the map (\ref{thermaltopoincare}), the coordinate transformation (\ref{bulkconftransf}) takes the form
\begin{equation}
\begin{cases}
\xi=\frac{8h^2-\pi^2z^2}{8h^2+\pi^2z^2}\textrm{e}^{i\frac{\pi}{h}\zeta}\\
\bar{\xi}=\frac{8h^2-\pi^2z^2}{8h^2+\pi^2z^2}\textrm{e}^{-i\frac{\pi}{h}\bar{\zeta}}\\
\eta=\frac{4\sqrt{2}\pi h z}{8h^2+\pi^2z^2}\textrm{e}^{i\frac{\pi}{2h}(\zeta-\bar{\zeta})}
\end{cases}
\label{thermalcoord}
\end{equation}
The functions $L$, $\bar{L}$ in equation (\ref{lfunctions}) are given by $L(\zeta)=\bar{L}(\bar{\zeta})=-b_{th}^2$, leading to the thermal $AdS_3$ metric (\ref{btzthermalmetrics}). The bulk domain in Poincar\'e-AdS coordinates is now given by
\begin{equation}
    \textrm{e}^{-\frac{2\pi^2}{h}}\leq \xi\bar{\xi}+\eta^2\leq \textrm{e}^{\frac{2\pi^2}{h}} \hspace{0.5cm} \land \hspace{0.5cm} \Im (\xi)\leq\frac{RT}{\sqrt{1-R^2T^2}}\eta
    \label{thermalpoincaredomain}
\end{equation}
where the surfaces of the two hemispheres at $\xi\bar{\xi}+\eta^2=\exp(-2\pi^2/h)$ and $\xi\bar{\xi}+\eta^2=\exp(2\pi^2/h)$ are identified. The connected ETW brane sits on the plane $\Im (\xi)=RT/\sqrt{1-R^2T^2}\eta$ and intersects the two concentric hemispheres. For $T=0$, the brane cuts the hemispheres in half, and the bulk domain is given by half of the region between them. Increasing the tension to positive values, the plane of the brane tilts towards positive values of $\Im(\xi)$ and more than half of the region between the hemispheres is part of the domain. For $T\to T_c$ the whole region is retained. Decreasing the tension to negative values, the plane of the brane tilts towards negative values of $\Im(\xi)$ and less than half of the region between the hemispheres is part of the domain. For $T\to -T_c$, the domain approaches the empty set (see Figure \ref{fig:thermal_annulusc}).\footnote{We remind that we restrict to negative values of the tension $T>T_*>-T_c$, where no brane intersections occur in the BTZ phase.}

\subsection{Mapping to the original $\lambda$ coordinates}
\label{sec:mapback}

\begin{figure}[h]
    \centering
    \includegraphics[width=.75\textwidth]{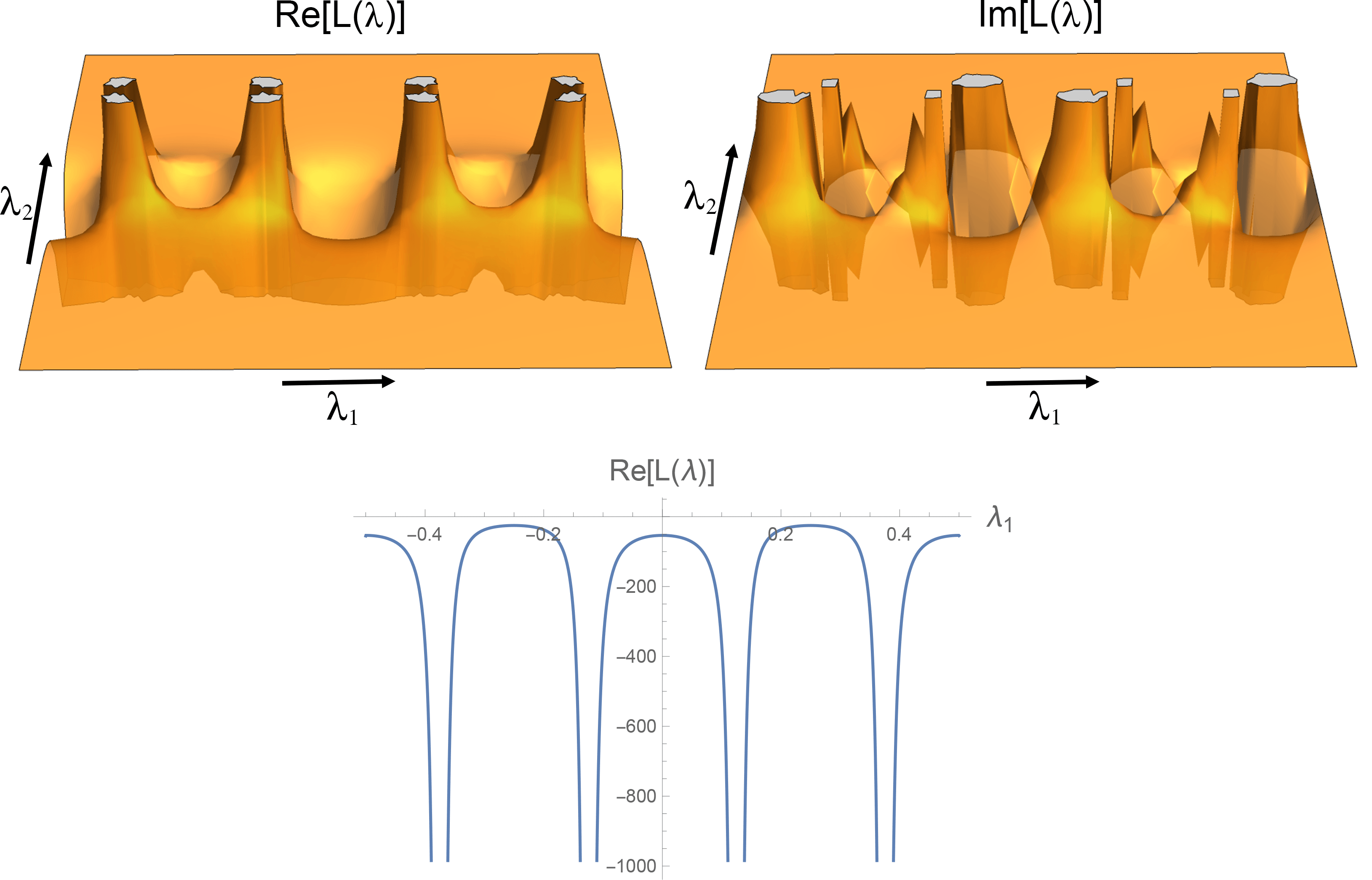}
    \caption{Top: Real (left) and imaginary (right) parts of $L(\lambda)$ in the thermal $AdS_3$ phase. The metric (\ref{mappedmetric}) is singular at the values of $\lambda$ corresponding to the endpoints of the slits (i.e. $\lambda_2=0$ and $\lambda_1=\{-L/2+\ell/2, -\ell/2, \ell/2, L/2-\ell/2\}$) for any value of $z_\lambda$, and approaches a constant value away from the $\lambda_2=0$ slice where the measurement is performed. Bottom: Real part of $L(\lambda)$ in the thermal $AdS_3$ phase as a function of $\lambda_1$ on the $\lambda_2=0$ slice. The singularities at the endpoints of the slits (here $\ell=s=0.25$) are well visible.}
    \label{thermallfunctions}
\end{figure}

Now that we have obtained the domains in Poincar\'e coordinates, we could map them back to the original $\lambda$, $\bar{\lambda}$ coordinates to obtain the bulk dual of the theory on the cylinder with slits. In order to do that, we must first express $\xi$, $\bar{\xi}$, $\eta$ as a function of $\lambda$, $\bar{\lambda}$, $z_\lambda$\footnote{We indicate here the radial coordinate $z$ appearing in equation (\ref{bulkconftransf}) with $z_\lambda$ to distinguish it from the $z$ in BTZ/thermal $AdS_3$ coordinates.} using the coordinate transformation (\ref{bulkconftransf}). In this case, $f$ is given by the composed map leading from the $\lambda$ coordinates to the Poincar\'e coordinates $\xi$:
\begin{equation}
    f_{btz}(\lambda)=\exp\left\{\frac{h}{2}\left[\frac{\sn^{-1}\left(\frac{\tan(\pi\lambda/L)}{\tan(\pi\ell/(2L))},k^2\right)}{\mathcal{K}(k^2)}-1\right]\right\}
\end{equation}
for the BTZ phase and 
\begin{equation}
     f_{th}(\lambda)=\exp\left\{i\frac{\pi}{2}\left[\frac{\sn^{-1}\left(\frac{\tan(\pi\lambda/L)}{\tan(\pi\ell/(2L))},k^2\right)}{\mathcal{K}(k^2)}-1\right]\right\}
\end{equation}
for the thermal $AdS_3$ phase. Using equation (\ref{lfunctions}) we can then compute the functions $L(\lambda)$, $\bar{L}(\bar{\lambda})$, which uniquely determine the metrics (\ref{mappedmetric}) in the two phases. The explicit expression for $L(\lambda)$, $\bar{L}(\bar{\lambda})$ are involved and we will not report them here, but the real and imaginary parts of $L(\lambda)$ are plotted in Figures \ref{btzlfunctions} and \ref{thermallfunctions}. As expected, the metrics are singular at the values of $\lambda$ corresponding to the endpoints of the slits, i.e. $\lambda_2=0$ and $\lambda_1=\{-L/2+\ell/2, -\ell/2, \ell/2, L/2-\ell/2\}$. In particular, the metric is singular for these values of $\lambda_1,\lambda_2$ for any value of $z_\lambda$, and regular everywhere else. A plot of $\Re [L(\lambda)]$ for $\lambda_2=0$ emphasizing this feature is also reported in Figures \ref{btzlfunctions} and \ref{thermallfunctions} ($\Im[L(\lambda)]=0$ for $\lambda_2=0$). 

The bulk domains in $\lambda$ coordinates are now implicitly given by the conditions (\ref{BTZpoincaredomain}) and (\ref{thermalpoincaredomain}) for the two phases. However, the singularities arising in our setup prevent us from numerically obtaining such domains in a consistent way, which would anyway not be of great physical significance because the singularities in the metric would not allow us to have a sensible interpretation of the resulting bulk spacetimes. Nonetheless, we used this mapping procedure to confirm numerically our expectations for the image of the $\lambda_2=0$ slice in BTZ/thermal $AdS_3$ geometries, as we have discussed in Section \ref{sec:connectivity}. We also remark that this procedure can be employed in more generic situations to build the bulk dual of arbitrary BCFT setups. It would be interesting to carry out this final mapping of the domains in a regularized version of our setup, where the slits have a finite height and the $\lambda_2=0$ slice is therefore not singular. We leave an analysis of such setup to future work.

\newpage
\bibliographystyle{jhep}
\bibliography{references}

\end{document}